\documentclass{aa}  
\usepackage{array}
\usepackage{graphicx}
\usepackage{amsmath}
\usepackage{natbib}
\usepackage[version=3]{mhchem}
\usepackage[colorlinks, linkcolor=blue, citecolor=blue]{hyperref}
\usepackage{txfonts}
\titlerunning{Structure of photodissociation fronts revealed by high-J CO emission lines}

\newcommand{\um}{\,$\mu$m\,}
\newcommand{\hii}{H\,{\sc ii}\,}

\begin{document}

\title{Structure of photodissociation fronts in star-forming regions revealed by observations of high-J CO emission lines with Herschel\thanks{{\it Herschel } is an ESA space observatory with science instruments provided by European-led Principal Investigator consortia and with important participation from NASA.}} 

\author{
C.~Joblin\inst{1} \and
E.~Bron\inst{2,3} \and
C.~Pinto\inst{4} \and
P.~Pilleri\inst{1} \and
F.~Le Petit\inst{3} \and
M.~Gerin\inst{3} \and
J.~Le Bourlot\inst{3,5} \and
A.~Fuente\inst{6} \and
O.~Berne\inst{1} \and
J.~R.~Goicoechea\inst{2} \and
E.~Habart \inst{7} \and
M.~K\"ohler \inst{7} \and
D.~Teyssier\inst{8} \and
Z.~Nagy \inst{9} \and
J.~Montillaud\inst{10} \and
C.~Vastel\inst{1} \and
J.~Cernicharo\inst{2} \and
M.~R\"ollig\inst{9} \and
V.~Ossenkopf-Okada (WADI team) \inst{9} \and
E.~A.~Bergin (HEXOS team) \inst{11}
}

\institute{ IRAP, Universit\'e de Toulouse, CNRS, UPS, CNES, 9 Av. colonel Roche, BP 44346, 31028 Toulouse Cedex 4, France\\
\email{christine.joblin@irap.omp.eu}
\and
Instituto de Fisica Fundamental (CSIC), Calle Serrano 121-123, 28006, Madrid, Spain
\and
LERMA, Observatoire de Paris, PSL Research University, CNRS, Sorbonne Universit\'es, UPMC Univ. Paris 06, F-92190, Meudon, France
\and
Aix-Marseille Universit\'e, CNRS, LAM (Laboratoire d'Astrophysique de Marseille) UMR 7326, 13388 Marseille, France
\and
Universit\'e Paris-Diderot, Paris, France
\and
Observatorio Astron\'omico Nacional, Apdo. 112, 28803 Alcal\'a de Henares, Madrid, Spain
\and
Institut d'Astrophysique Spatiale (IAS), Universit\'e Paris Sud \& CNRS, 91405 Orsay, France
\and
European Space Astronomy Centre, ESA, PO Box 78, 28691 Villanueva de la Ca\~nada, Madrid, Spain
\and
I. Physikalisches Institut der Universit\"at zu K\"oln, Z\"ulpicher Strasse 77, 50937 K\"oln, Germany
\and 
Institut Utinam, CNRS UMR 6213, OSU THETA, Universit\'e de Franche-Comt\'e, 41bis avenue de l'Observatoire, 25000 Besan\c{c}on, France
\and
Department of Astronomy, University of Michigan, 311 West Hall, 1085 S. University Avenue, Ann Arbor, MI 48109, USA
}

\date{Working draft}

\abstract
{
In bright photodissociation regions (PDRs) associated to massive star formation, the presence of dense "clumps" that are immersed in a less dense interclump medium is often proposed to explain the difficulty of models to account for the observed gas emission in high-excitation lines.
}
{
We aim at presenting a comprehensive view of the modeling of the 
CO rotational ladder in PDRs, including the high-J lines that trace warm molecular gas at PDR interfaces.}
{
We observed the $^{12}$CO and $^{13}$CO ladders in two prototypical PDRs, the Orion Bar and NGC 7023 NW using the instruments onboard {\it Herschel }.
We also considered line emission from key species in the gas cooling of PDRs (C$^+$, O, H$_2$) and other tracers of PDR edges such as OH and CH$^+$. All the intensities are collected from Herschel observations, the literature and the Spitzer archive and are analyzed using the Meudon PDR code.
}
{
A grid of models was run to explore the parameter space of only two parameters: thermal gas pressure and a global scaling factor that corrects for approximations in the assumed geometry. 
We conclude that the emission in the high-J CO lines, which were observed up to J$_{up}$=23  in the Orion Bar (J$_{up}$=19 in NGC\,7023), can only originate from small structures of typical thickness of a few $10^{-3}\,\mathrm{pc}$ and
at high thermal pressures ($P_{\mathrm{th}}\sim 10^8\,\mathrm{K~cm}^{-3}$). 
}
{
Compiling data from the literature, we found that the gas thermal pressure increases with the intensity of the UV radiation field given by $G_0$, following a trend in line with recent simulations of the photoevaporation of illuminated edges of molecular clouds.
This relation can help rationalising the analysis of high-J CO emission in massive star formation and provides an observational constraint for models that study stellar feedback on molecular clouds.
}
%
\keywords{(ISM:) photon-dominated region (PDR) - ISM:individual objects:Orion Bar - ISM:individual objects:NGC7023 - ISM: lines and bands
- Infrared, Submillimeter: ISM - Molecular processes}

\maketitle

\section{Introduction}
\label{Sec:intro}

Photodissociation regions (PDRs) are key regions in the study of the interstellar medium. They are the interfaces between molecular gas, where stars form, and the surrounding galactic medium \citep[see review by][]{Hollenbach99}. Dissociating UV photons produced by young stars are absorbed in PDRs by dust and gas allowing a transition from the atomic phase to the molecular phase. Understanding the structure and physical and chemical processes in PDRs is necessary to constrain stellar feedback on molecular clouds, but also to be able to interpret observations of distant galaxies where the contribution of unresolved PDRs dominate the IR spectrum. Intense emission from fine-structure lines of C$^+$, O, C, as well as H$_2$ rotational and rovibrational transitions and CO rotational transitions can be observed in PDRs. It is now admitted that emission in these atomic and molecular lines is mainly induced by the heating of the gas by UV photons from nearby massive stars involving the photo-electric effect on grains \citep{Bakes94, Weingartner01} as well as the collisional de-excitation of H$_2$ excited by UV pumping \citep{Sternberg89, Burton90, Rollig06}. Several stationary PDR models have been developed to analyse the emission in these lines \citep{Tielens85, Sternberg89, Kaufman99, LePetit06} and to constrain the chemical and physical processes that take place in PDRs. Over the years, these models that simulate radiative transfer, chemistry, thermal processes of the gas and dust have progressed in their description of the microphysics and chemical processes at play in these regions \citep{Rollig07}. Today, some of these codes as the Meudon PDR code, can simulate very detailed micro-physical processes as input and output line and continuum radiative transfer, non local pumping by infrared photons and detailed surface chemistry with stochastic processes \citep{Goicoechea07, Gonzalez08, LeBourlot12, Bron14, Bron16}.

The analysis of line emission in PDRs is intimately connected to considerations on the morphology of these regions. Earlier observations of a few mid-/high-J CO lines from ground-based submillimeter and airborne far-IR observations have suggested that PDR interfaces reach high temperatures and densities both in the atomic and molecular gas, which requires the interface to be clumpy or filamentary \citep{Stutzki88}. This led \cite{Burton90} to propose a clumpy PDR model in which dense clumps ($n_\textrm{H}$ $\sim$ $10^6$ - $10^7$ cm$^{-3}$) are embedded in a less dense interclump medium ($n_\textrm{H}$ $\sim$ $10^3$ - $10^5$ cm$^{-3}$). In following studies on the bright PDRs M17 SW, Orion Bar and Carina, clumpy PDR models were used to analyse quite successfully the observations \citep{Meixner92, Hogerheijde95, Kramer08,Andree-Labsch17}. 

The Herschel Space Observatory, with its three instruments HIFI, SPIRE and PACS \citep{Pilbratt10}, has opened the possibility to observe, more systematically and continuously  in wavelengths, the warm molecular gas in galactic and extragalactic sources by covering all CO excitation lines from J$_{up}$ = 4 to J$_{up}$ = 50. This allows us to build full CO spectral line energy distributions (SLEDs)  including high-J levels. Such CO SLEDs have been particularly studied in star-forming regions in order to provide information on the energetic processes acting in these objects.
This includes low- and high-mass protostars \citep{Yildiz10, Visser12, Manoj13, Karska13, Goicoechea15} and a couple of PDRs associated with young stars of high or intermediate mass \citep{Koehler14, Stock15}. Studies were also performed on the Galactic Center with observations towards Sgr A* \citep{Goicoechea13}, as well as in the Sgr B2 cores, B2(M) and B2(N) \citep{Etxaluze13}. In external galaxies, CO SLEDs have been obtained for Seyfert galaxies, starburst galaxies, (ultra)luminous infrared galaxies ((U)LIRGs) and AGNs \citep{Rangwala11, Hailey12, Greve14, Kamenetzky14, Mashian15, Rosenberg15, Wu15}.

Modelling CO SLEDs in protostars or in active regions of galaxies is complicated by the fact that mechanical heating due to shocks (supernova explosions or stellar winds) is likely to be a source of energy powering this CO emission \citep[see in particular][]{Kazandjian12, Kazandjian15}.
In PDRs, UV photons are expected to be the major actor and these CO lines offer the possibility to further constrain the gradients in gas density and temperature, as well as the underlying excitation processes. Using a PDR model, \cite{Stock15} analysed the full CO SLEDs of two PDRs generated by massive star formation, S~106 and IRAS~23133+6050. Their best results were obtained with a two-component model comprising high-density clumps (n$_\textrm{H}$ $\sim 10^6$\,cm$^{-3}$) immersed in a strong far-UV radiation field \citep[G$_0$ $\sim 10^5$ in units of the Habing field;][]{Habing68} and an interclump medium at lower density and irradiation field (n$_\textrm{H}$ $\sim 10^4$\,cm$^{-3}$, G$_0$ $\sim 10^4$). However, the high value derived for the UV field on the clump (a factor of 10 higher compared to the interclump medium) is striking. 

In this work, we try to bring some new insights into this subject. We present a study of two prototypical PDRs, NGC~7023 NW and the Orion Bar. The Orion Bar is probably the most studied PDR \citep{Tielens93}.
NGC~7023 NW is well known to exhibit bright H$_2$ filaments at the interface with the illuminating star \citep{Lemaire96}. The two objects have been extensively studied in the past but often using a limited set of tracers. Here, we take benefit of the Herschel HIFI, SPIRE and PACS data. In order to include further tracers of the warm/hot molecular gas, we
complete this dataset with ancillary data coming from ground-based instruments as well as by ISO and Spitzer space missions. The full data sets include mid-/high-J CO, H$_2$, CH$^+$, [CII], [OI], [CI], HD, OH and HCO$^+$ lines. We analyse these observations using the latest version of the Meudon PDR code \citep{LePetit06}. Our goal is to get better insights on the structure of the irradiated interface at the border of PDR and to determine whether UV photons alone can explain the high-J excited CO lines observed in PDRs.

The article is organised as follows. In Sect.\,\ref{Sec:obj}, we compile information from the literature on the two PDRs of interest. The observations are described in Sect.\,\ref{Sec:obs}, including new Herschel data as well as complementary data gathered from the literature or archives. The CO observations are described in Sect.\,\ref{Sec:ObsData}, including both line profiles and intensity ladders. In Sect.\,\ref{Sec:Models}, we describe the Meudon PDR model and the procedure used to fit the observational data, and present our modeling results. We used isobaric models to mimick the density gradient at the interface. In Sect.\,\ref{Sec:Discussion}, we revisit the impact of our analysis on the interpretation of the emission in the mid-/high-J CO lines and conclude on the presence of sharp PDR interfaces at high thermal pressure that were likely shaped by the UV radiation field.

\begin{figure} [ht]
\begin{center}
\includegraphics[width=0.45\textwidth]{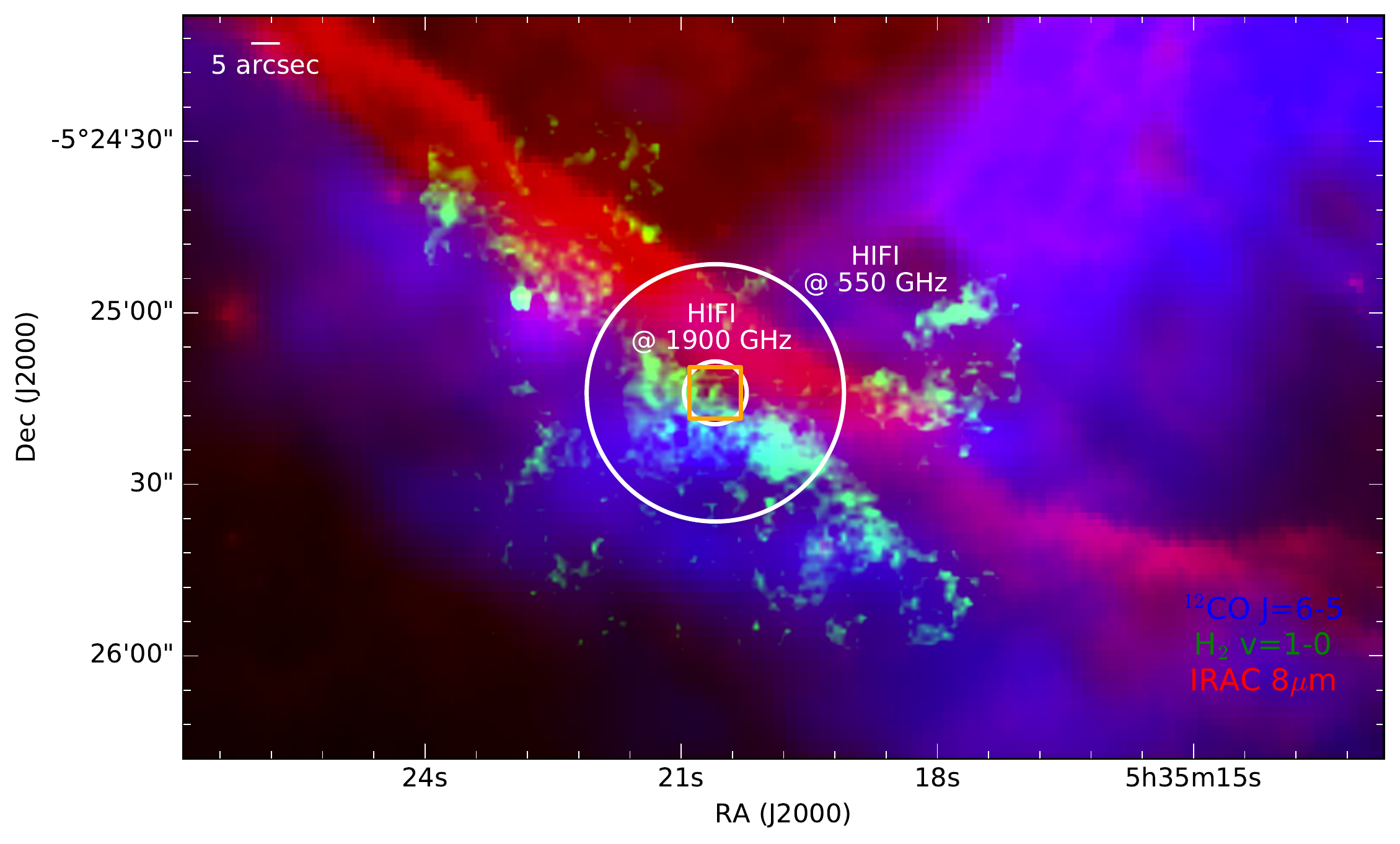}
\includegraphics[width=0.45\textwidth]{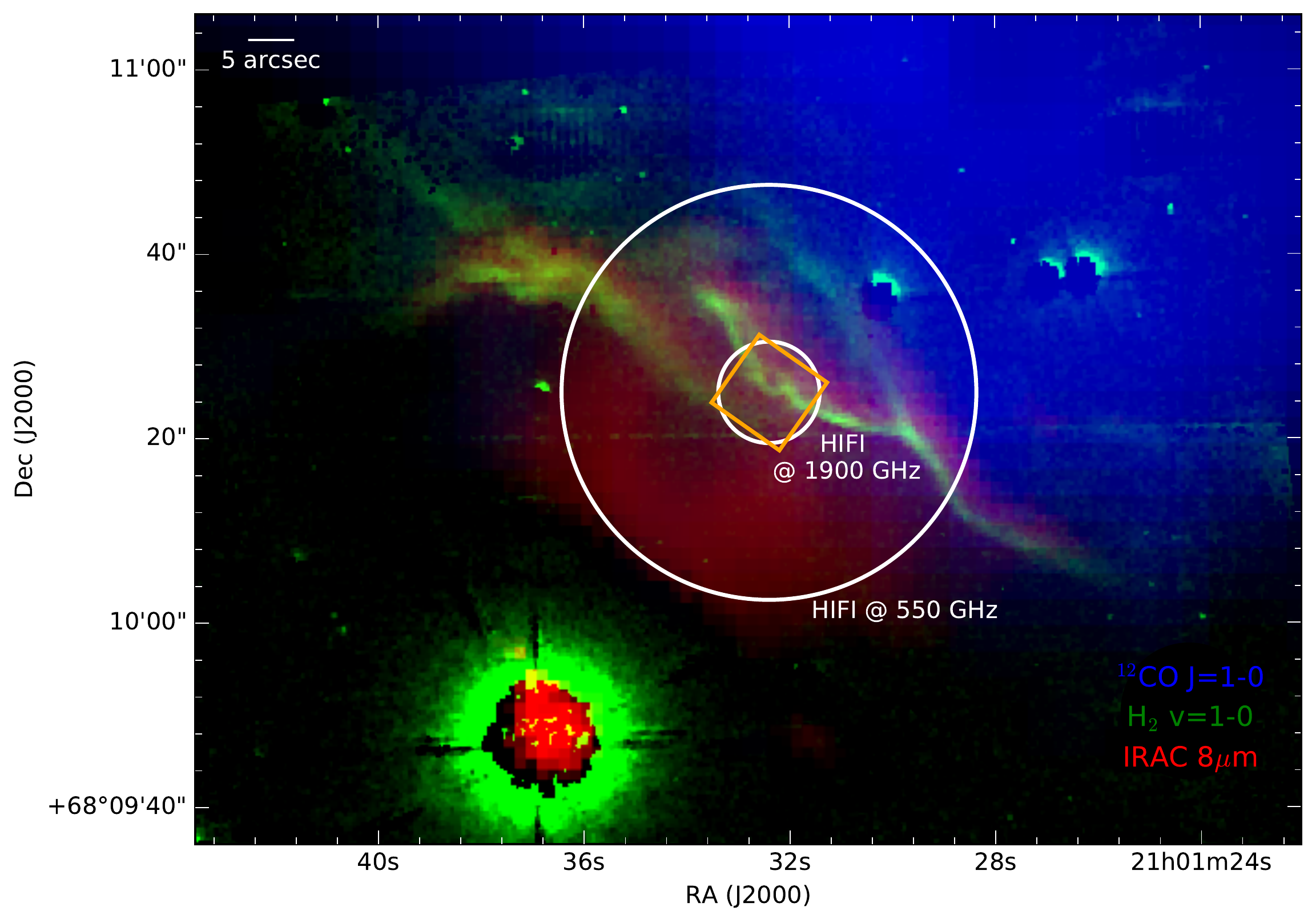}
\end{center}
\caption{Composite images of the Orion Bar (top) and NGC 7023 (bottom). Red indicates the 8 $\mu$m emission observed with {\it Spitzer}. Green shows the vibrationally excited emission of H$_2$ from \cite{Lemaire96} and \cite{VanDerWerf96} for NGC 7023 and Orion Bar, respectively. Blue shows the $^{12}$CO emission, J = 6-5 for the Orion Bar \citep{Lis03} and $J=1-0$ for NGC 7023 \citep{Gerin98}. 
The circles represent the HPBW of {\it Herschel} at 550 GHz and 1900 GHz. 
The square indicates the position of the central spaxel of the PACS observations.  
\label{Fig:Images_OrionBar_N7023}}
\end{figure}

\begin{figure} [ht]
\begin{center}
\includegraphics[width=0.45\textwidth]{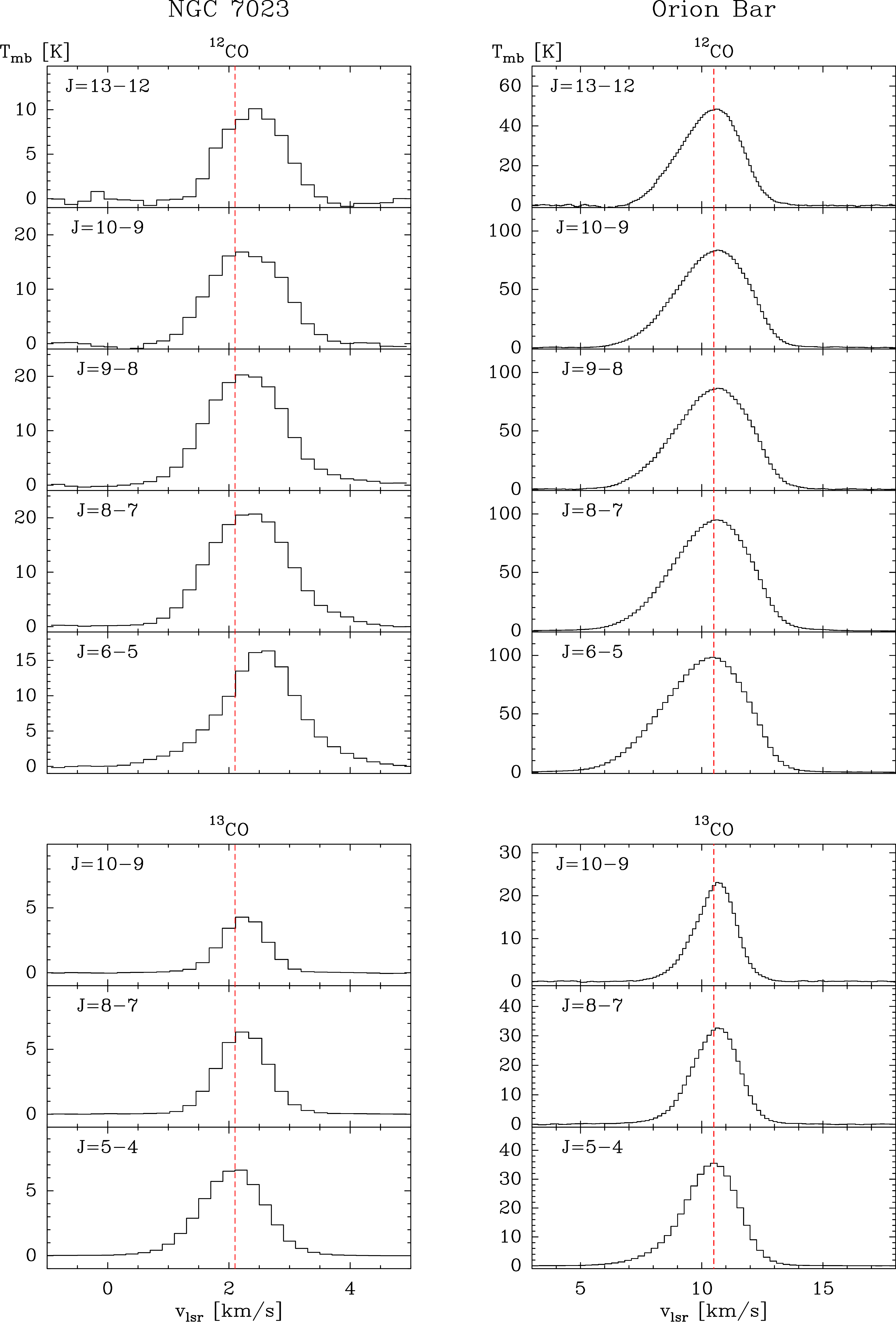}
\end{center}
\caption{$^{12}$CO and $^{13}$CO lines observed with HIFI toward NGC 7023 (left) and the Orion Bar (right). The vertical red line indicates the systemic velocity of the source.  \label{Fig:Line_profiles_CO_13CO}}
\end{figure}

\section{The prototypical PDRs: NGC~7023 and Orion Bar}
\label{Sec:obj}

\subsection{NGC~7023}\label{sec:Object_NGC7023}
NGC 7023 is a reflection nebula in the Cepheus constellation illuminated by HD~200775 [RA(2000) = 21h01m36.9s~; Dec(2000) = +68$^\circ$09'47.8"], a spectroscopic binary system (B3Ve and B5, see \citealt{Alecian08}). Its distance from the Sun was measured by Hipparcos at $430_{-90}^{+160}\,\mathrm{pc}$ in the 1997 catalogue \citep{VanDenAncker97}. The new reduction by \cite{VanLeeuwen07} gave $520\pm150\,\mathrm{pc}$, whereas \cite{Benisty13} have proposed a distance of $320\pm51\,\mathrm{pc}$ based on a study of the orbital parameters of the spectroscopic binary system HD~200775, which we adopt in the following.  At this distance, 1\arcsec corresponds to a physical length of $1.6\times10^{-3}$pc. 

\cite{Chokshi88} observed the [CII] 158\,$\mu$m and [OI] 63\,$\mu$m lines in NGC~7023 and derived at the emission peak located 50\arcsec~NW from the star a UV field of $G_0 = 2600$ and a density of $n_\textrm{H} \sim 4\times 10^3\,\mathrm{cm}^{-3}$. Later observations have shown that the star formation process has shaped a cavity inside the molecular cloud with walls consisting of dense PDRs, at the north-west (NW), at the south and at the east \citep{Fuente92, Rogers95, Fuente98, Gerin98}. Very bright thin filaments were revealed by high angular resolution images in the Extended Red Emission (ERE), vibrationally excited H$_2$ emission lines \citep{Sellgren92, Lemaire96, Witt06} and in HCO$^+$ millimeter lines \citep{Fuente96}. These structures also correspond to enhanced emission in polycyclic aromatic hydrocarbon (PAH) emission \citep{An03, Berne07}, in [O I] (63 and 145\,$\mu$m) and [CII] (158\,$\mu$m)  lines \citep{Bernard-Salas15}, and in warm CO and dust emission \citep{Rogers95, Gerin98, Koehler14, Bernard-Salas15}. From these observations emerges a picture of the NGC~7023 NW morphology in which the PDR interface is made of filamentary structures at high-density, $n_\textrm{H} \sim 10^5-10^6\,\mathrm{cm}^{-3}$ \citep{Martini97, Lemaire96, Fuente96}, which are embedded in a more diffuse gas with $n_\textrm{H} \sim 10^3-10^4\,\mathrm{cm}^{-3}$ \citep{Chokshi88,Rogers95}. 
This filamentary structure is observed at small spatial scales of 0.004 pc or less. Whereas it is composed of detached filaments or the result of compressed sheets is still unclear  \citep{Lemaire96, Fuente96}. 
Because of its geometry, brightness and proximity, the NW PDR of NGC~7023 turned out to be one of the best sites to study the physical and chemical processes taking place in a PDR.

\subsection{The Orion Bar}

The Orion Bar PDR lies $\sim 2 \arcmin$ SE of  the Trapezium stars cluster: $\theta^1$ Orionis C, A and E 
\citep{SimonDiaz06, Allers05},  a cluster of O  and B stars that creates a \hii region that penetrates into the parent molecular cloud.
The UV intensity impinging on the PDR has been estimated to be $G_0 = 1-4\times10^4$ in Habing units \citep{Tielens85, Marconi98}. 
The distance of the Orion nebula has been measured with great precision by \cite{Menten07} from trigonometric parallax, yielding a value of $414\pm7\,\mathrm{pc}$. An angular distance of 1 arsecond corresponds therefore to $2\times10^{-3}$ pc.
Because of its proximity and edge-on geometry, the Bar is one of the most studied PDRs. It is the prototype PDR associated with massive-star formation, which can be used as a template for more distant regions including extragalactic studies.
\cite{Hogerheijde95} reported spatial observations of the Bar in rotational transitions from a variety of molecules and concluded that the morphology of the molecular emission is mainly due to the geometry of the Bar that changes from face-on to almost edge-on and then face-on. The bright PDR corresponds to the edge-on part. The overall observed spatial stratification of the Bar was found to require an average density of at least $5\times 10^4$ cm$^{-3}$ in order to account for the observed offsets of ionization front and molecular lines \citep[see also][]{Wyrowski97}. Most models of the molecular emission in the Orion Bar used high-density clumps (n$_\textrm{H} \sim 10^6-10^7$ cm$^{-3}$) embedded in a lower density gas ($n_\textrm{H} = 5\times10^4-10^5$ cm$^{-3}$). Clumps were introduced first to explain the emission of excited lines of CO and warm H$_2$ emission \citep{Parmar91,Tauber94, VanDerWerf96} but were also found to be necessary to model the HCO$^+$, HCN \citep{YoungOwl00} and OH \citep{Goicoechea11} emission at the PDR interface.

The interface with the \hii region is of special interest if one wants to study the feedback of star formation on the molecular cloud.
\cite{Omodaka94} have discussed that the Bar has been shaped by shock compression related to the expansion of the \hii region.
\cite{Giard94} mapped the 3.3\,\um PAH emission and revealed a clumpy structure of the gas down to scales of a few 10$^{-3}$\,pc behind the ionization front and in front of the molecular interface traced by vibrational H$_2$ \citep{Tielens93}. This is also well illustrated in Fig.\,\ref{Fig:Images_OrionBar_N7023} that combines the IRAC 8\um map dominated by PAH emission with the H$_2$ map of \cite{VanDerWerf96}.
\cite{Fuente96_Orion} observed the molecular interface and concluded that it consists of a corrugated dense layer ($n_\textrm{H} \sim 10^6$ cm$^{-3}$). \cite{Walmsley00} proposed that it can be described by a filament using a cylindrical shape. \cite{Andree-Labsch17} showed that this shape can be excluded due to the visible shadowing pattern seen in the maps.
Recently, the Orion Bar has been observed in CO and HCO$^+$ by ALMA \citep{Goicoechea16}. The sharp edge ($\sim$ 2 \arcsec) at which the CO and HCO$^+$ emission start to occur coincides with the bright H$_2$ vibrational emission. The H/H$_2$ and C$^+$/C/CO transitions should therefore be very close at this interface, which is incompatible with stationary PDR models at $n_\textrm{H} \sim 5\times 10^4$ cm$^{-3}$.


\begin{figure*} [ht]
\begin{center}
\includegraphics[width=0.45\textwidth]{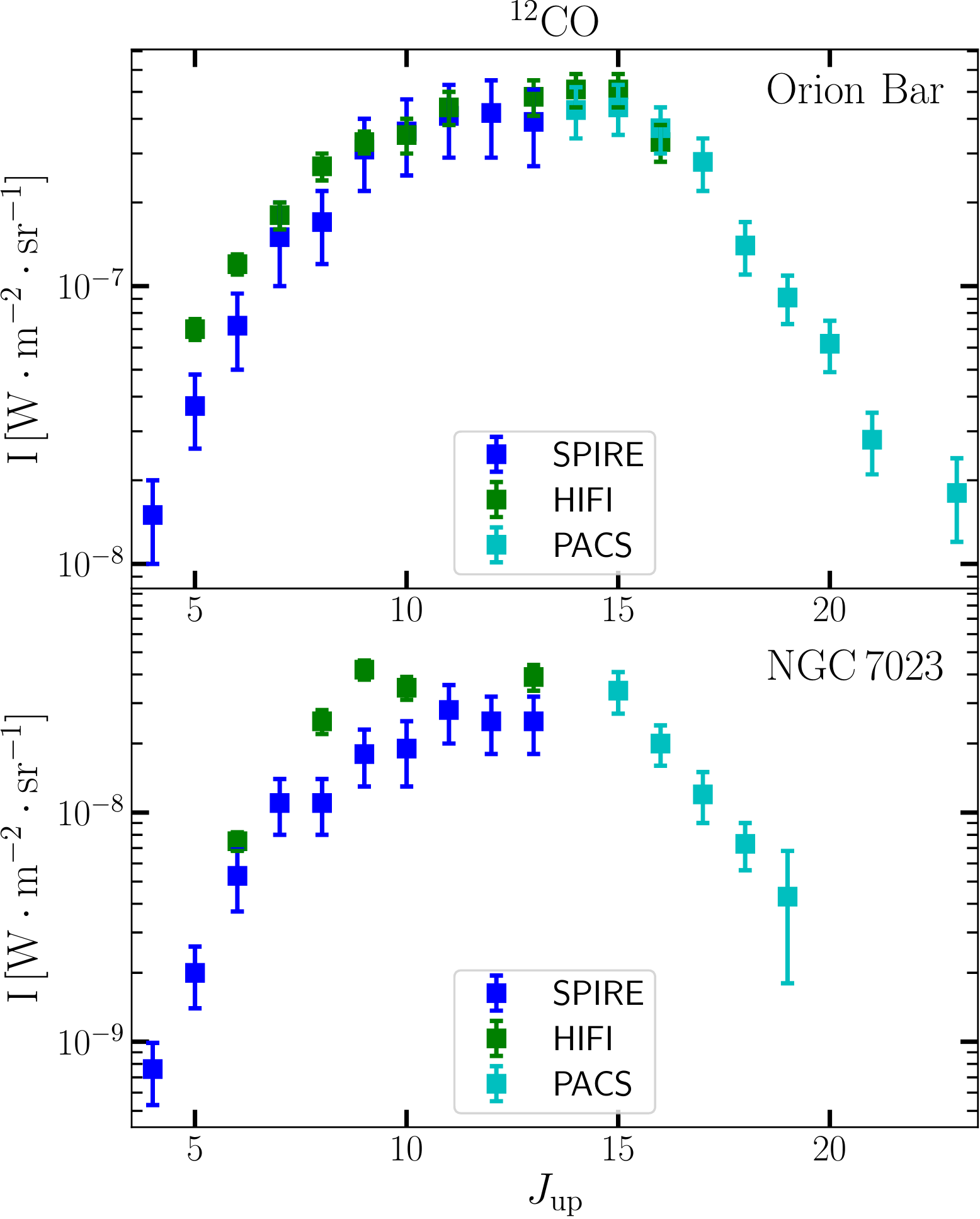}
\includegraphics[width=0.45\textwidth]{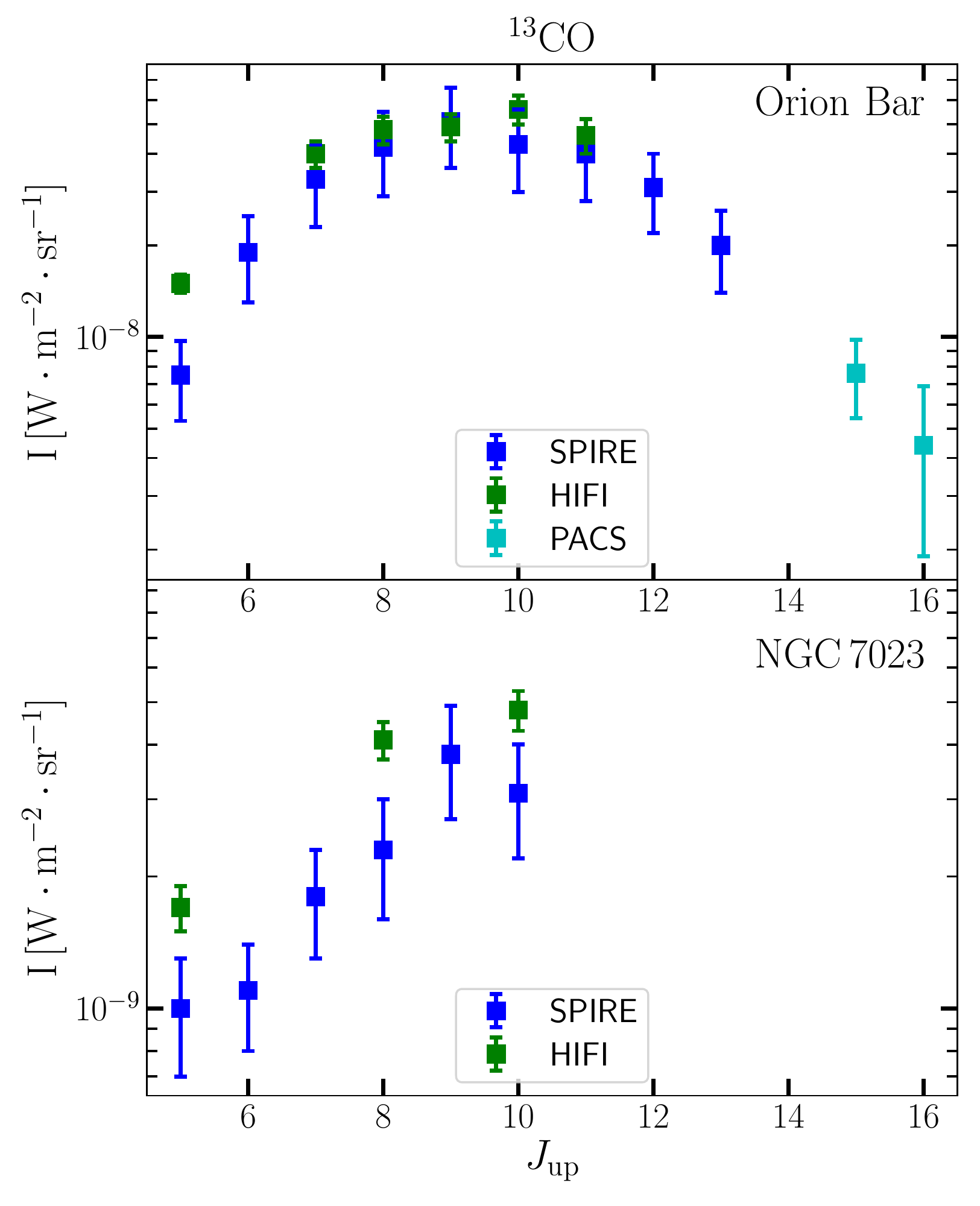}
\end{center}
\caption{Observed intensities of $^{12}$CO (left) and $^{13}$CO (right) in the Orion Bar (top panel) and NGC~7023~NW (bottom panel). \label{Fig:12_CO_ladder}}
\end{figure*}

\section{Observations}
\label{Sec:obs}

The observations of NGC 7023 NW were programmed in the framework of the WADI  \citep{Ossenkopf11} Guaranteed Time Key Program (GTKP) and were centered on the bright H$_2$ filaments, a position referred to as the H$_2$ peak \cite[][$\alpha_{2000}$=$21^{\mathrm{h}}01^{\mathrm{m}}32.4^{\mathrm{s}}, \delta_{2000}$=$+68^{\circ}10\arcmin25.0\arcsec$]{Joblin10}. The Orion Bar was observed at the CO$^+$ peak position \citep[][$\alpha_{2000}$=$5^{\mathrm{h}}35^{\mathrm{m}}20.61^{\mathrm{s}}, \delta_{2000}$=$-5^{\circ}25\arcmin14.0\arcsec$; cf. Fig.\,\ref{Fig:Images_OrionBar_N7023}]{Stoerzer95} as part of the HEXOS GTKP \citep{Bergin10}. Earlier studies based on these data have discussed OH emission \citep{Goicoechea11} as well as CH$^{+}$ and SH$^{+}$ lines \citep{Nagy13}.
The WADI and HEXOS programs gathered spectroscopic data using HIFI and PACS instruments. We also use in this article spectroscopic data from SPIRE that were obtained on the same objects as part of the SAG\,4 GTKP \citep{Habart10,Koehler14}. Data from the litterature and archives were also collected in particular for H$_2$. The full data sets, which are used in this study, are reported in Tables\,\ref{Tab:NGC7023_raw_data} and \ref{Tab:Orion_raw_data} for NGC\,7023 and the Orion Bar, respectively.

\subsection{Herschel PACS observations}
The PACS range spectroscopy observations of NGC 7023 and the Orion Bar were reduced using HIPE version 10. 
We used the standard pipeline to extract the spectrum from the central spaxel for the blue and red channels, including defringing. 
We applied the correction for point sources, considering that the high-J CO emission originates from dense structures of arcsec size and therefore smaller than the spaxel size, which is 3.2$\times$3.2\,arcsec$^2$  (cf. Sec.\ref{Sect:obs_corrections}). This is still an approximation since the observed filamentary interface extends over several spaxels but there is no correction available for a semi-extended source.
We checked that the intensities of the lines of interest remain very similar when performing the reduction with a more recent version of HIPE (e.g. v 13). 

Line flux was calculated by fitting the lines observed on the central spaxel with a Gaussian function. Error bars were calculated by quadratically summing the different sources of errors: calibration error and error from the Gaussian fit associated with spectral rms. We used a  20\% error for the flux calibration, which is intermediate between the values of 10-30\% used in previous studies \citep{Bernard-Salas12, Bernard-Salas15, Okada13}.
\subsection{Herschel HIFI observations}

The HIFI observations were obtained using the Wide Band Spectrometer, that provides a spectral resolution of 1.1 MHz ($\sim0.6$ km\,s$^{-1}$ at 550 GHz). The half power beam width (HPBW) varies between 9\arcsec at high frequency (1900 GHz) and 39\arcsec at low frequency (550 GHz). 

For NGC~7023, the observations were focused on specific frequency ranges, and cover a number of CO lines, CH$^+$ (J=1-0 and 2-1), HCO$^+$ (J=6-5) and the [CII] lines. The data reduction for NGC 7023 was straightforward and consisted in the subtraction of a linear baseline for each scan and then in averaging all the scans, including both the H and V polarisations. The data reduction procedure  for the [CII] line is described in \citet{Ossenkopf13}.  
In the case of the Orion Bar, our observations consist in a spectral survey  performed at the CO$^+$ peak presented in \citet{Nagy13} and further discussed in \citet{Nagy17}.   

Line intensities for both sources were calculated by using the beam efficiencies of \cite{Roelfsema12} and by performing a Gaussian fit to the line profiles. \cite{Ossenkopf13} argued that $T_{\mathrm{A}}$ is more appropriate for extended emission, and $T_{\mathrm{mb}}$ for point sources. Emission in some lines is extended (e.g. [CII]), whereas it is not in others (e.g. high-J CO). In this work, we use the mean value of $T_{\mathrm{A}}$ and $T_{\mathrm{mb}}$ because the bright interface is in between point-source and extended. In addition, it is not practical to have a different treatment for every individual line.
Finally, the fit error being negligible ($<1\%$), our observational uncertainty is defined as the quadratic sum of the spectral rms and the flux calibration error, which is reported in \cite{Roelfsema12}.

A sample of the observed $^{12}$CO and $^{13}$CO  lines observed with Herschel-HIFI for both sources is shown in  Fig.\,\ref{Fig:Line_profiles_CO_13CO}. 

\subsection{Herschel SPIRE observations}
We have used the SPIRE FTS fully sampled maps to extract complementary data at the CO$^+$ peak. 
The data reduction is presented in details in \cite{Koehler14} for NGC~7023, and in \cite{Parikka17} 
for the Orion Bar. 
One important aspect is the use of the super-resolution method SUPREME (Ayasso et al., in prep.) 
to achieve higher spatial resolution than standard SPIRE observations, more specifically 
11.9$\arcsec$ at 200\,$\mu$m, 19.0$\arcsec$ at 400\,$\mu$m and 24.1$\arcsec$ at 600\,$\mu$m. 
The total error on the integrated line intensities was estimated to be 30\%, which includes the calibration 
uncertainties and the line fitting errors \citep[see][for more information]{Koehler14}.

\subsection{Additional data}

We used additional observations to further constrain our models. For NGC 7023, the pure rotational lines of H$_2$ were  observed with the low spectral resolution modules of Spitzer-IRS \citep{Werner04a, Houck04} and with ISO-SWS \citep{Degraauw96}. For the Spitzer data \citep{Werner04a}, we used the CUBISM reduction tool \citep{Smith07} including the slit-loss correction function for extended sources, considering that the filamentary interfaces in Orion Bar and NGC 7023 are narrow but extended sources; their width is indeed marginally resolved at the Spitzer spatial resolution. The line fluxes were then extracted by fitting Gaussian profiles to the spectra extracted toward the H$_2$ peak position over 4 pixels. 
Uncertainties take into account fit and calibration errors. The same lines were observed with ISO, and we used the intensity values reported in \citet{Fuente99}. The intensities for the vibrationally-excited lines of H$_2$, observed at the CFHT and the Perkins Telescope, are taken from \citet{Lemaire96}, \citet{Lemaire99} and \citet{Martini99}, respectively. Finally, we used the  HCO$^+$ $J=1-0$ observations obtained at the Plateau de Bure Interferometer \citep{Fuente96}, integrating line intensities between 1 and 5 km\,s$^{-1}$.

For Orion Bar, we used pure rotational H$_2$ lines from ISO-SWS data \citep[Bertoldi, private communication;][]{Habart04} and ground-based TEXES data \citep{Allers05}. The latter intensities are slightly lower compared to the ISO measurements. This can be  due to the fact that  the two instruments observed at different positions. The ISO-SWS data provides a position closer to the CO$^+$ peak but lower spatial resolution. For vibrationally excited H$_2$, we used data obtained with the BEAR instrument at the CFHT (Joblin, Maillard, Noel, unpublished) and the observations from \cite{VanDerWerf96}.


\begin{table*} [ht]
\begin{center}
\caption{Observed data for NGC 7023, and dilution factor $\Omega$. The reported intensities have not been corrected for beam dilution. \label{Tab:NGC7023_raw_data}} 
\scriptsize 
\begin{tabular}{l l r | l l l l |l}
\hline \hline
\multicolumn{3}{c}{Line} & \multicolumn{5}{c}{Observation data sets $[\mathrm{W} \mathrm{m}^{-2} \mathrm{sr}^{-1}]$} \\
\hline
Species                  & Transition       &    \multicolumn{1}{c}{Position}                  & SPIRE                 & HIFI                      & PACS                  & others                          & $\Omega$ (dilution factor) \\
\hline
$^{12}\mathrm{CO}$   &           &                         &                       &                           &                       &                                 &       \\
                     &  J=4-3    &    461.041 GHz          & $7.6 \pm 2.3\,(-10)$  & -                         & -                     & -                               & 0.05  \\
                     & J=5-4     &    576.268 GHz          & $2.0 \pm 0.6\,(-9)$   & -                         & -                     & -                               & 0.07  \\
                     & J=6-5    &    691.473 GHz          & $5.3 \pm 1.6\,(-9)$   & $7.5 \pm 0.7\,(-9)$       & -                     & -                               & 0.08  \\ 
                     & J=7-6     &    806.652 GHz          & $1.1 \pm 0.3\,(-8)$   & -                         & -                     & -                               & 0.10  \\
                     & J=8-7     &    921.800 GHz          & $1.1 \pm 0.3\,(-8)$   & $2.5 \pm 0.3\,(-8)$       & -                     & -                               & 0.11  \\
                     & J=9-8     &   1036.912 GHz          & $1.8 \pm 0.5\,(-8)$   & $4.2 \pm 0.4\,(-8)$       & -                     & -                               & 0.12  \\
                     & J=10-9    &   1151.985 GHz          & $1.9 \pm 0.6\,(-8)$   & $3.5 \pm 0.4\,(-8)$       & -                     & -                               & 0.14  \\
                     & J=11-10   &   1267.014 GHz          & $2.8 \pm 0.8\,(-8)$   & -                         & -                     & -                               & 0.15  \\
                     & J=12-11   &   1381.995 GHz          & $2.5 \pm 0.7\,(-8)$   & -                         & -                     & -                               & 0.17  \\
                     & J=13-12   &   1496.923 GHz          & $2.5 \pm 0.7\,(-8)$   & $3.9 \pm 0.5\,(-8)$       & -                     & -                               & 0.18  \\
                     & J=15-14   &   1726.603 GHz          & -                     & -                         & $3.4 \pm 0.7\,(-8)$   & -                               & 0.21  \\
                     & J=16-15   &   1841.345 GHz          & -                     & -                         & $2.0 \pm 0.4\,(-8)$   & -                               & 0.23  \\
                     & J=17-16   &   1956.018 GHz          & -                     & -                         & $1.2 \pm 0.3\,(-8)$   & -                               & 0.28  \\
                     & J=18-17   &   2070.616 GHz          & -                     & -                         & $7.3 \pm 1.7\,(-9)$   & -                               & 0.28  \\
                     & J=19-18   &   2185.135 GHz          & -                     & -                         & $4.3 \pm 2.5\,(-9)$   & -                               & 0.28  \\
$^{13}\mathrm{CO}$   &           &                         &                       &                           &                       &                                 &       \\
                     & J=5-4     &    550.926 GHz          &  $1.0 \pm 0.3\,(-9)$  &  $1.7 \pm 0.2\,(-9)$      & -                     & -                               & 0.07  \\
                     & J=6-5     &    661.067 GHz          &  $1.1 \pm 0.3\,(-9)$  &  -                        & -                     & -                               & 0.08  \\
                     & J=7-6     &    771.184 GHz          &  $1.8 \pm 0.5\,(-9)$  &  -                        & -                     & -                               & 0.09  \\
                     & J=8-7     &    881.273 GHz          &  $2.3 \pm 0.7\,(-9)$  &  $4.1 \pm 0.4\,(-9)$      & -                     & -                               & 0.11  \\
                     & J=9-8     &    991.329 GHz          &  $3.8 \pm 1.1\,(-9)$  &  -                        & -                     & -                               & 0.12  \\
                     & J=10-9    &   1101.350 GHz          &  $3.1 \pm 0.9\,(-9)$  &  $4.8 \pm 0.5\,(-9)$      & -                     & -                               & 0.13  \\
$\mathrm{CH}^+$      &           &                         &                       &                           &                       &                                 &       \\
                     & J=1-0     &    835.137 GHz          & -                     & $1.0 \pm 0.1\,(-9)$       & -                     & -                               & 0.10  \\
                     & J=2-1     &   1669.281 GHz          & -                     & $5.5 \pm 0.8\,(-9)$       & $6.6 \pm 2.5\,(-9)$   & -                               & 0.28  \\
                     & J=3-2     &   2501.440 GHz          & -                     & -                         & $5.6 \pm 2.1\,(-9)$   & -                               & 0.28  \\
$\mathrm{HCO}^+$     &           &                         &                       &                           &                       &                                 &       \\
                     & J=1-0     &     89.188 GHz          & -                     & -                         & -                     & $4.5 \pm 1.3\,(-12)^\mathrm{a}$ & 0.49  \\
                     & J=6-5     &    535.062 GHz          & -                     & $8.2 \pm 0.7\,(-11)$      & -                     & -                               & 0.06  \\
$\mathrm{C}^+$       &           &                         &                       &                           &                       &                                 &       \\
& $^2P_{3/2}$ - $^2P_{1/2}$      & $157.68\,\mu\mathrm{m}$ & -                     & $7.6 \pm 1.1\,(-7)$       & $7.3 \pm 1.5\,(-7)$   & $9.9 \pm 2.0\,(-7)^\mathrm{b}$  & 0.28  \\
$\mathrm{C}$         &           &                         &                       &                           &                       &                                 &       \\
             & $^3P_1$ - $^3P_0$  & 492.161 GHz   &  $2.8 \pm 0.8\,(-10)$ & -                         & -                     & -                               & 0.10  \\
$\mathrm{O}$         &           &                         &                       &                           &                       &                                 &       \\
             & $^3P_0$ - $^3P_1$  & $145.53\,\mu\mathrm{m}$    & -                     & -                         & $4.0 \pm 0.8\,(-7)$   & $3.8 \pm 0.8\,(-7)^\mathrm{b}$  & 0.28  \\
             & $^3P_1$ - $^3P_2$  & $ 63.18\,\mu\mathrm{m}$    & -                     & -                         & -                     & $1.8 \pm 0.4\,(-6)^\mathrm{b}$  & 0.28  \\
$\mathrm{HD}$        &           &                         &                       &                           &                       &                                 &       \\
                     & J=0-1     & $112.07\,\mu\mathrm{m}$ & -                     & -                         & $2.7\pm2.2\,(-9)$     & -                               & 0.28  \\
\hline
$\mathrm{H}_2$       &           &                         & ISO-SWS               &  {\it Spitzer}                  & CFHT $^\mathrm{c}$    & Perkins Telescope$^\mathrm{d}$  & -     \\
                     & 0-0 S(0)  & $28.22\,\mu\mathrm{m}$  & $3.4\pm1.0\,(-8)$     & -                         & -                     & -                               & 0.10  \\
                     & 0-0 S(1)  & $17.03\,\mu\mathrm{m}$  & $2.1\pm0.4\,(-7)$     & $2.0_{-0.3}^{+0.6}\,(-7)$ & -                     & -                               & 0.10$^{\mathrm{ISO}}$ / 0.20$^{\mathrm{{\it Spitzer} }}$   \\
                     & 0-0 S(2)  & $12.28\,\mu\mathrm{m}$  & $2.4\pm0.6\,(-7)$     & $5.5_{-1.7}^{+2.0}\,(-7)$ & -                     & -                               & 0.10$^{\mathrm{ISO}}$ / 0.55$^{\mathrm{{\it Spitzer} r}}$   \\
                     & 0-0 S(3)  & $ 9.66\,\mu\mathrm{m}$  & $4.1\pm1.0\,(-7)$     & $6.9_{-1.5}^{+3.1}\,(-7)$ & -                     & -                               & 0.10$^{\mathrm{ISO}}$ / 0.55$^{\mathrm{{\it Spitzer} }}$   \\
                     & 0-0 S(4)  & $ 8.02\,\mu\mathrm{m}$  & $1.5\pm0.4\,(-7)$     & -                         & -                     & -                               & 0.10$^{\mathrm{ISO}}$ / 0.55$^{\mathrm{{\it Spitzer} }}$   \\
                     & 0-0 S(5)  & $ 6.91\,\mu\mathrm{m}$  & $2.6\pm0.4\,(-7)$     & $4.6 \pm 1.4\,(-7)$       & -                     & -                               & 0.10$^{\mathrm{ISO}}$ / 0.55$^{\mathrm{{\it Spitzer} }}$   \\
                     & 1-0 S(1)  & $ 2.12\,\mu\mathrm{m}$  & -                     & -                         & $2.1\pm0.21\,(-7)$    & -                               & 1                                                   \\
                     & 1-0 S(2)  & $ 2.03\,\mu\mathrm{m}$  & -                     & -                         & $7.6\pm1.7\,(-8)$     & -                               & 1                                                   \\
         & 2-1 S(1) / 1-0 S(1)   &                         & -                     & -                         &  -                    & 0.29                            &                                                     \\
\hline
\end{tabular}\\
$^\mathrm{a}$ Fuente et al. (1996) - $^\mathrm{b}$ Bernard-Salas et al. (2015)  -  $^\mathrm{c}$ Lemaire et al. 1996, 1999 - $^\mathrm{d}$ Martini et al. 1999
\end{center}
\end{table*}

\newpage 

\begin{table*} [ht!]
\begin{center}
\caption{Observed data for the Orion Bar and dilution factor $\Omega$. The reported intensities have not been corrected for beam dilution.\label{Tab:Orion_raw_data}}
\scriptsize 
\begin{tabular}{l l r | l l l l | l}
\hline \hline
\multicolumn{3}{c}{Line} & \multicolumn{5}{c}{Integrated intensity $[\mathrm{W} \mathrm{m}^{-2} \mathrm{sr}^{-1}]$} \\
\hline
Species                  & Transition       &    \multicolumn{1}{c}{Position}            & SPIRE                         & HIFI                  & PACS                  & others                          &$\Omega$ \\
\hline
$^{12}\mathrm{CO}$                           &                 &                               &                       &                       &                                 & \\
& J=4-3                                      &     461.041 GHz & $1.5  \pm 0.5\,(-8)$          & -                     & -                     & -                               & 0.05 \\
& J=5-4                                      &     576.268 GHz & $3.7  \pm 1.1\,(-8)$          & $7.0  \pm 0.6\,(-8)$  & -                     & -                               & 0.07 \\
& J=6-5                                     &     691.473 GHz & $7.2  \pm 2.2\,(-8)$          & $1.2  \pm 0.1\,(-7)$  & -                     & -                               & 0.08\\
& J=7-6                                      &     806.652 GHz & $1.5  \pm 0.5\,(-7)$          & $1.8  \pm 0.2\,(-7)$  & -                     & -                               & 0.10\\
& J=8-7                                      &     921.800 GHz & $1.7  \pm 0.5\,(-7)$          & $2.7  \pm 0.3\,(-7)$  & -                     & -                               & 0.11\\
& J=9-8                                      &    1036.912 GHz & $3.1  \pm 0.9\,(-7)$          & $3.3  \pm 0.3\,(-7)$  & -                     & -                               & 0.12\\
& J=10-9                                     &    1151.985 GHz & $3.6  \pm 1.1\,(-7)$          & $3.5  \pm 0.5\,(-7)$  & -                     & -                               & 0.14\\
& J=11-10                                    &    1267.014 GHz & $4.1  \pm 1.2\,(-7)$          & $4.4  \pm 0.6\,(-7)$  & -                     & -                               & 0.15\\
& J=12-11                                    &    1381.995 GHz & $4.2  \pm 1.3\,(-7)$          & -                     & -                     & -                               & 0.17\\
& J=13-12                                    &    1496.923 GHz & $3.9  \pm 1.2\,(-7)$          & $4.8  \pm 0.7\,(-7)$  & -                     & -                               & 0.18\\
& J=14-13                                    &    1611.793 GHz &-                              & $5.1  \pm 0.7\,(-7)$  &  $4.3  \pm 0.9\,(-7)$ & -                               & 0.20\\
& J=15-14                                   &    1726.603 GHz & -                             & $5.1  \pm 0.7\,(-7)$  &  $4.4  \pm 0.9\,(-7)$ & -                               &0.21\\
& J=16-15                                   &    1841.345 GHz & -                             & $3.3  \pm 0.5\,(-7)$  &  $3.7  \pm 0.7\,(-7)$ & -                               &0.23\\
& J=17-16                                    &    1956.018 GHz & -                             &-                      &  $2.8  \pm 0.6\,(-7)$ & -                               &0.28\\
& J=18-17                                    &    2070.616 GHz & -                             &-                      &  $1.4  \pm 0.3\,(-7)$ & -                               &0.28\\
& J=19-18                                    &    2185.135 GHz & -                             &-                      &  $9.1  \pm 1.8\,(-8)$ & $1.1 \pm 0.2\,(-7)^\mathrm{a}$  &0.28\\
& J=20-19                                    &    2299.570 GHz & -                             &-                      &  $6.2  \pm 1.3\,(-8)$ & -                               &0.28\\
& J=21-20                                    &    2413.917 GHz & -                             &-                      &  $2.8  \pm 0.7\,(-8)$ & -                               &0.28 \\
& J=23-22                                    &    2642.330 GHz & -                             &-                      &  $1.8  \pm 0.6\,(-8)$ & -                               &0.28\\
$^{13}\mathrm{CO}$                           &                 &                               &                       &                       &                                 & \\
& J=5-4                                      &     550.926 GHz & $7.5 \pm 2.2\,(-9)$           & *$1.5 \pm 0.1\,(-8)$  & -                     & -                               & 0.07\\
& J=6-5                                      &     661.067 GHz & $1.9 \pm 0.6\,(-8)$           &-                      & -                     & -                               & 0.08\\
& J=7-6                                      &     771.184 GHz & $3.3 \pm 1.0\,(-8)$           & $4.0 \pm 0.4\,(-8)$   & -                     & -                               & 0.09\\
& J=8-7                                      &     881.273 GHz & $4.2 \pm 1.3\,(-8)$           & $4.8 \pm 0.5\,(-8)$   & -                     & -                               & 0.11\\
& J=9-8                                      &     991.329 GHz & $5.1 \pm 1.5\,(-8)$           & $4.9 \pm 0.5\,(-8)$   & -                     & -                               & 0.12\\
& J=10-9                                     &    1101.350 GHz & $4.3 \pm 1.3\,(-8)$           & $5.6 \pm 0.6\,(-8)$   & -                     & -                               & 0.13\\
& J=11-10                                    &    1211.330 GHz & $4.0 \pm 1.2\,(-8)$           & $4.6 \pm 0.6\,(-8)$   & -                     & -                               & 0.15\\
& J=12-11                                    &    1321.265 GHz & $3.1 \pm 0.9\,(-8)$           &-                      & -                     & -                               & 0.16\\
& J=13-12                                    &    1431.153 GHz & $2.0 \pm 0.6\,(-8)$           &-                      & -                     & -                               & 0.17\\
& J=15-14                                    &    1650.767 GHz &-                              &                       & $7.6 \pm 2.2\,(-9)$   & -                               & 0.20\\
& J=16-15                                    &    1760.486 GHz &-                              &-                      & $4.4 \pm 2.5\,(-9)$   & -                               & 0.22\\
$\mathrm{CH}^+$                              &                 &                               &                       &                       &                                 & \\
& J=1-0                                      &     835.137 GHz & -                             & $1.30 \pm 0.13\,(-8)$ & -                     & -                               & 0.10\\
& J=2-1                                      &    1669.281 GHz & -                             & $4.32 \pm 0.60\,(-8)$ & -                     & -                               & 0.28\\
& J=3-2                                      &    2501.440 GHz & -                             & -                     & $3.4 \pm 0.8\,(-8)$   & -                               & 0.28\\
& J=4-3                                      &    3330.630 GHz & -                             & -                     & $3.3 \pm 0.8\,(-8)$   & -                               & 0.28\\
& J=5-4                                      &    4155.872 GHz & -                             & -                     & $2.8 \pm 1.0\,(-8)$   & -                               & 0.28\\
& J=6-5                                      &    4976.201 GHz & -                             & -                     & $1.9 \pm 1.1\,(-8)$   & -                               & 0.28\\
$\mathrm{OH}$                                &                 &                               &                       &                       &                                 & \\
& $^2\Pi_{3/2}\,\,J=5/2^+-3/2^-$             & $119.4416\,\mu\mathrm{m}$ & -                   & -                     & $7.8 \pm 1.6\,(-8)$   & -                               & 0.28 \\
& $^2\Pi_{3/2}\,\,J=5/2^--3/2^+$             & $119.2345\,\mu\mathrm{m}$ & -                   & -                     & $6.6 \pm 1.4\,(-8)$   & -                               & 0.28 \\
& $^2\Pi_{1/2}-^2\Pi_{3/2}\,\,J=1/2^+-3/2^-$ & $ 79.1792\,\mu\mathrm{m}$ & -                   & -                     & $6.0 \pm 2.2\,(-8)$   & -                               & 0.28 \\
& $^2\Pi_{1/2}-^2\Pi_{3/2}\,\,J=1/2^--3/2^+$ & $ 79.1712\,\mu\mathrm{m}$ & -                   & -                     & $6.6 \pm 2.3\,(-8)$   & -                               & 0.28 \\
& $^2\Pi_{1/2}\,\,J=3/2^--1/2^+$             & $163.3962\,\mu\mathrm{m}$ & -                   & -                     & $1.4 \pm 0.3\,(-8)$   & -                               & 0.28 \\
& $^2\Pi_{1/2}\,\,J=3/2^+-1/2^-$             & $163.0153\,\mu\mathrm{m}$ & -                   & -                     & $1.3 \pm 0.3\,(-8)$   & -                               & 0.28 \\
& $^2\Pi_{3/2}\,\,J=7/2^--5/2^+$             & $ 84.5967\,\mu\mathrm{m}$ & -                   & -                     & $3.1 \pm 0.9\,(-8)$   & -                               & 0.28 \\
& $^2\Pi_{3/2}\,\,J=7/2^+-5/2^-$             & $ 84.4203\,\mu\mathrm{m}$ & -                   & -                     & $3.4 \pm 1.0\,(-8)$   & -                               & 0.28 \\
& $^2\Pi_{3/2}\,\,J=9/2^+-7/2^-$             & $ 65.2789\,\mu\mathrm{m}$ & -                   & -                     & $0.5 \pm 0.7\,(-8)$   & -                               & 0.28 \\
& $^2\Pi_{3/2}\,\,J=9/2^--7/2^+$             & $ 65.1318\,\mu\mathrm{m}$ & -                   & -                     & $1.3 \pm 0.7\,(-8)$   & -                               & 0.28 \\
$\mathrm{HD}$                                &                           &                     &                       &                       &                                 &      \\
& J=1-0                                      & $112.07\,\mu\mathrm{m}$   & -                   & -                     & $0 \pm 4.0\,(-9)$     & -                               & 0.28 \\
& J=2-1                                      & $ 56.23\,\mu\mathrm{m}$   & -                   & -                     & $3.1 \pm 1.1\,(-8)$   & -                               & 0.28 \\
$\mathrm{C}^+$                               &                           &                     &                       &                       &                                 &      \\
            & $^2P_{3/2}$ - $^2P_{1/2}$  & $157.68\,\mu\mathrm{m}$   & -                   & $5.5\pm 0.8\,(-6)$    & -                     & $7.5 \pm 1.5\,(-6)^\mathrm{b}$  & 0.28 \\
$\mathrm{C}$                                 &                           &                     &                       &                       &                                 &      \\
            & $^3P_1$ - $^3P_0$              & 492.161 GHz      & -                   & $2.9\pm 0.3 (-9)$     & -                     & -                               & 0.10 \\
            & $^3P_2$ - $^3P_1$              & 809.342 GHz     & -                   & $2.3\pm 0.2 (-9)$     & -                     & -                               & 0.06 \\
$\mathrm{O}$                                 &                           &                     &                       &                       &                                 &      \\
            & $^3P_0$ - $^3P_1$              & $145.53\,\mu\mathrm{m}$      & -                   & -                     & -                     & $6.0 \pm 1.2\,(-6)^\mathrm{b}$  & 0.28 \\
            & $^3P_1$ - $^3P_2$              & $63.18\,\mu\mathrm{m}$       & -                   & -                     & $5.4\pm 1.1\,(-5)$    & $5.0 \pm 1.0\,(-5)$             & 0.28 \\
\hline
$\mathrm{H}_2$&                              &                           & ISO SWS$^\mathrm{c}$          & IRTF $^\mathrm{d}$    & CFHT$^\mathrm{e}$     &                        &      \\
& 0-0 S(0)                                   & $28.22\,\mu\mathrm{m}$    & $0.9 \pm 0.3\,(-7)$           & -                     & -                     &                        & 0.10 \\
& 0-0 S(1)                                   & $17.03\,\mu\mathrm{m}$    & $6.5 \pm 1.3\,(-7)$           & $8.5\pm0.3\,(-7)$     & -                     &                        & 0.10 \\
& 0-0 S(2)                                   & $12.28\,\mu\mathrm{m}$    & $3.7\pm0.9\,(-7)$             & $6.8\pm0.2\,(-7)$     & -                     &                        & 0.10 \\
& 0-0 S(3)                                   & $ 9.66\,\mu\mathrm{m}$    & $6.0\pm1.5\,(-7)$             & -                     & -                     &                        & 0.10 \\
& 0-0 S(4)                                   & $ 8.02\,\mu\mathrm{m}$    & $2.8\pm0.7\,(-7)$             & $4.1\pm0.2\,(-7)$     & -                     &                        & 0.10 \\
& 0-0 S(5)                                   & $ 6.91\,\mu\mathrm{m}$    & $6.4\pm1\,(-7)$               & -                     & -                     &                        & 0.10 \\
& 1-0 S(1)                                   & $ 2.12\,\mu\mathrm{m}$    &                               & -                     & $5.8\,(-7)$           & $3.6 (-7) ^\mathrm{f}$ & 1    \\
& 2-1 S(1)                                   & $ 2.25\,\mu\mathrm{m}$    &                               & -                     & $1.2\,(-7)$           &                        & 1    \\
\hline
\end{tabular}\\
$^\mathrm{a}$Parikka et al. (2017). 
$^\mathrm{b}$Bernard Salas et al. (2012) - $^\mathrm{c}$ Bertoldi, private comm., Habart et al. (2004) - $^\mathrm{d}$Allers et al. (2005) - $^\mathrm{e}$ Joblin, Maillard, Noel, unpublished BEAR data - $^\mathrm{f}$ van der Werf et al. (1996). 
\end{center}
\end{table*}

\newpage

\section{Observation results}
\label{Sec:ObsData}

\subsection{Combination of observational data}\label{Sect:obs_corrections}
For the purpose of comparing the observations with PDR models (cf. Sect.\,\ref{Sec:Models}), we need to combine the results obtained with  different instruments. This is particularly challenging because of different calibrations, beam sizes and observational techniques. In this section, we discuss the procedure we used to deal with these differences. 

\paragraph{Beam dilution factors}
The data used in this work gather line intensities that were measured using different beam sizes. This raises the question of the morphology of the emitting structures and to which extend these structures fill the different beams. In NGC~7023, the bright NW interface is composed of a filamentary structure with a spatial characteristic width of a few arcsecs as clearly seen in vibrationally excited H$_2$ emission \citep{Lemaire96}. From these maps one can derive a relevant width of 2\arcsec\ for the bright PDR interface that is centered toward the H$_2$ peak. In Orion Bar, the sharp molecular edge has been observed in vibrational and rotational H$_2$ transitions \citep{Tielens93,VanDerWerf96,Walmsley00, Allers05}. These observations suggest a narrow (few arcsec) and patchy interface that is also seen in  ALMA maps obtained at a spatial resolution between 1 and 5\,$\arcsec$ \citep{Goicoechea16,Goicoechea17}. The ALMA data show that emission lines of some ions such as HCO$^+$ 4-3 and SH$^+$ 1-0 are located in the narrow layer given by vibrationally excited H$_2$. Although the morphology is more complex than a single structure, we assume in the following that the observed emission from warm molecular tracers arises from a 2\arcsec\ filamentary interface, consistently with the case of NGC\,7023.
To derive dilution factors, we therefore assumed in both PDRs that the emitting structure is a filament that follows the interface with infinite length and a 2\arcsec\ thickness. For each observation, we calculated the fractional coverage of the beam by this filament. This is a simplistic procedure but the best we can do to overcome the lack of spatial information. The values of the derived dilution factors $\Omega$ are reported in Tables~\ref{Tab:NGC7023_raw_data} and \ref{Tab:Orion_raw_data}. The observed intensities were then divided by this factor $\Omega$ in order to be compared with the models.

\paragraph{Cross-calibration factors}
In addition to the dilution factors we have to apply scaling factors on some of the data sets. The $^{12}\mathrm{CO}$ ladder for NGC~7023 (see the data in Fig.\,\ref{Fig:12_CO_ladder}) reveals discrepancies between the different instruments that cannot be compensated by our dilution factors.
In the following, we consider that HIFI fluxes are the references and scale the intensities from SPIRE and PACS. As common lines are observed by SPIRE and HIFI, we simply searched for the scaling factor giving the least square error, and divided all line intensities observed by SPIRE (including species other than CO) by a factor of 0.54 as a result. PACS observations can not be directly compared, but the rotational diagram of $^{12}$CO reveals an unphysical jump between the PACS and SPIRE/HIFI lines. We determined the factor giving the best linear alignment of the PACS observations with the SPIRE/HIFI lines on this rotational diagram, and divided all line intensities observed by PACS by a factor of 1.3 as a result. 
In the case of the H$_2$ observations in NGC\,7023 it is not possible to satisfactorily merge the Spitzer and ISO data. This is due to the fact that H$_2$ emission is quite extended and the ISO beam contains emission from other regions than just the filament of interest. 
Spitzer values are taken as references and we divided all ISO observations by a factor of 2.54 to get the best agreement between the two data sets.
In the case of the Orion Bar, we find no obvious reason for such adjustments. In particular no systematic discrepancy is visible in the CO ladder (cf. Fig.\,\ref{Fig:12_CO_ladder}). Thus, no such adjustment was applied to the observations of the Orion Bar.

Finally, after correcting for beam dilution and cross-calibration factors, we combined the line intensities from the different instruments by simply taking an average of the different observations of a given line. We computed error bars on this mean value as the interval between the minimum and maximum values in the error ranges of the different instruments. 

The original (uncorrected) data of the different instruments are presented in Tables~\ref{Tab:NGC7023_raw_data} and \ref{Tab:Orion_raw_data} and shown on Fig.\,\ref{Fig:12_CO_ladder} for $^{12}$CO and $^{13}$CO.
Tables\,\ref{Tab:NGC7023_obs_model} and \ref{Tab:Orion_obs_model} present the data after correction of dilution, cross calibration and averaging of the different instruments. These corrected and averaged data are used in all figures except Fig.\,\ref{Fig:12_CO_ladder}.

\begin{figure*}
\begin{center}
\includegraphics[width=0.45\textwidth]{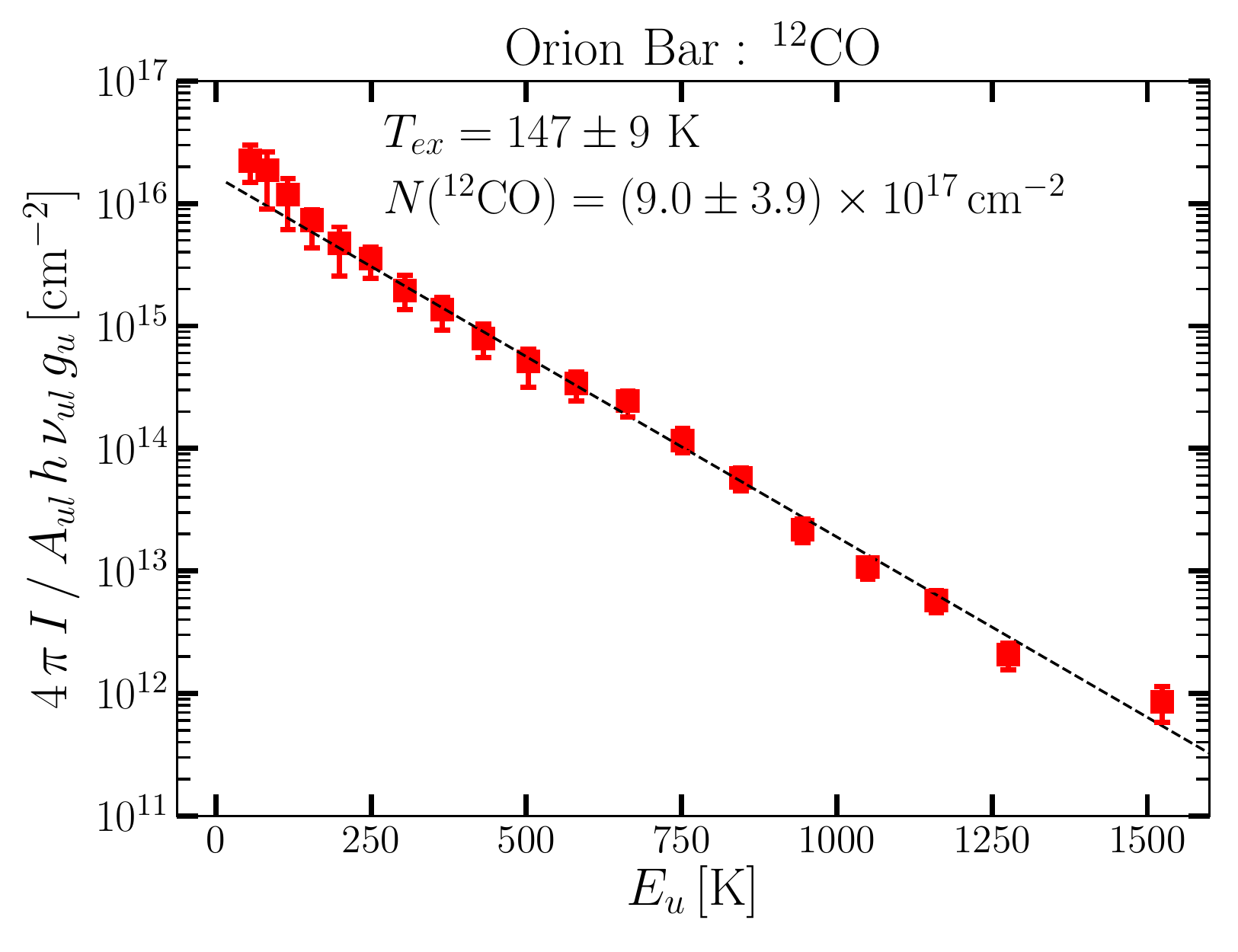}
\includegraphics[width=0.45\textwidth]{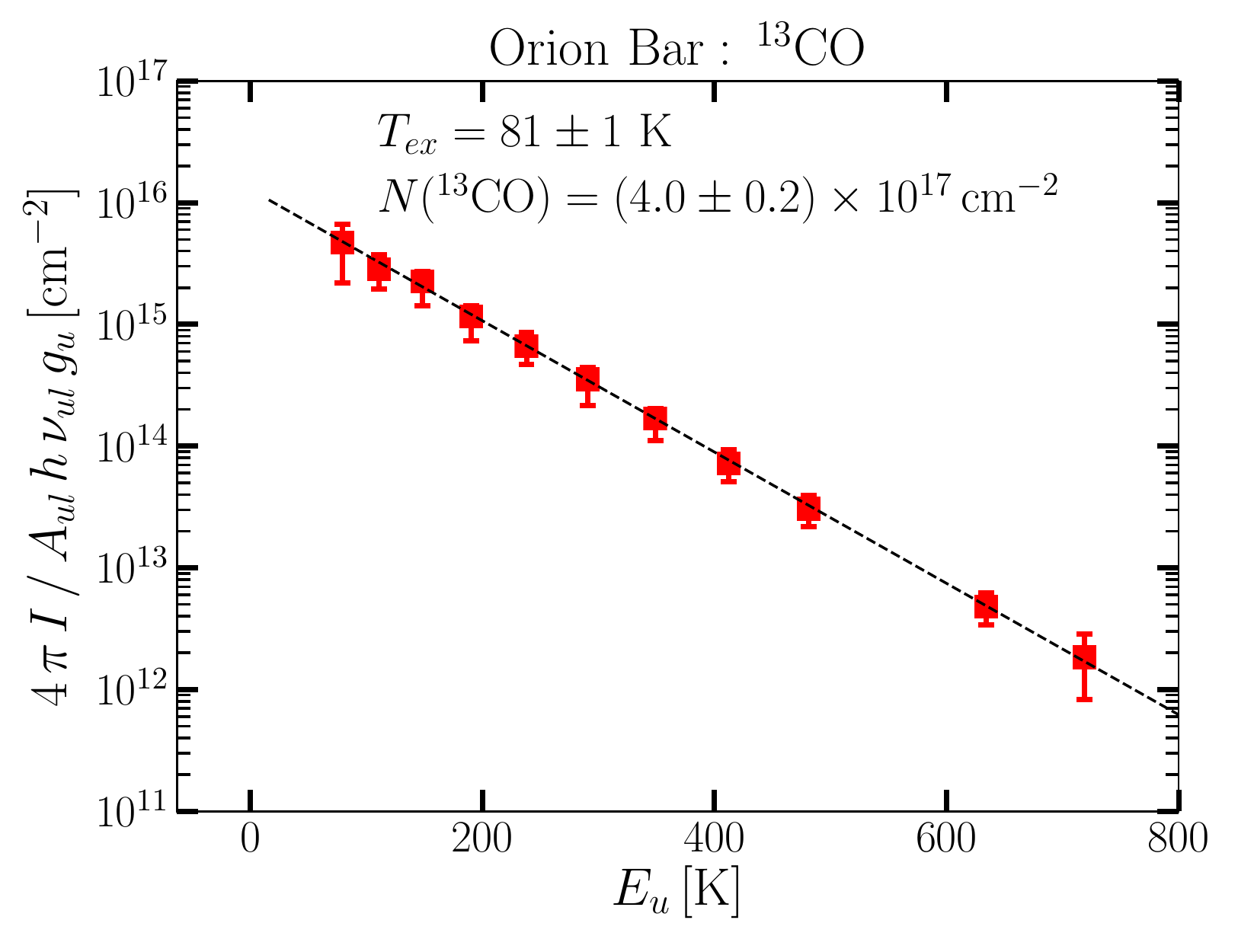}\\
\includegraphics[width=0.45\textwidth]{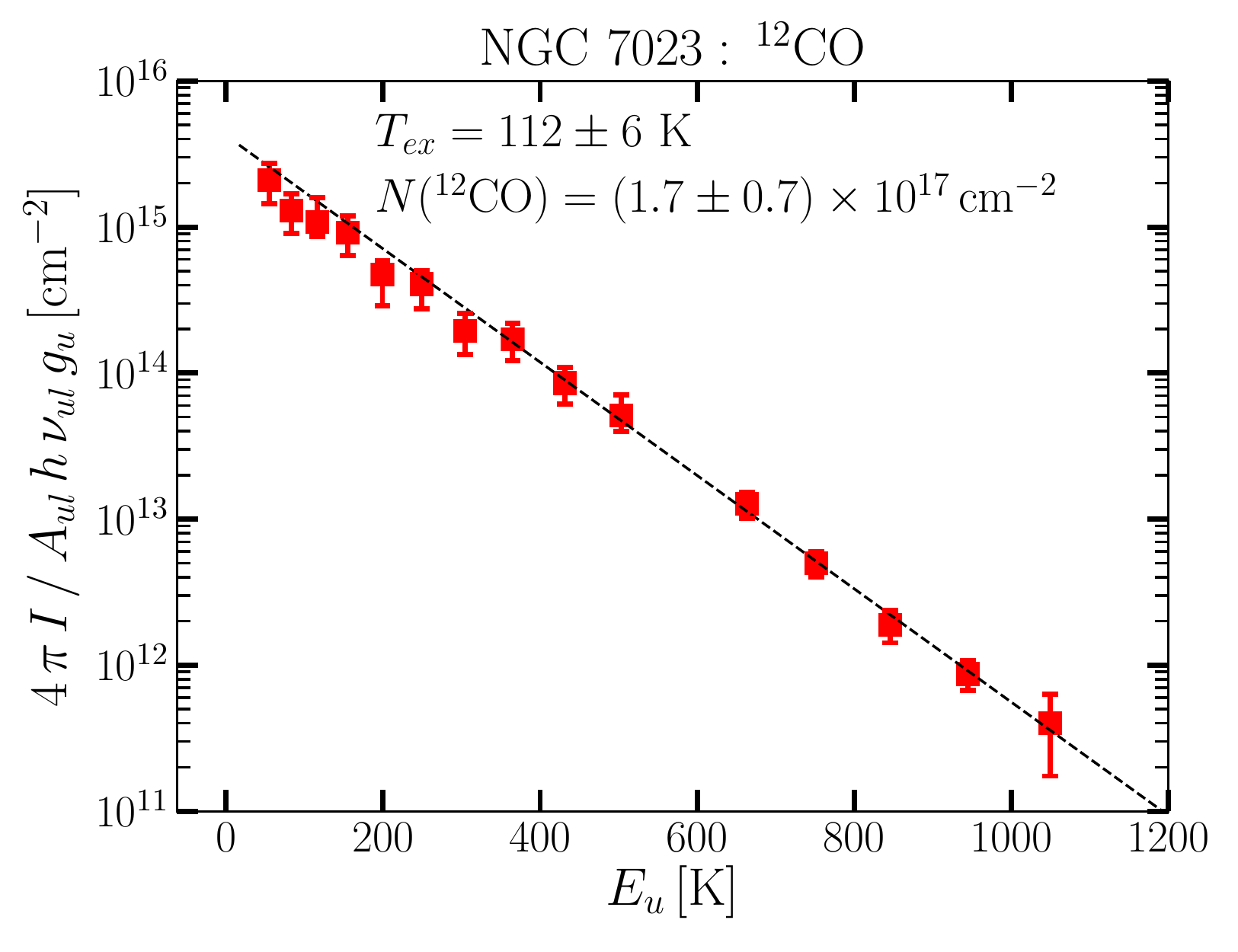}
\includegraphics[width=0.45\textwidth]{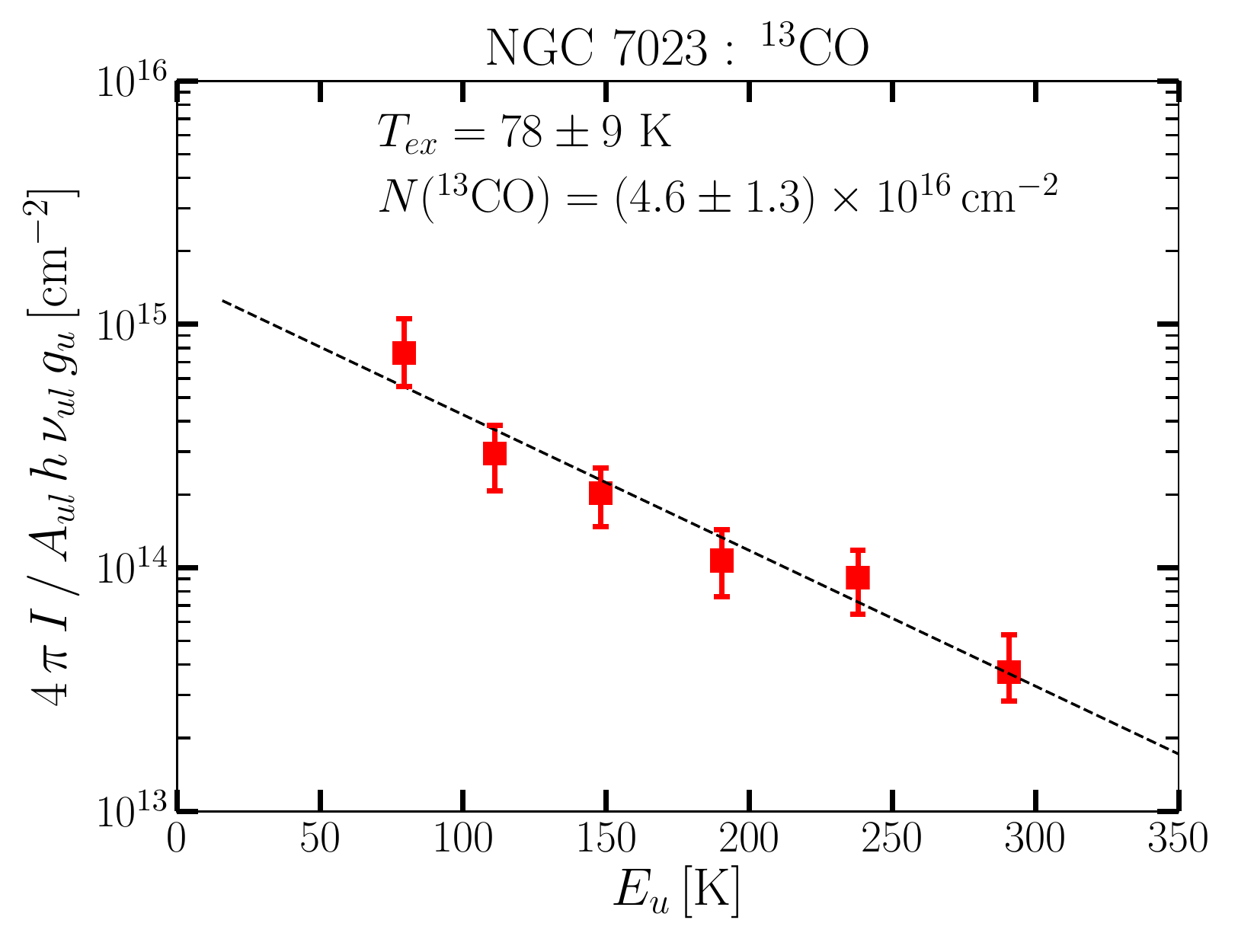}
\end{center}
\caption{Rotational diagram of $^{12}$CO (left) and $^{13}$CO (right) lines observed in the Orion Bar (top) and NGC\,7023 (bottom) PDRs. For $^{12}$CO, the excitation temperature and total column density are computed in the range J$_{up}$=[15, 23] (Orion Bar), and J$_{up}$=[15, 19] (NGC 7023). For $^{13}$CO, the lines used were J$_{up}$=[5, 16] (Orion Bar), and J$_{up}$=[5, 10] (NGC 7023). \label{Fig:12CO_rot_diag}}
\end{figure*}

\begin{figure}
\begin{center}
\includegraphics[width=0.4\textwidth]{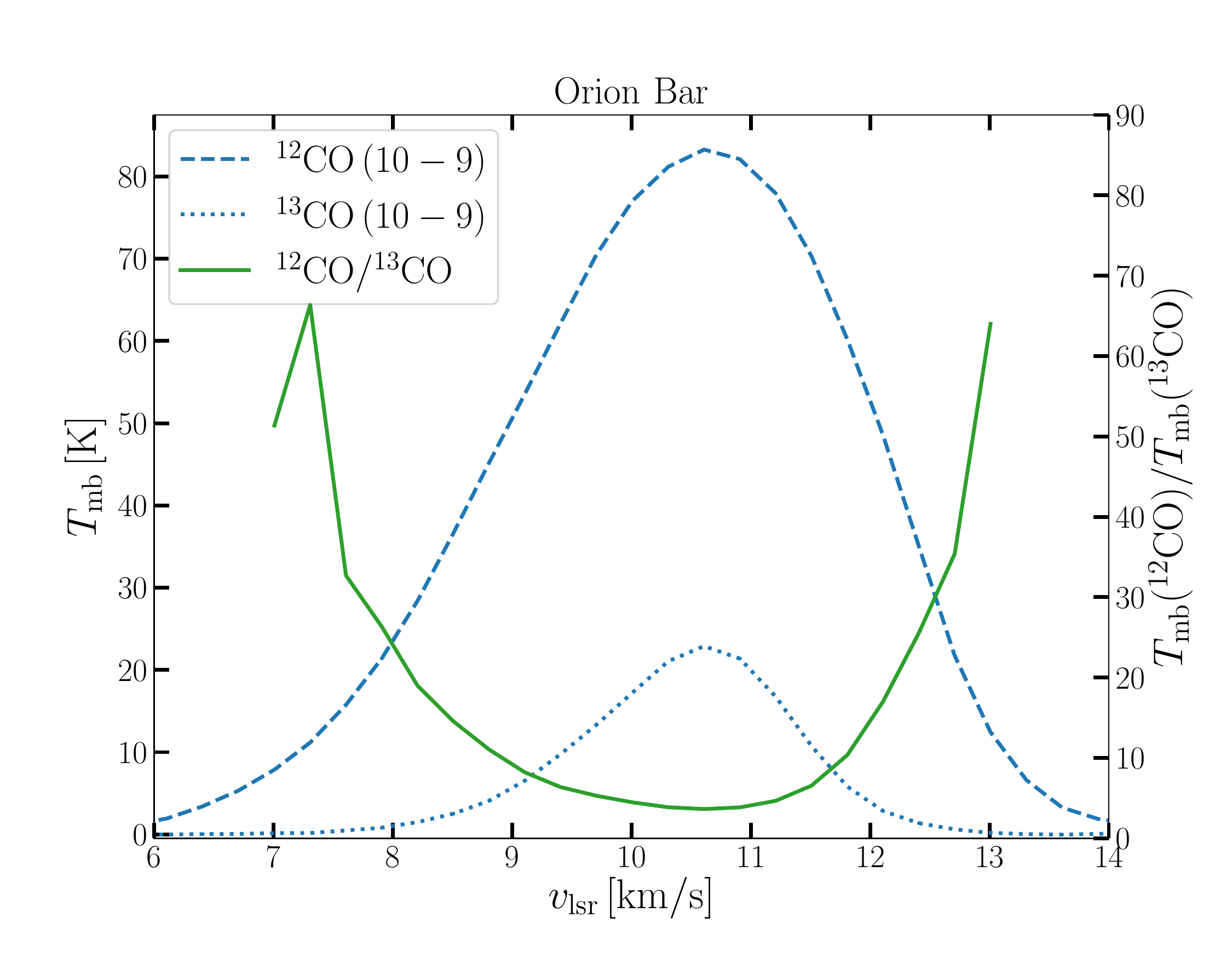}
\includegraphics[width=0.4\textwidth]{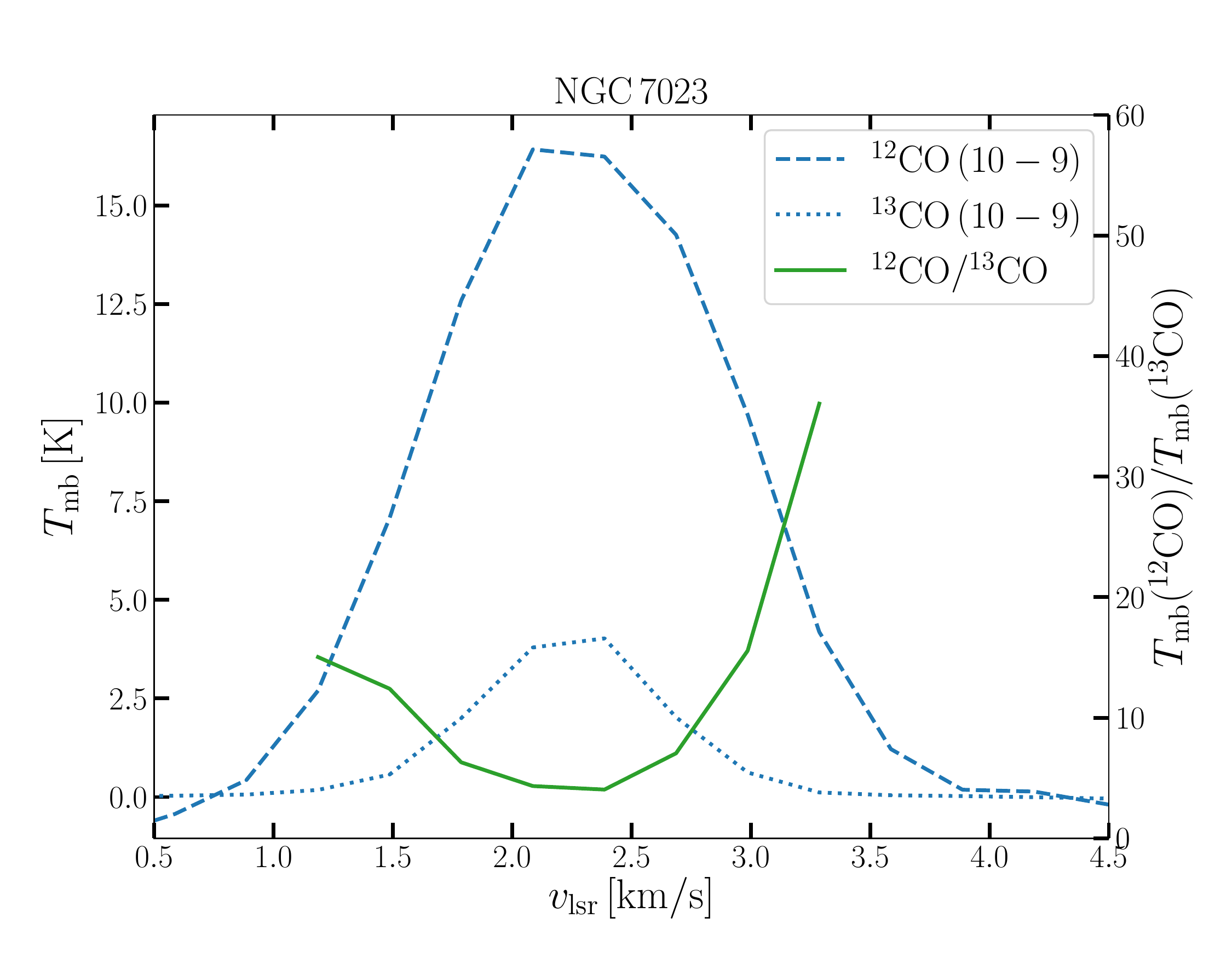}
\end{center}
\caption{Line profile for the J=10--9 transition of  $^{12}$CO (dashed line) and $^{13}$CO (dotted line) in the Orion Bar (top panel) and NGC~7023~NW (bottom panel). The relative intensity of the $^{12}$CO to $^{13}$CO lines across the velocity profiles is also shown (plain line). \label{Fig:12CO_13CO_profile}}
\end{figure}

\subsection{The CO rotational diagrams}
\label{Sec:CO_rot}

Figure\,\ref{Fig:12CO_rot_diag}  presents the rotational diagrams and LTE fits of the $^{12}$CO and $^{13}$CO observations. The rotational diagrams present column density estimates in the upper rotational levels,  $N_u$, without corrections for opacity effects  with $N_u/g_u = \frac{4\,\pi\,I}{A_{ul}\,h\nu_{ul}\,g_u}$ where I is the observed intensity, $A_{ul}$ and $\nu_{ul}$ are the Einstein coefficient and frequency of the transition from the upper to lower levels and $g_u$ is the degeneracy of the upper level. As such, the derived column denities are lower limits when the lines are optically thick. In the case of $^{12}$CO, the fit was restricted to the highest excitation lines, starting at J$_{up}$=15, in order to obtain information on the warmest gas.  In the Orion Bar, we derived an excitation temperature of  $147\pm9\,\mathrm{K}$  and a column density of N($^{12}$CO)=($9.0\pm3.9)\times10^{17}\,\mathrm{cm}^{-2}$. In NGC\,7023, the excitation temperature is $112\pm6\,\mathrm{K}$ with N($^{12}$CO)=$(1.7\pm0.7)\times10^{17}\,\mathrm{cm}^{-2}$. We note that the column density of warm CO is higher (at least a factor of 4) in the Orion Bar as compared to NGC\,7023. The $^{12}$CO rotational temperature is also somewhat higher in the Orion Bar. Both facts favour the detection of higher-J CO transitions in Orion Bar relative to NGC\,7023.
We can compare these results with previous studies. \cite{Koehler14} studied the CO lines measured with SPIRE in NGC~7023. Using the non-LTE radiative transfer code RADEX \citep{VanDerTak07}, they derived a kinetic temperature in the range  65-130\,K and a column density of N($^{12}$CO)=$2-3\times10^{18}\,\mathrm{cm}^{-2}$. They concluded that the emitting structure has a dilution factor of 0.1. All these results are compatible with ours if we keep in mind our strategy to gain contrast on the hottest molecular interface, including taking into account higher-J CO lines. Our derived rotational temperature of 141\,K  for the Bar is also consistent with the kinetic temperature in the range 100--150\,K that was reported by \cite{Nagy17} in the analysis of the Herschel/HIFI spectral line survey of the Orion Bar that was mentioned earlier. It is also compatible with the lower limit of the kinetic temperature derived by \cite{Goicoechea16} at the dissociation front.

The $^{13}$CO diagrams reveal gas at an average rotational temperature of $\sim$80\,K, which is cooler than the  $^{12}$CO temperature discussed above. In addition, Figure\,\ref{Fig:12CO_13CO_profile} reports the value of the $^{12}$CO over $^{13}$CO line intensity ratio across the velocity profile of the $^{12}$CO 10-9 line.
This ratio is found to be weak ($\sim 4$) near the line center, whereas it reaches the isotopic $^{12}$C/$^{13}$C ratio value of $\sim 50$ in the wings. This is characteristic of strong opacity effects.
 \cite{Koehler14} calculated the optical depth of the $^{12}$CO lines in NGC\,7023 using the RADEX code and concluded that the lines are optically thick up to J$_{up}\sim13 - 14$. This justifies why we selected only lines from J$_{up}$=15 and higher in our fit of the $^{12}$CO rotational diagrams (cf. Fig.\,\ref{Fig:12CO_rot_diag}). 

The above results strongly suggest that CO emission in the beam stems from a two-component medium:  an extended cool/warm component and the hot and sharp interface. In $^{13}$CO, the former component is well seen, whereas it is partly hidden due to optical depth effects in $^{12}$CO emission. This results in $^{12}$CO emission looking like stemming mainly from the hot component.  On the opposite, the hot component is difficult to observe in the high-J lines of $^{13}$CO due to the low expected signal.  As an example, in  the Orion Bar, an intensity ratio of 5.6 can be derived from the observed J=16-15 and J=20-19 $^{12}$CO lines (see Tab.\,\ref{Tab:Orion_raw_data}). By applying the same factor to the $^{13}$CO lines, we can predict an intensity of 8\,10$^{-10}$\,W\,m$^{-2}$\,sr$^{-1}$ for the J=20-19 $^{13}$CO line, which cannot be detected considering that it is a factor of 3 weaker relative to the error bar of the J=16-15 $^{13}$CO line.
In conclusion, only high-J $^{12}$CO lines can be used to characterise the physical conditions at the warm and bright PDR interface.

\section{Models}
\label{Sec:Models}

\subsection{The Meudon PDR model}
We compared our observations to PDR models using an updated version of the Meudon PDR code \citep[][https://ism.obspm.fr]{LePetit06}. 
This 1D PDR code simulates the physical and chemical processes at stationary state in a plane-parallel slab of gas and dust. At each position in the cloud, the code computes the temperatures of gas and grains, the chemical densities and, for the most important species, the non-LTE level populations. Then a post-treatment gives access to column densities and line intensities. 

Several atomic and molecular data have been updated. In particular, it is worthwhile to notice for the present study that this version of the PDR code implements the CO-H$_2$ collision rates of \cite{Yang10} and the CO-H collision rates of \cite{Balakrishnan02}. The chemical network includes 213 species linked by 5067 gas-phase chemical reactions (except H$_2$ formation on grains). Table\,\ref{Tab:InputParam} summarizes the elemental abundances used in this paper. The formation reaction of H$_2$ on grains plays a critical role in the computation of the chemical structure of PDRs. Several prescriptions are available in the Meudon PDR code: a mean formation rate of $3\times10^{-17} \sqrt{T/100}\,\mathrm{cm}^{3} \,\mathrm{s}^{-1}$ based on Copernicus and FUSE H$_2$ observations in diffuse clouds \citep{Jura74, Gry02}, Langmuir-Hinshelwood (LH) and Eley-Rideal (ER) mechanisms treated with a rate equation formalism \citep{LeBourlot12} and a new stochastic approach that considers the impact of grain temperature fluctuations on H$_2$ formation (both LH and ER) in a master equation formalism \citep{Bron14,Bron16}. In this paper, we use the \citet{LeBourlot12} formalism\footnote{We do not use the most sophisticated treatment at our disposal, \cite{Bron14}, for computing time reasons.}. 

The code considers two external sources of energy, the radiation field and cosmic rays. The external radiation field can be the combination of a beamed stellar radiation field that simulates neighboring stars and an isotropic ambient radiation field. This isotropic component is composed of the \cite{Mathis83} field for the far UV to IR part, scaled by a factor,
and black bodies that simulate dust IR emission and the cosmic microwave background. The Meudon PDR code allows us to use different radiation fields on each side of the cloud. We call front side of the PDR the one illuminated by the stars responsible for the PDR and back side the other one. 
For NGC 7023, we used a beamed stellar radiation field built from the Kurucz stellar spectra \citep{Kurucz93} as explained in \cite{Pilleri12}. This leads to a value of the UV intensity at the edge of the NW PDR of G$_0$ = 2600 in Habing units for a distance of d$\simeq$\,0.143\,pc between the star and the dissociation front. In addition, the scattered light in the surrounding region was determined to represent about 4\% of the direct stellar light using the scattered light measured at 475 nm by the Hubble Space Telescope \citep{Witt06}. 
We thus assume that the back side of the PDR is illuminated by an isotropic radiation field with $G_0 = 100$, making the simplifying assumption that there is no dependence of the scattering with wavelength. This back side illumination was found to contribute marginaly to the line intensities. For the Orion Bar, several estimations of the UV flux have been proposed in the literature. 
As discussed by \cite{Allers05}, the estimated UV flux intensity depends on several parameters such 
as the inclination of the Bar. 
Based on \cite{Tielens85} and \cite{Marconi98}, we fixed the radiation field impinging on the PDR so that, at the edge of the PDR, $G_0 = 2\times10^4$ in Habing units. We also assumed an ambient UV field illuminating the back side of the cloud with 10\% of the front side illumination.
Finally, the cosmic ray flux is introduced as a cosmic-ray ionization rate $\zeta$ of H$_2$ molecules. 
As we lack information about this flux in NGC 7023 and Orion Bar PDRs, we use an intermediate value of $\zeta = 5 \times 10^{-17}\,\mathrm{s}^{-1}$ that lies between estimations in diffuse and dense gas \citep{Indriolo15, Padovani13, LePetit04, McCall99}.
We checked that an increase of this ionization rate by a factor 10 has no impact on CO line intensities. 

The determination of photo-reaction rates requires to compute the specific intensity in the UV at each position in the cloud. The Meudon PDR code solves the radiative transfer equation on a wavelength grid (from the UV to the radio domain) at each position, including absorption and scattering by dust as well as absorption in lines or in the continuum for some chemical species. For instance, continuum absorption by C photo-ionization turns out to have a non-negligible impact on the PDR structure \citep{Rollins12} and is considered, as well as H line absorptions. For other species (as H$_2$ and CO) line self-shielding is computed using the \cite{Federman79} approximation. Absorption by grains is implemented using parametrized extinction curves (\citealt{Fitzpatrick86}). In PDRs, extinction curves are usually flatter in the far UV than the mean Galactic extinction curve. That is characteristic of larger than standard grain sizes. The adopted extinction curve is the one of HD~38087 in \cite{Fitzpatrick90} and $R_\textrm{V} = 5.62$ as measured in NGC~7023 by \citealt{Witt06}, close to the value determined for Orion Bar, 5.5 \citep{Marconi98}. We also assume a column density to reddening ratio $N_\textrm{H}/\textrm{E(B-V)} = 1.05\times10^{22}\,\mathrm{cm}^{-2} \, \mathrm{mag}^{-1}$, an intermediate value between the standard value $5.8\times10^{21}$ and the value determined for $\rho$ Oph, $15.4\times10^{21}\,\mathrm{cm}^{-2} \, \mathrm{mag}^{-1}$ \citep{Bohlin78}. 
Grains are simulated as a mixture of spherical amorphous carbonaceous and silicate grains following a MRN size distribution \citep{Mathis77}, with minimum and maximum radii $3\times10^{-7}$ and $3\times10^{-5}$ cm, and with a dust-to-gas mass ratio of 0.01. The dust scattering properties are from \cite{Laor93}. 

At each position, the code computes the equilibrium gas temperature from the balance of total heating and cooling rates. The heating mechanisms considered are the photo-electric effect on grains, cosmic rays heating, and exothermic chemical reactions. Gas-grain collisions and H$_2$ vibrational de-excitation can heat or cool the gas depending on the local physical conditions. For the photo-electric effect, we use the \cite{Bakes94} prescription. Cooling rates are obtained by computing the radiative emission of the main coolants (15 species included), among which C$^+$, O, C, CO and its isotopologues. For these species, non-LTE level populations are computed taking into account collisional excitations and de-excitations, spontaneous emission, non-local radiative pumping by line and continuum photons, and chemical formation and destruction in specific levels as described in \cite{Gonzalez08}.

Finally, several prescriptions can be used to define the gas density profile as a function of depth. In this work, we tried two prescriptions: constant density models and constant pressure models. Parameters used in the models of NGC 7023 and Orion Bar are summarized in Tab. \ref{Tab:InputParam}. The modeling strategy for the two objects is described in the following sections. 

\begin{table}[h!]
\center
\caption{Input parameters used in the Meudon PDR code.}
\label{Tab:InputParam}
\begin{tabular}{llll}
\hline
\hline
Parameter                & Value                   &  Unit                          & Note    \\
\hline
\multicolumn{4}{c}{free parameters}\\
\hline
$P_{\mathrm{th}}$ NGC 7023             & $10^8$           & $\mathrm{K~cm}^{-3}$                  & best fit    \\
$P_{\mathrm{th}}$ Orion Bar            & $2.8\times10^8$  & $\mathrm{K~cm}^{-3}$                  & best fit    \\
\hline
\multicolumn{4}{c}{fixed parameters}\\
\hline
$G_0$ NGC 7023           & \emph{2600}             & Habing                         & (1)         \\
$G_0$ Orion Bar          & $2\times 10^4$          & Habing                         & (2),(3)     \\
$A_\textrm{V}^{tot}$     & 10                      & mag                            &             \\
Flux of cosmic-rays      & $5\times10^{-17}$       & $\mathrm{s}^{-1}$ per H$_2$    &             \\
Dust extinction          & HD 38087                &                                & (4)         \\
R$_\textrm{V}$           & 5.62                    &                                & (5)         \\
N$_\textrm{H}$ / E(B-V)  & $1.05\times 10^{22}$    & $\mathrm{cm}^{-2}\,\,\mathrm{mag}^{-1}$  & see text \\
Mass grain / Mass gas    & $0.01$                  &                                &              \\ 
Grain size distribution  & $\propto a^{-3.5}$      &                                & (6)          \\
min radius of grains     & $3\times 10^{-7}$       & cm                             &          \\
max radius of grains     & $3\times 10^{-5}$       & cm                             &          \\
\hline
\multicolumn{4}{c}{elementary abundances}\\
\hline
He                       & 0.1                   & &    \\
C                        & $1.32\times 10^{-4}$  & &(7)   \\
O                        & $3.19\times 10^{-4}$  & &(8)   \\                     
S                        & $1.86\times 10^{-5}$  & &(7)   \\
N                        & $7.50\times 10^{-5}$  & &(9)   \\
D/H                      & $1.5\times 10^{-5}$   & &(10)   \\
$^{12}$C/$^{13}$C        &  $50$                 & &(11)   \\
\hline
\end{tabular}
\tablebib{(1) For the NGC~7023 PDR model, we adopt a beamed stellar radiation field as described in the text. If converted in a Habing scaling factor, this corresponds to $G_0$ = 2600, (2) \cite{Tielens85}, (3) \cite{Marconi98}, (4) \cite{Cardelli89}, (5) \cite{Witt06}, (6) \cite{Mathis77}, (7) \cite{Savage96}, (8) \cite{Meyer98}, (9) \cite{Meyer97}, (10) \cite{Oliveira06}, (11) Intermediate value from observations in Orion \citep{Demyk07, Ossenkopf13, Haykal14}.}
\end{table}

\subsection{Fitting strategy}

Since we focus our analysis on the physical conditions at the PDR edge, we investigate the models that best account for the emission of tracers specific to this region, such as high-J $^{12}$CO and H$_2$ lines.
We investigated different scenarios. First, we ran PDR models at constant density but they proved to be incompatible with the observations. In particular, they were unable to reach the observed excitation temperature of high-J CO levels. We found that a density change across the PDR was necessary, with the density increasing with PDR depth. As earlier suggested by \cite{Marconi98}, constant pressure models were found to provide a satisfying density gradient due to the temperature drop when going towards the inside of the cloud. 

Both PDRs appear as bright narrow interfaces ($\sim$ 2\arcsec) in vibrational H$_2$ emission (cf. Fig.\,\ref{Fig:Images_OrionBar_N7023}), which could be the result of an overdense surface layer seen roughly edge-on. 
We thus adopt for our models a geometry in which the PDR is observed 
with a high viewing angle of $60^{\circ}$; this angle being defined with $0^{\circ}$ being face-on and $90^{\circ}$ edge-on. The value of $60^{\circ}$ gives an approximation of a nearly edge-on PDR and is the maximum inclination that can be used to derive line intensities in the 1D PDR Meudon code. The uncertainty on this angle could lead to an additional scaling factor on all line intensities. We thus allowed for a free global scaling factor $f$ on the model intensities when fitting the model. A value of this factor larger than 1 would indicate a more edge-on configuration. In addition, this factor can correct for systematic errors we made on the assumed geometry, in particular viewing angle and the dilution factors.

We searched for the best fitting model in a grid of isobaric models for varying values of the thermal pressure $P_{\mathrm{th}}$ and global scaling factor $f$. In both cases, we used the observed high-J $^{12}$CO lines from J$_{up}$=11, the rotational H$_2$ lines (S(0) to S(5)) and CH$^+$ (1-0 to 6-5 in Orion Bar and 1-0 to 3-2 in NGC 7023) rotational lines.  We thus have 17 (NGC 7023) or 23 (Orion Bar) observational constraints that the best model must simultaneously reproduce  with two free-parameters, $P_{\mathrm{th}}$ and $f$. As these tracers only constrain the warm molecular layer of the PDR, we fixed a total $A_\textrm{V}$ value of 10 ($N_\textrm{H}\sim 2\times10^{22}$\,cm$^{-2}$ for a face-on geometry and a factor 2 higher assuming $60^{\circ}$ inclination) in our models. We then compare the results of the best model with other available observations, i.e. lines from $^{13}$CO, C$^+$, O, C, vibrational H$_2$, HD, and in addition HCO$^+$ in NGC 7023 and OH in the Orion Bar. 

To provide an idea on how much the observations constrain the thermal pressure, we also report in the following the results obtained by using a pressure that differs by a factor 1.5 (lower and higher) from the pressure obtained in the best fit model. For these cases, the scaling factor was adjusted to provide the best fit.

\subsection{Model results}
\label{Sec:model_results}

\begin{table}
\caption{Comparison of models and observations (combined) for NGC 7023. Values in parentheses are powers of ten. \label{Tab:NGC7023_obs_model}} 

\scriptsize 
\begin{tabular}{l l| l |l}
\hline \hline
\multicolumn{2}{c}{Line} & Observed intensity & Model prediction \\
\multicolumn{2}{c}{} & $(\mathrm{W}\mathrm{m}^{-2}\mathrm{sr}^{-1})$ & $(\mathrm{W}\mathrm{m}^{-2}\mathrm{sr}^{-1})$\\
 &   &   &   \\
\hline
$^{12}\mathrm{CO}$   &             &                              &                 \\
                     & J=4-3       & $2.8 \pm 0.9\,(-8)$            & $4.6(-8)$     \\
                     & J=5-4       & $5.3 \pm 1.6\,(-8)$            & $9.1(-8)$     \\
                     & J=6-5       & $1.1_{-0.2}^{+0.5}\,(-7)$      & $1.5(-7)$     \\ 
                     & J=7-6       & $2.0 \pm 0.6\,(-7)$            & $2.3(-7)$     \\
                     & J=8-7       & $2.0_{-0.8}^{+0.5}\,(-7)$       & $3.1(-7)$     \\
                     & J=9-8       & $3.1_{-1.0}^{+0.7}\,(-7)$      & $3.9(-7)$     \\
                     & J=10-9      & $2.5 \pm 0.8\,(-7)$            & $4.3(-7)$     \\
                     & J=11-10     & $3.5 \pm 1.0\,(-7)$            & $4.2(-7)$     \\
                     & J=12-11     & $2.7 \pm 0.8\,(-7)$            & $3.6(-7)$     \\
                     & J=13-12     & $2.4_{-0.5}^{+0.9}\,(-7)$      & $2.7(-7)$     \\
                     & J=15-14     & $1.2 \pm 0.2\,(-7)$            & $1.9(-7)$     \\
                     & J=16-15     & $6.5 \pm 1.3\,(-8)$            & $7.0(-8)$     \\
                     & J=17-16     & $3.3 \pm 0.8\,(-8)$            & $4.1(-8)$     \\
                     & J=18-17     & $2.0 \pm 0.5\,(-8)$            & $2.3(-8)$     \\
                     & J=19-18     & $1.2 \pm 0.7\,(-8)$            & $1.4(-8)$     \\
$^{13}\mathrm{CO}$   &             &                              &                \\
                     & J=5-4       &  $2.6_{-0.7}^{+1.0}\,(-8)$   & $1.1(-8)$ \\
                     & J=6-5       &  $2.5 \pm 0.8\,(-8)$         & $1.5(-8)$   \\
                     & J=7-6       &  $3.7 \pm 1.0\,(-8)$         & $1.7(-8)$    \\
                     & J=8-7       &  $3.8_{-1.1}^{+1.3}\,(-8)$   & $1.7(-8)$    \\
                     & J=9-8       &  $5.8 \pm 1.7\,(-8)$         & $1.6(-8)$    \\
                     & J=10-9      &  $4.0_{-1.0}^{+1.7}\,(-8)$         & $1.3(-8)$   \\
$\mathrm{CH}^+$      &              &                             &                  \\
                     & J=1-0        & $1.0 \pm 0.1\,(-8)$          & $4.2(-9)$       \\
                     & J=2-1        & $1.9_{-0.8}^{+0.6}\,(-8)$    & $1.4(-8)$       \\
                     & J=3-2        & $1.5 \pm 0.6\,(-8)$          & $1.5(-8)$       \\
$\mathrm{HCO}^+$     &              &                             &                  \\
                     & J=1-0        & $9.2 \pm 2.7\,(-12)$        & $4.1(-12)$       \\
                     & J=6-5        & $1.4 \pm 0.1\,(-9)$          & $3.1(-10)$      \\
$\mathrm{C}^+$       &              &                             &                  \\
                     & $157.68\,\mu\mathrm{m}$  & $2.7_{-1.2}^{+1.5}\,(-6)$   & $6.7(-7)$        \\
$\mathrm{C}$         &              &                             &                  \\
         & $609.13\,\mu\mathrm{m}$     & $5.1 \pm 1.5\,(-9)$         & $5.1(-9)$        \\
$\mathrm{O}$                        &                             &                  \\
         & $145.53\,\mu\mathrm{m}$     & $1.2_{-0.3}^{+0.4}\,(-6)$   & $1.6(-6)$        \\
         & $63.18\,\mu\mathrm{m}$      & $6.4 \pm 1.4\,(-6)$         & $4.3(-5)$        \\
$\mathrm{HD}$                       &                             &                  \\
         & $112.07\,\mu\mathrm{m}$     & $7.5\pm 6.1\,(-9)$          & $4.1(-9)$        \\
\hline
$\mathrm{H}_2$       &              &                             &                \\
                     & 0-0 S(0)     & $1.3 \pm 0.4\,(-7)$         & $1.1(-7)$   \\ 
                     & 0-0 S(1)     & $9.1_{-2.4}^{+3.8}\,(-7)$   & $9.7(-7)$  \\
                     & 0-0 S(2)     & $9.7_{-3.3}^{+3.9}\,(-7)$   & $1.3(-6)$  \\
                     & 0-0 S(3)     & $1.4_{-0.7}^{+0.6}\,(-6)$   & $3.2(-6)$  \\
                     & 0-0 S(4)     & $5.9 \pm 1.6\,(-7)$         & $7.4(-7)$  \\
                     & 0-0 S(5)     & $9.2_{-3.4}^{+2.4}\,(-7)$   & $9.3(-7)$  \\
                     & 1-0 S(1)     & $2.1\pm0.2\,(-7)$           & $1.4(-7)$  \\
                     & 1-0 S(2)     & $7.6\pm1.7\,(-8)$           & $4.7(-8)$  \\
                     & 2-1 S(1) / 1-0 S(1)  & 0.29                & $0.14$     \\
\hline
\end{tabular}\\
\end{table}

\begin{table}
\caption{Comparison of models and observations (combined) for the Orion Bar. Values in parentheses are powers of ten. \label{Tab:Orion_obs_model}} 

\scriptsize 
\begin{tabular}{l l| l| l l l }
\hline \hline
\multicolumn{2}{c}{Line} & Observed intensity & Model prediction\\
\multicolumn{2}{c}{} & $(\mathrm{W}\mathrm{m}^{-2}\mathrm{sr}^{-1})$ & $(\mathrm{W}\mathrm{m}^{-2}\mathrm{sr}^{-1})$\\
\hline
$^{12}\mathrm{CO}$ &                                       &                              &                     \\
                   & J=4-3                                 & $3.0  \pm 1.0\,(-7)$         & $9.4(-8)$           \\
                   & J=5-4                                 & $7.6_{-3.9}^{+3.2}\,(-7)$    & $1.9(-7)$           \\
                   & J=6-5                                 & $1.2_{-0.6}^{+0.4}\,(-6)$    & $3.4(-7)$           \\
                   & J=7-6                                 & $1.6_{-0.7}^{+0.4}\,(-6)$    & $5.2(-7)$           \\
                   & J=8-7                                 & $2.0_{-0.9}^{+0.7}\,(-6)$    & $7.4(-7)$           \\
                   & J=9-8                                 & $2.7_{-0.8}^{+0.7}\,(-6)$    & $9.8(-7)$           \\
                   & J=10-9                                & $2.5 \pm 0.8\,(-6)$          & $1.2(-6)$           \\
                   & J=11-10                               & $2.8_{-0.9}^{+0.7}\,(-6)$    & $1.4(-6)$           \\
                   & J=12-11                               & $2.5  \pm 0.8\,(-6)$         & $1.5(-6)$        \\
                   & J=13-12                               & $2.4_{-0.9}^{+0.6}\,(-6)$    & $1.4(-6)$           \\
                   & J=14-13                               & $2.3_{-0.7}^{+0.6}\,(-6)$    & $1.3(-6)$           \\
                   & J=15-14                               & $2.3_{-0.6}^{+0.5}\,(-6)$    & $1.1(-6)$           \\
                   & J=16-15                               & $1.5_{-0.3}^{+0.4}\,(-6)$    & $8.4(-7)$           \\
                   & J=17-16                               &  $1.0  \pm 0.2\,(-6)$        & $6.4(-7)$           \\
                   & J=18-17                               &  $5.0  \pm 1.1\,(-7)$        & $4.6(-7)$           \\
                   & J=19-18                               &  $3.2  \pm 0.6\,(-7)$        & $3.3(-7)$           \\
                   & J=20-19                               &  $2.2  \pm 0.5\,(-7)$        & $2.3(-7)$           \\
                   & J=21-20                               &  $1.0  \pm 0.3\,(-7)$        & $1.6(-7)$           \\
                   & J=23-22                               &  $6.4  \pm 2.1\,(-8)$        & $7.6(-8)$           \\
$^{13}\mathrm{CO}$ &                                       &                              &                     \\
                   & J=5-4                                 & $1.6_{-0.9}^{+0.7}\,(-7)$    & $2.2(-8)$           \\
                   & J=6-5                                 & $2.4 \pm 0.8\,(-7)$          & $3.5(-8)$           \\
                   & J=7-6                                 & $4.1_{-1.5}^{+0.8}\,(-7)$    & $4.5(-8)$           \\
                   & J=8-7                                 & $4.1_{-1.5}^{+0.9}\,(-7)$    & $5.2(-8)$           \\
                   & J=9-8                                 & $4.2_{-1.2}^{+1.3}\,(-7)$    & $5.3(-8)$           \\
                   & J=10-9                                & $3.8_{-1.5}^{+1.0}\,(-7)$    & $5.0(-8)$           \\
                   & J=11-10                               & $2.9_{-1.0}^{+0.6}\,(-7)$    & $4.5(-8)$           \\
                   & J=12-11                               & $1.9 \pm 0.6\,(-7)$          & $3.8(-8)$           \\
                   & J=13-12                               & $1.2 \pm 0.4\,(-7)$          & $3.2(-8)$           \\
                   & J=15-14                               & $3.8 \pm 1.1\,(-8)$          & $2.0(-8)$           \\
                   & J=16-15                               & $2.0 \pm 1.1\,(-8)$          & $1.5(-8)$           \\
$\mathrm{CH}^+$    &                                       &                              &                         \\
                   & J=1-0                                 & $1.3 \pm 0.1\,(-7)$          & $6.2(-8)$             \\
                   & J=2-1                                 & $1.5 \pm 0.2\,(-7)$          & $2.8(-7)$             \\
                   & J=3-2                                 & $1.2 \pm 0.3\,(-7)$          & $2.3(-7)$              \\
                   & J=4-3                                 & $1.2 \pm 0.3\,(-7)$          & $1.8(-7)$              \\
                   & J=5-4                                 & $1.0 \pm 0.4\,(-7)$          & $1.7(-7)$              \\
                   & J=6-5                                 & $6.8 \pm 3.9\,(-8)$          & $1.5(-7)$              \\
$\mathrm{OH}$      &                                       &                              &                        \\
                   & $119.4416\,\mu\mathrm{m}$  & $2.8 \pm 0.6\,(-7)$          & $6.3(-7)$              \\
                   & $119.2345\,\mu\mathrm{m}$  & $2.4 \pm 0.5\,(-7)$          & $5.5(-7)$              \\
                   &  $79.1792\,\mu\mathrm{m}$   & $2.1 \pm 0.8\,(-7)$          & $3.8(-7)$              \\
                   &  $79.171156\,\mu\mathrm{m}$ & $2.4 \pm 0.8\,(-8)$          & $4.2(-7)$              \\
                   &  $163.3962\,\mu\mathrm{m}$  & $5.0 \pm 1.1\,(-8)$          & $1.6(-7)$              \\
                   &  $163.0153\,\mu\mathrm{m}$  & $4.6 \pm 1.1\,(-8)$          & $1.9(-7)$              \\
                   &  $84.5967\,\mu\mathrm{m}$   & $1.1 \pm 0.3\,(-7)$          & $3.5(-7)$              \\
                   &  $84.4203\,\mu\mathrm{m}$   & $1.2 \pm 0.4\,(-7)$          & $2.3(-7)$              \\
                   &  $65.2789\,\mu\mathrm{m}$   & $1.8 \pm 2.5\,(-8)$          & $1.1(-7)$              \\
                   &  $65.1318\,\mu\mathrm{m}$   & $4.6 \pm 2.5\,(-8)$          & $2.1(-8)$              \\
$\mathrm{HD}$      &                                       &                              &                         \\
                   & $112.07\,\mu\mathrm{m}$         & $ < 1.4\,(-8)$               & $1.0(-8)$             \\
                   &  $56.23\,\mu\mathrm{m}$          & $1.1 \pm 0.4\,(-7)$          & $4.0(-8)$             \\
$\mathrm{C}^+$     &                                       &                              &                          \\
                   &  157.68$\,\mu$m                          & $2.3_{-0.6}^{+0.9}\,(-5)$    & $1.4(-6)$            \\
$\mathrm{C}$       &                                       &                              &                           \\
                   & $609.13\,\mu\mathrm{m}$                  & $2.9\pm 0.3 (-8)$            & $1.3(-8)$            \\
                   & $370.41\,\mu\mathrm{m}$                  & $3.8\pm 0.3 (-8)$            & $8.8(-8)$            \\
$\mathrm{O}$       &                                       &                              &                         \\
                   & $145.53\,\mu\mathrm{m}$                  & $2.1 \pm 0.4\,(-5)$          & $5.7(-6)$           \\
                   & $63.18\,\mu\mathrm{m}$                   & $1.9_{-0.4}^{+0.5}\,(-4)$    & $1.4(-4)$          \\
\hline
$\mathrm{H}_2$     &                                       &                              &                           \\
                   & 0-0 S(0)                              & $9.0 \pm 3.0\,(-7)$          & $2.1(-7)$              \\
                   & 0-0 S(1)                              & $3.7_{-2.9}^{+4.1}\,(-6)$    & $2.0(-6)$              \\
                   & 0-0 S(2)                              & $2.2_{-1.5}^{+2.4}\,(-6)$    & $2.8(-6)$              \\
                   & 0-0 S(3)                              & $6.0\pm1.5\,(-6)$            & $1.1(-5)$              \\
                   & 0-0 S(4)                              & $1.6_{-1.2}^{+1.9}\,(-6)$    & $3.7(-6)$              \\
                   & 0-0 S(5)                              & $6.4\pm 1.0\,(-6)$           & $8.2(-6)$              \\
                   & 1-0 S(1)                              & $4.7\pm 1.1\,(-7)$           & $2.8(-6)$              \\
                   & 2-1 S(1)                              & $1.2 \, (-7)$                & $5.6(-8)$              \\
\hline
\end{tabular}\\
\end{table}


\begin{figure*} [ht]
\begin{center}
\includegraphics[width=0.45\textwidth]{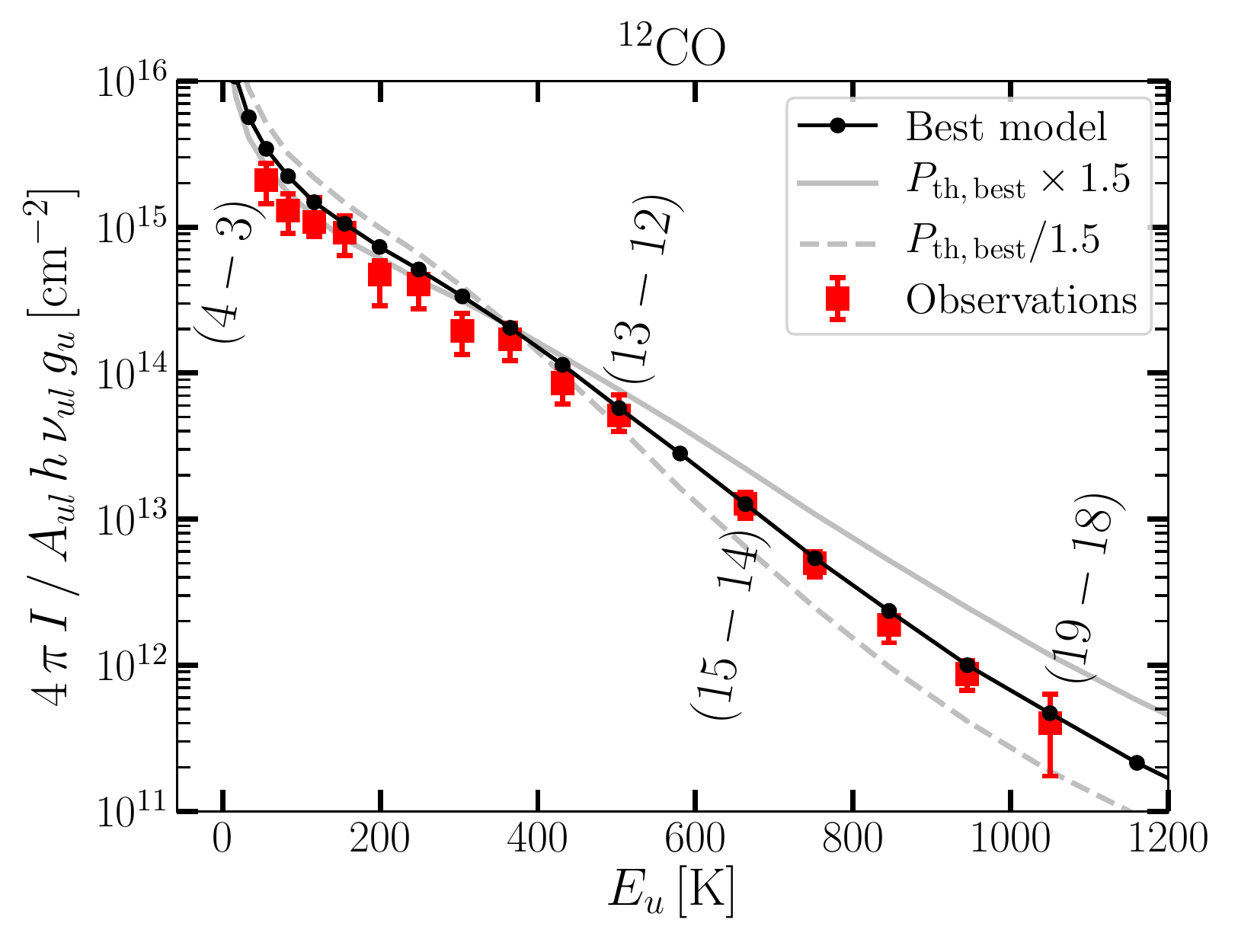}
\includegraphics[width=0.45\textwidth]{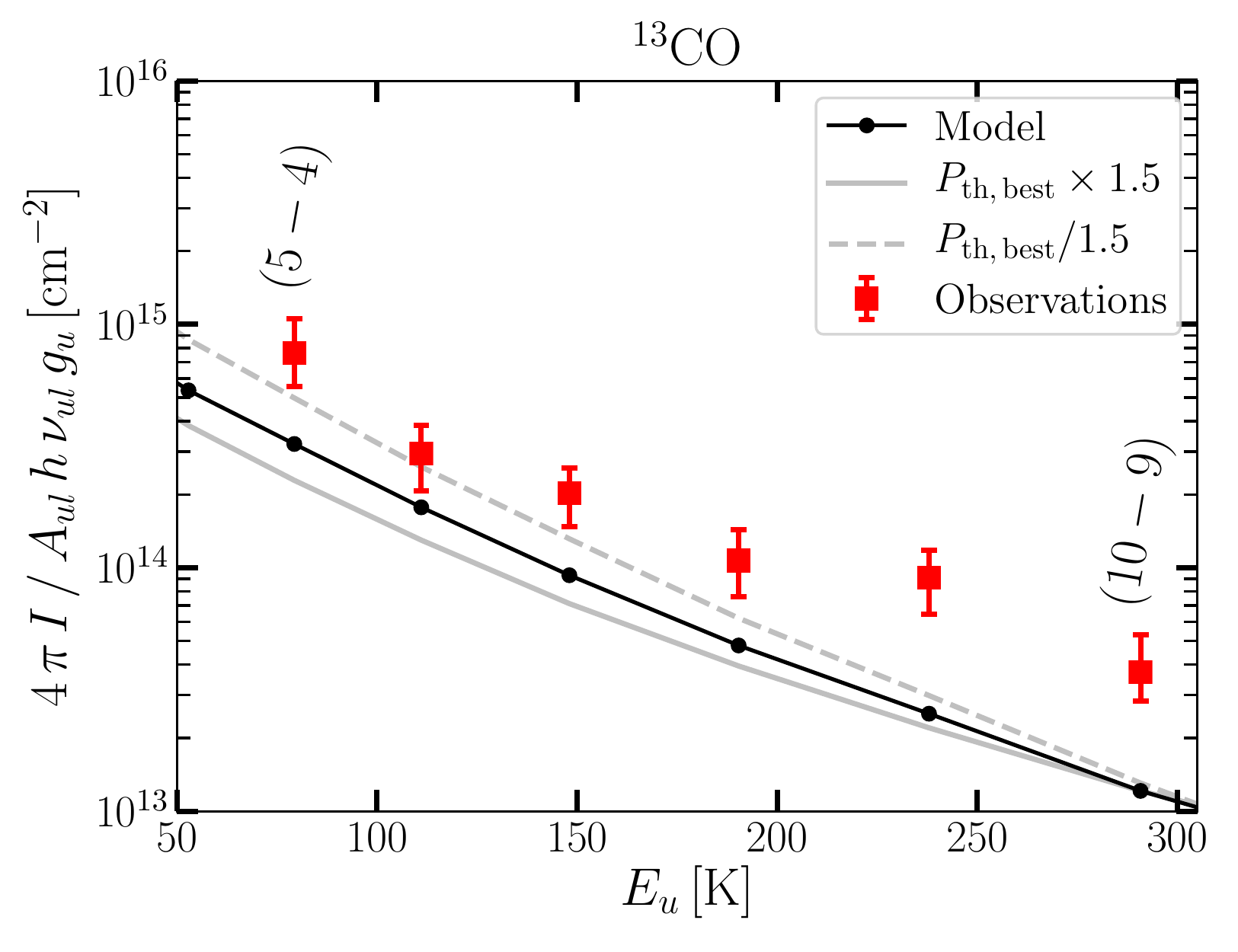}\\
\includegraphics[width=0.45\textwidth]{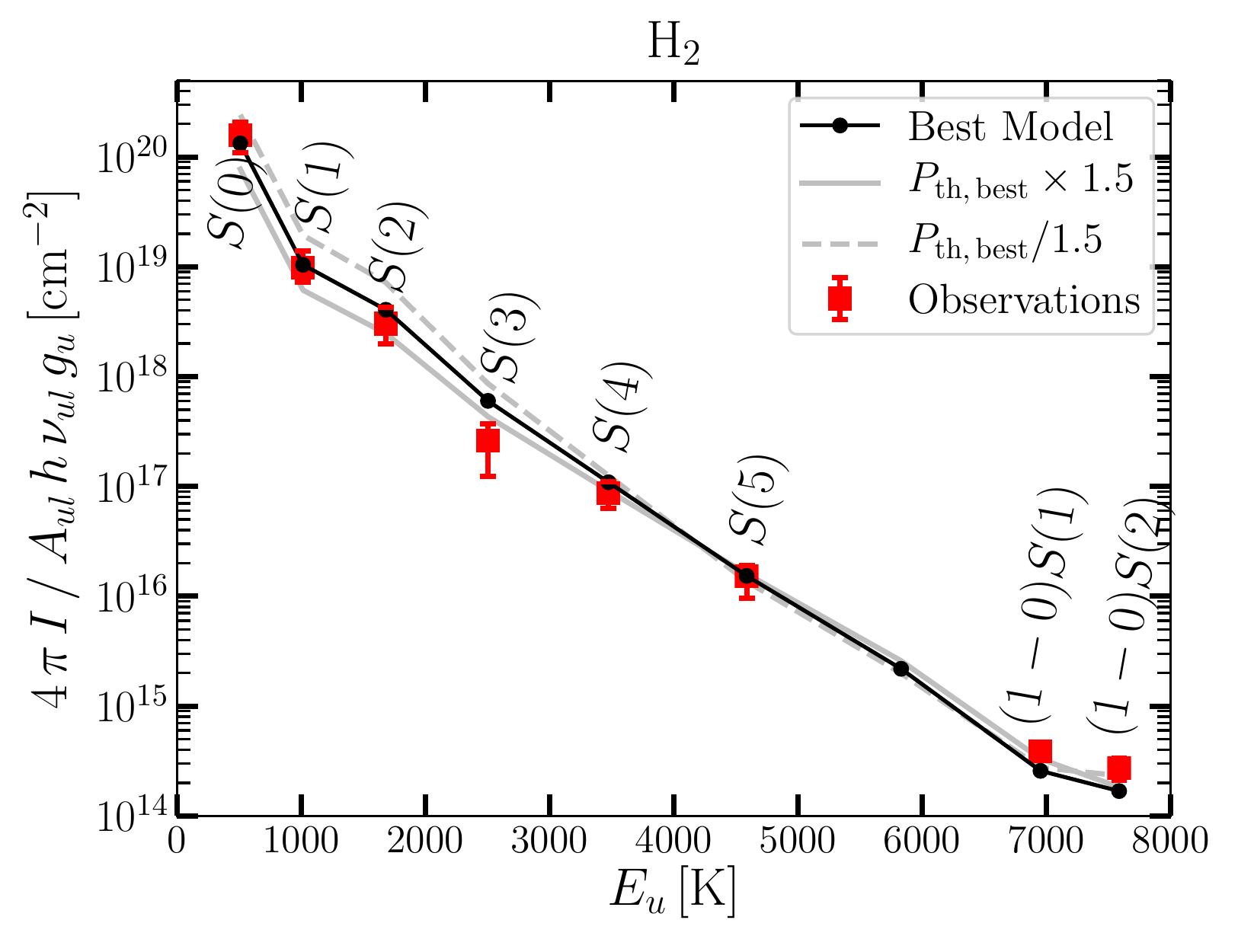}
\includegraphics[width=0.45\textwidth]{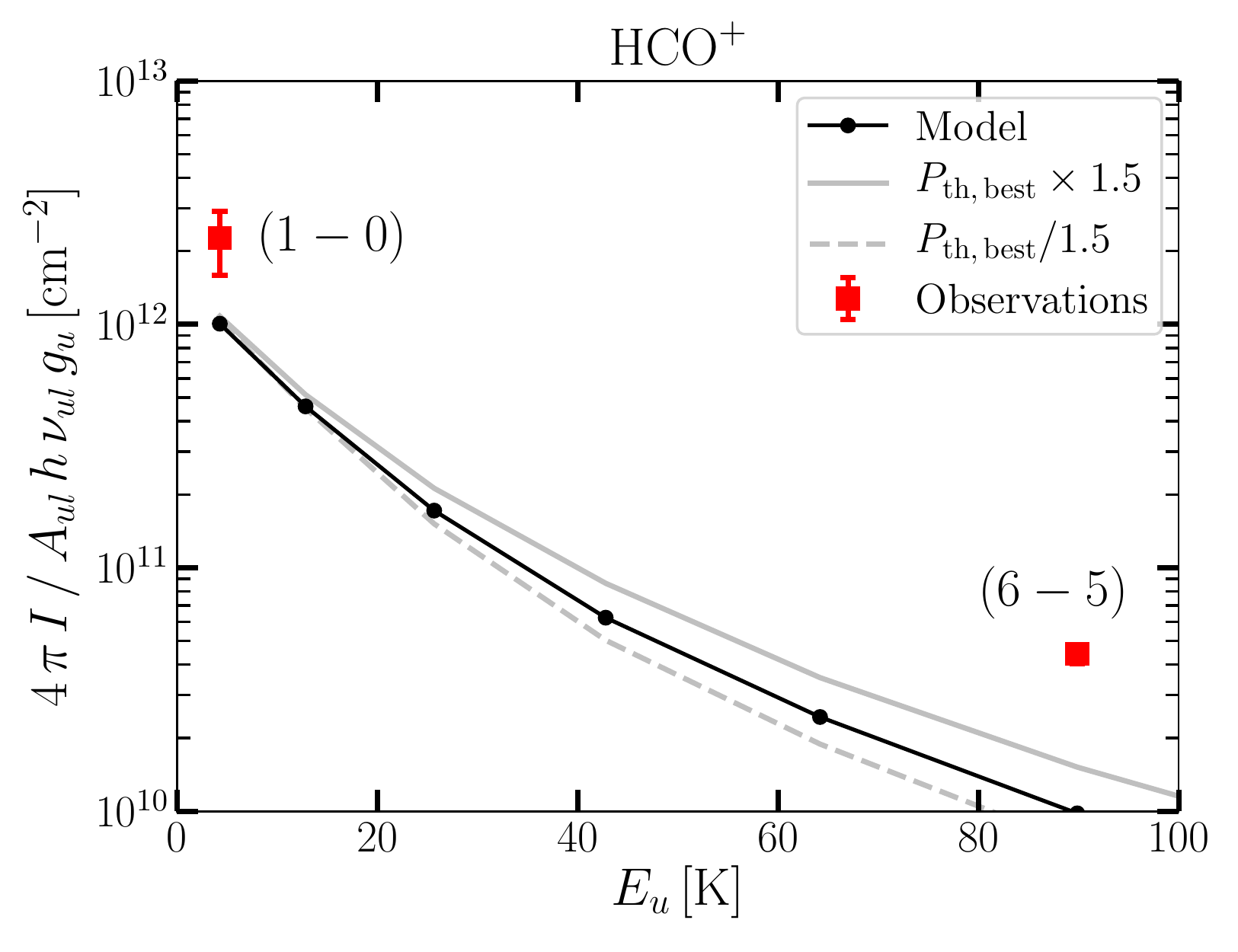}\\
\includegraphics[width=0.45\textwidth]{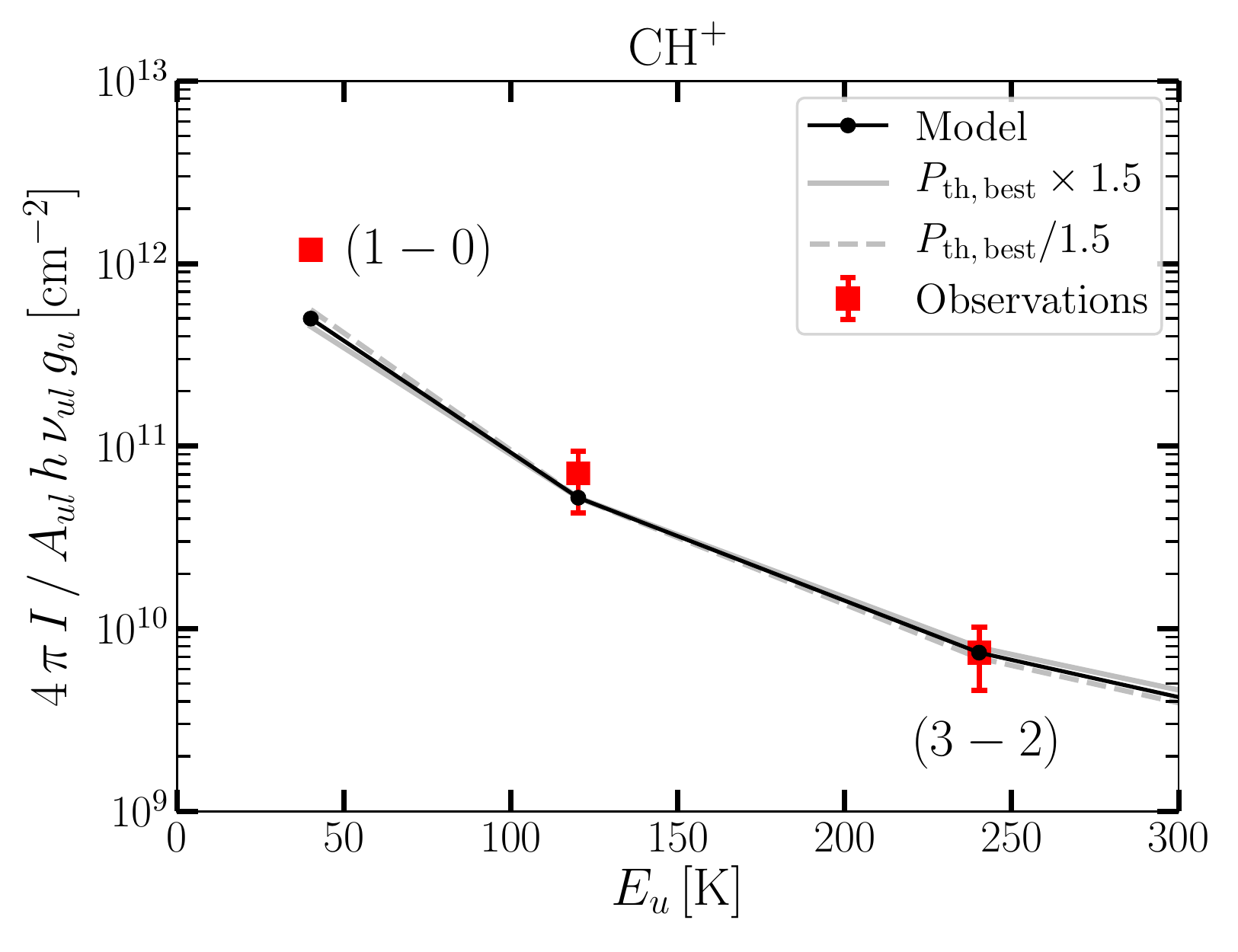}
\includegraphics[width=0.45\textwidth]{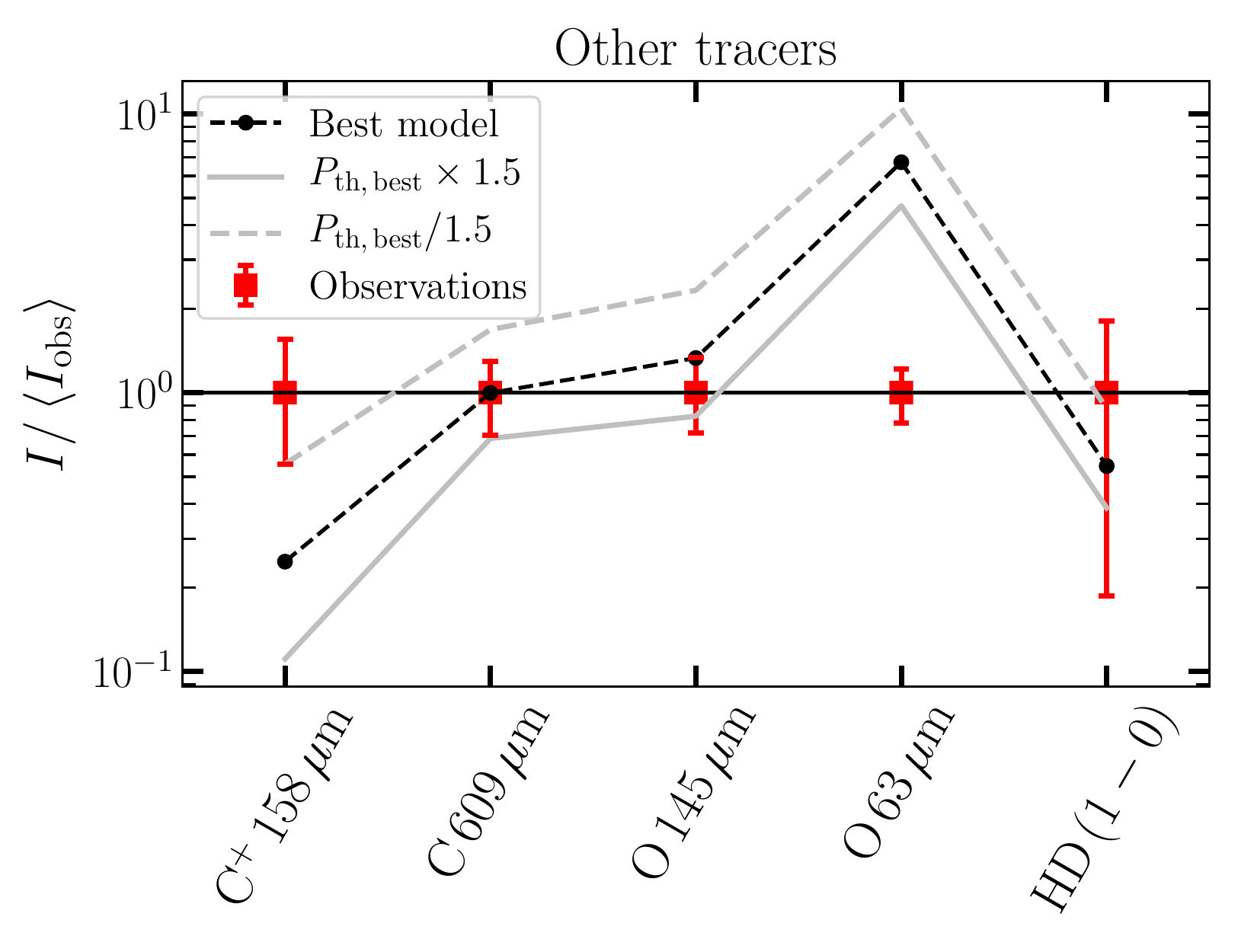}
\end{center}
\caption{Excitation diagram of the different tracers observed in NGC~7023 after dilution correction (red squares) and the best fit model (black, $P_{\mathrm{th}}=10^8$ K cm$^{-3}$, $A_\textrm{V}^{\mathrm{tot}}=10$, global scaling factor = 0.7). The best model has been chosen to optimise the fitting of the $^{12}$CO (high-J lines), H$_2$ and CH$^+$ lines only.  In the last panel, the intensity values are normalized by the mean observed value for each line. The grey lines show the obtained variability when the thermal pressure is divided (dashed lines) or multiplied (plain lines)  by a factor of 1.5. The best value for the scaling factor was found to be 1.2 for the model at $P_{\mathrm{th}}$/1.5 and 0.47 for the model at $P_{\mathrm{th}}$$\times$1.5.
\label{Fig:N7023_Best_model_12CO}}
\end{figure*}

\begin{figure*}
\begin{center}
\includegraphics[width=0.45\textwidth]{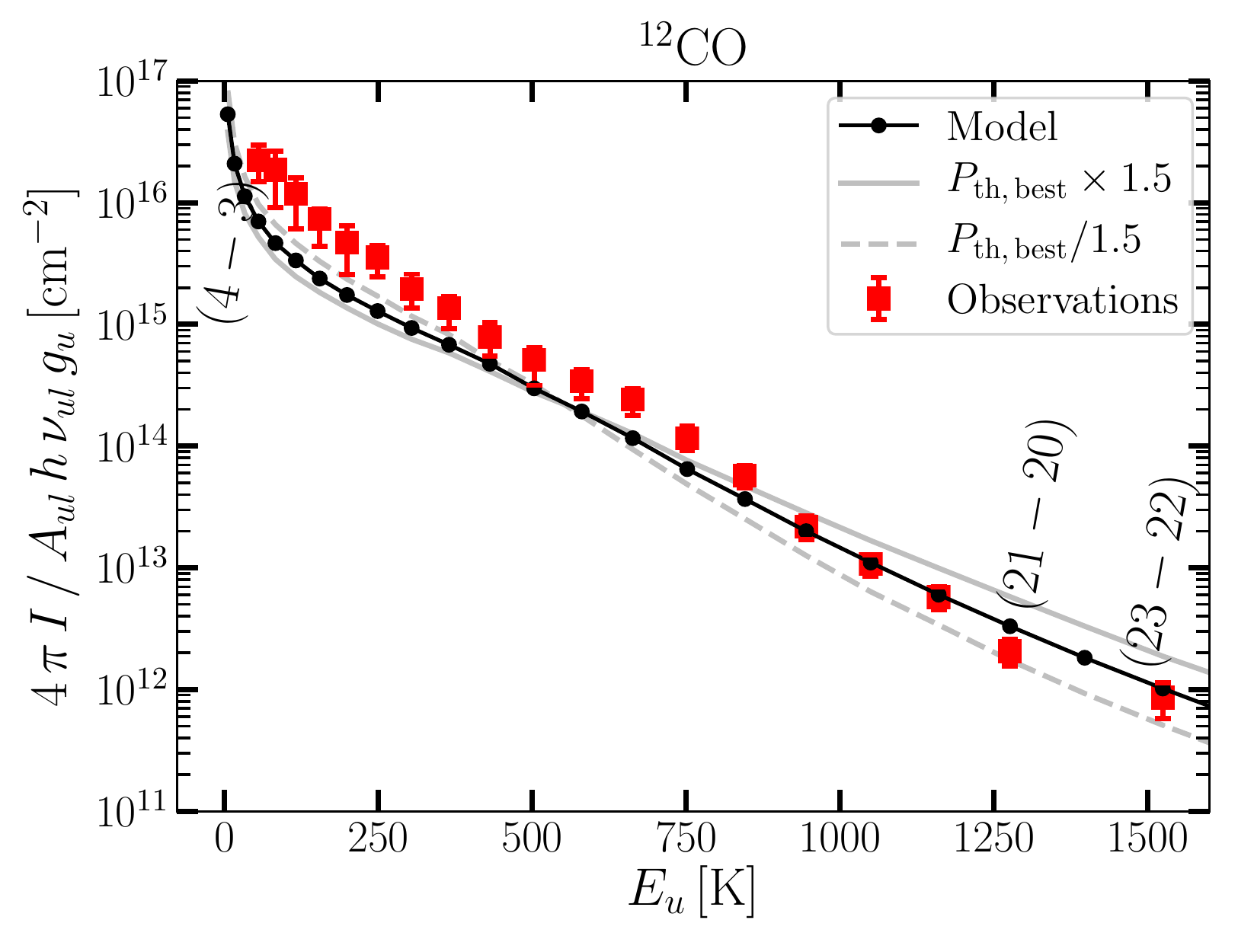}
\includegraphics[width=0.45\textwidth]{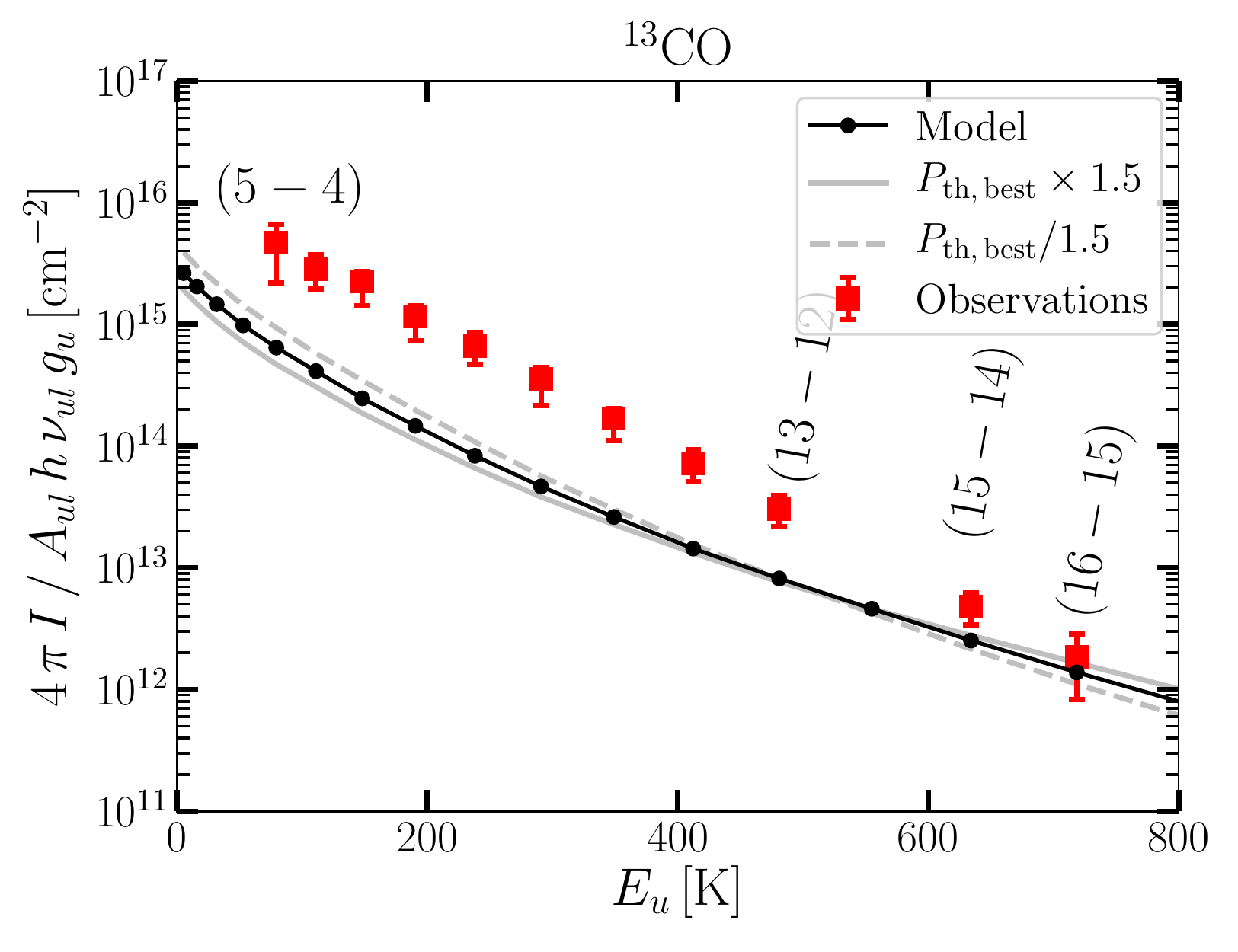}\\
\includegraphics[width=0.45\textwidth]{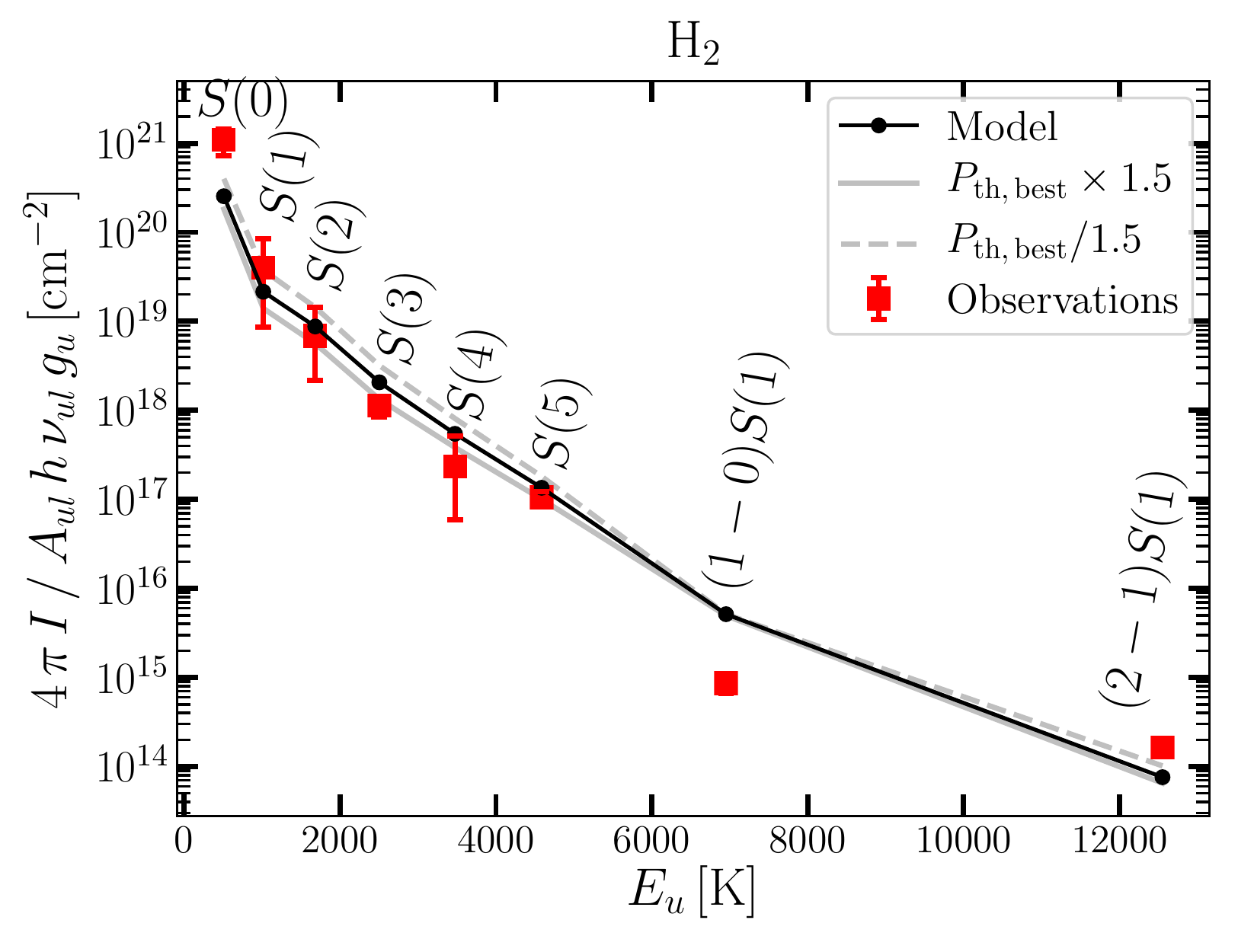}
\includegraphics[width=0.45\textwidth]{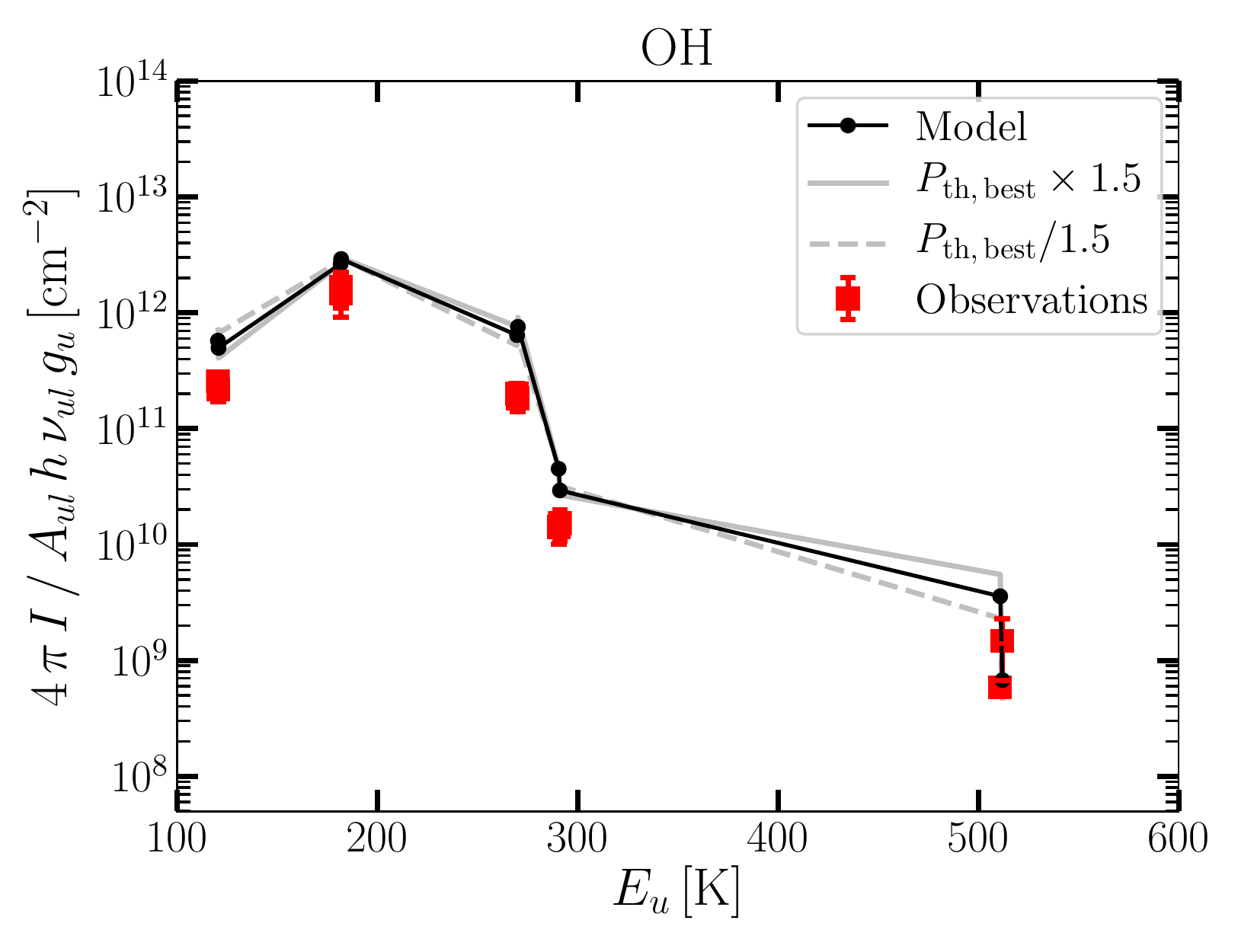}\\
\includegraphics[width=0.45\textwidth]{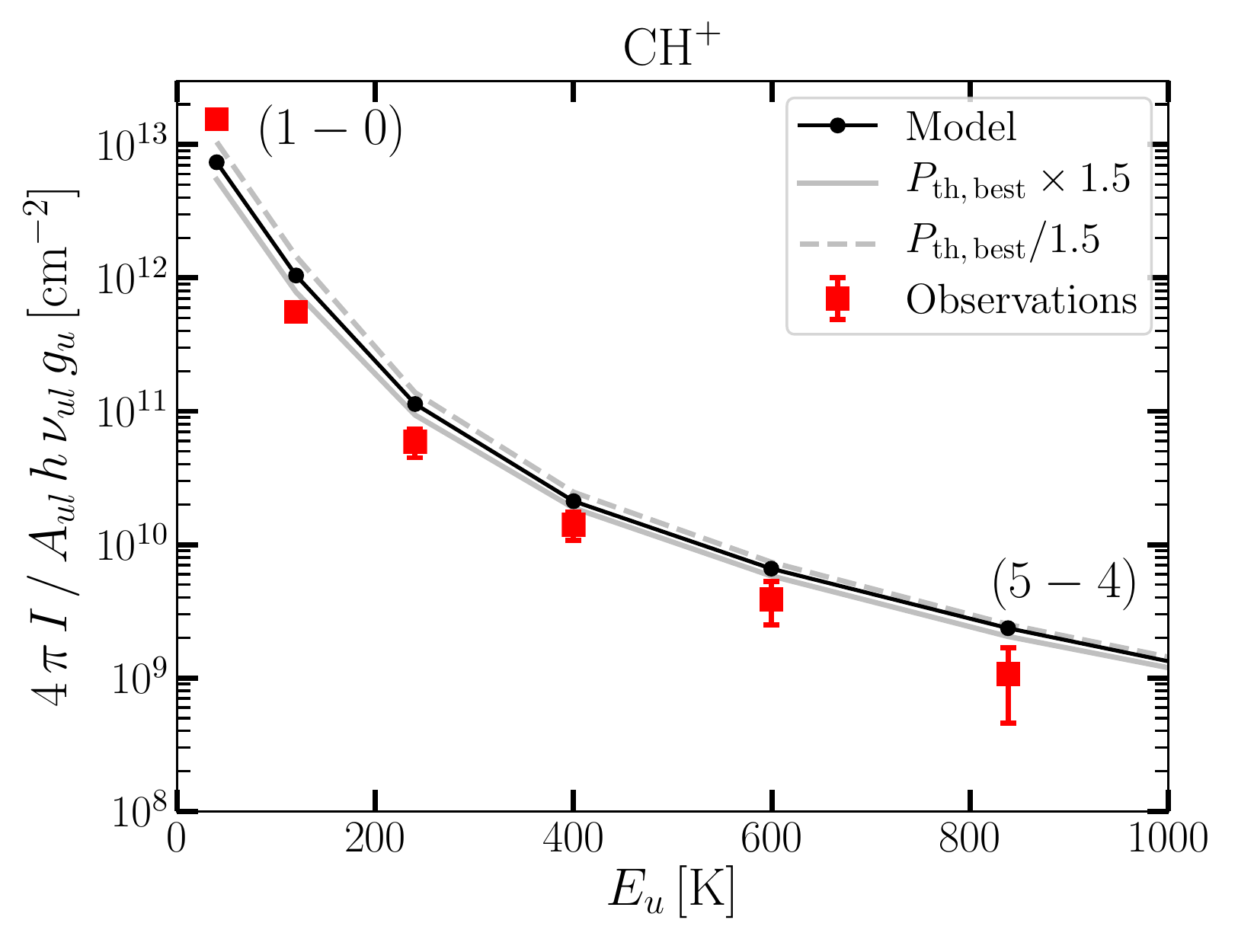}
\includegraphics[width=0.45\textwidth]{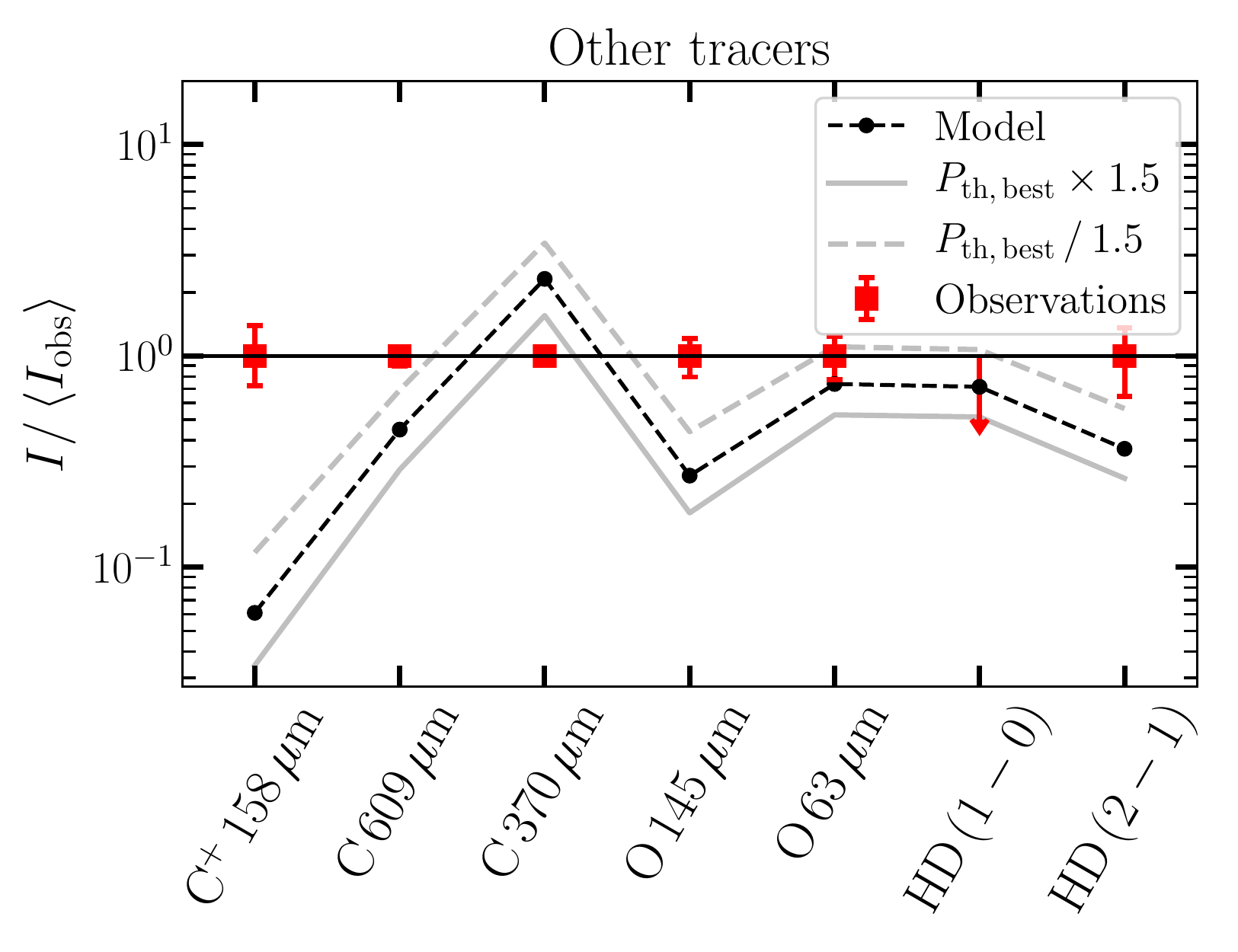}
\end{center}
\caption{Excitation diagram of the different tracers observed in the Orion Bar (red squares) and the best model (black, $P_{\mathrm{th}}=2.8\times10^8$ K cm$^{-3}$, $A_\textrm{V}^{tot} = 10$ and global scaling factor $f = 1.3$).  The best model has been chosen to optimise the fitting of the $^{12}$CO (high-J lines), rotational H$_2$ and CH$^+$ lines only.  In the last panel, the intensity values are normalized by the mean observed value for each line.  The grey lines show the obtained variability when the thermal pressure is divided (dashed lines) or multiplied (plain lines)  by a factor of 1.5. The best value for the scaling factor was found to be 2.0 for the model at $P_{\mathrm{th}}$/1.5 and 0.9 for the model at $P_{\mathrm{th}}$$\times$1.5.
\label{Fig:OrionBar_Best_model_12CO}}
\end{figure*}

\begin{figure*} [ht]
\sidecaption
  \includegraphics[width=12cm]{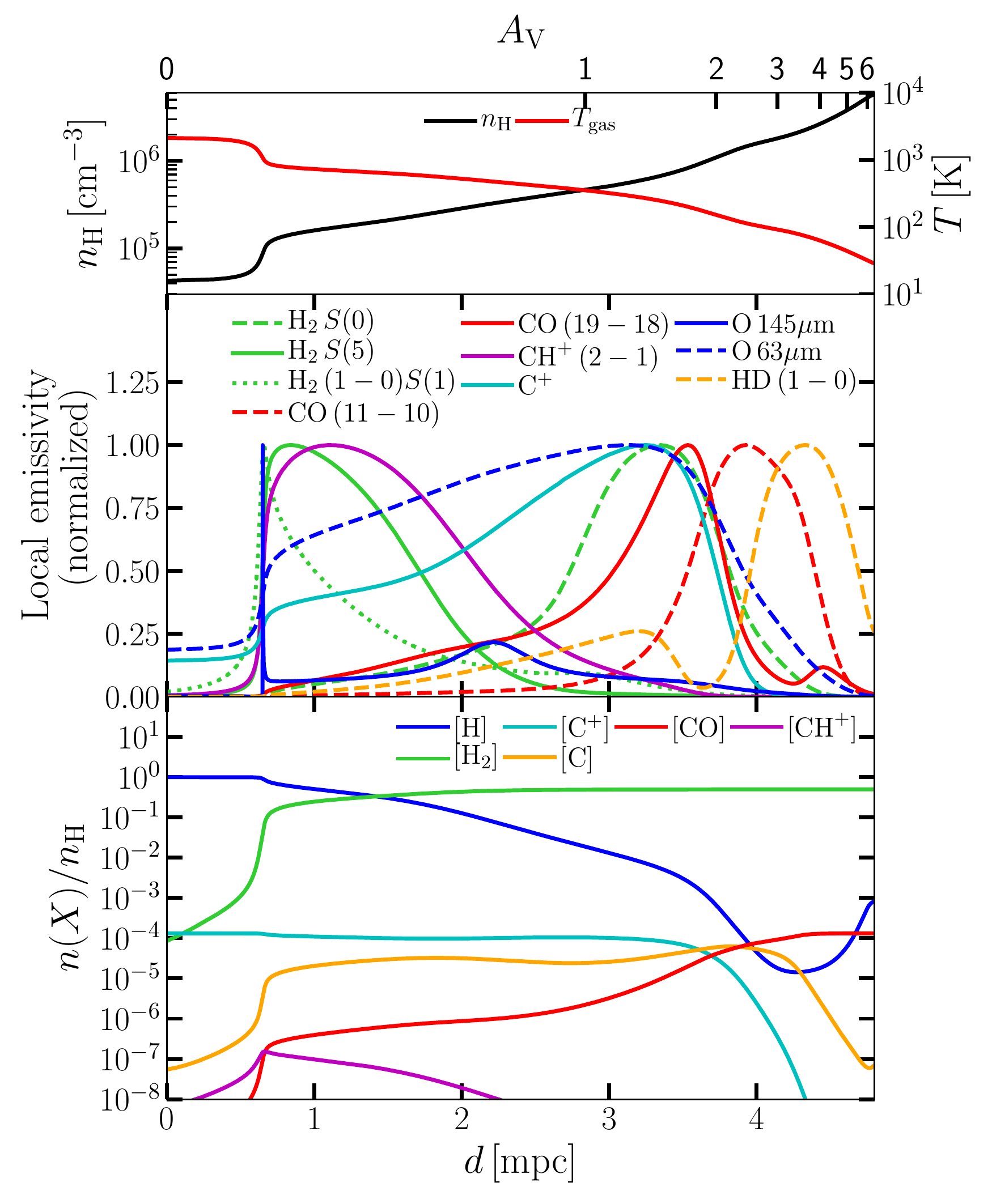}
\caption{NGC 7023 PDR model ($P_{\mathrm{th}}=10^8$ K cm$^{-3}$, global scaling factor = 0.7). Top: Evolution of the H nuclei number density and gas temperature with A$_\textrm{V}$ or distance (in 10$^{-3}$ pc). Center: Spatial profile of the local emissivities of the main tracers. The emissivities have been scaled so that their maximum in the cloud is 1. Bottom:  Spatial profiles of the abundances of the species of interest in the model.\label{Fig:N7023_Best_model_locem}
}
\end{figure*}

\begin{figure*}[ht!]
\sidecaption
  \includegraphics[width=12cm]{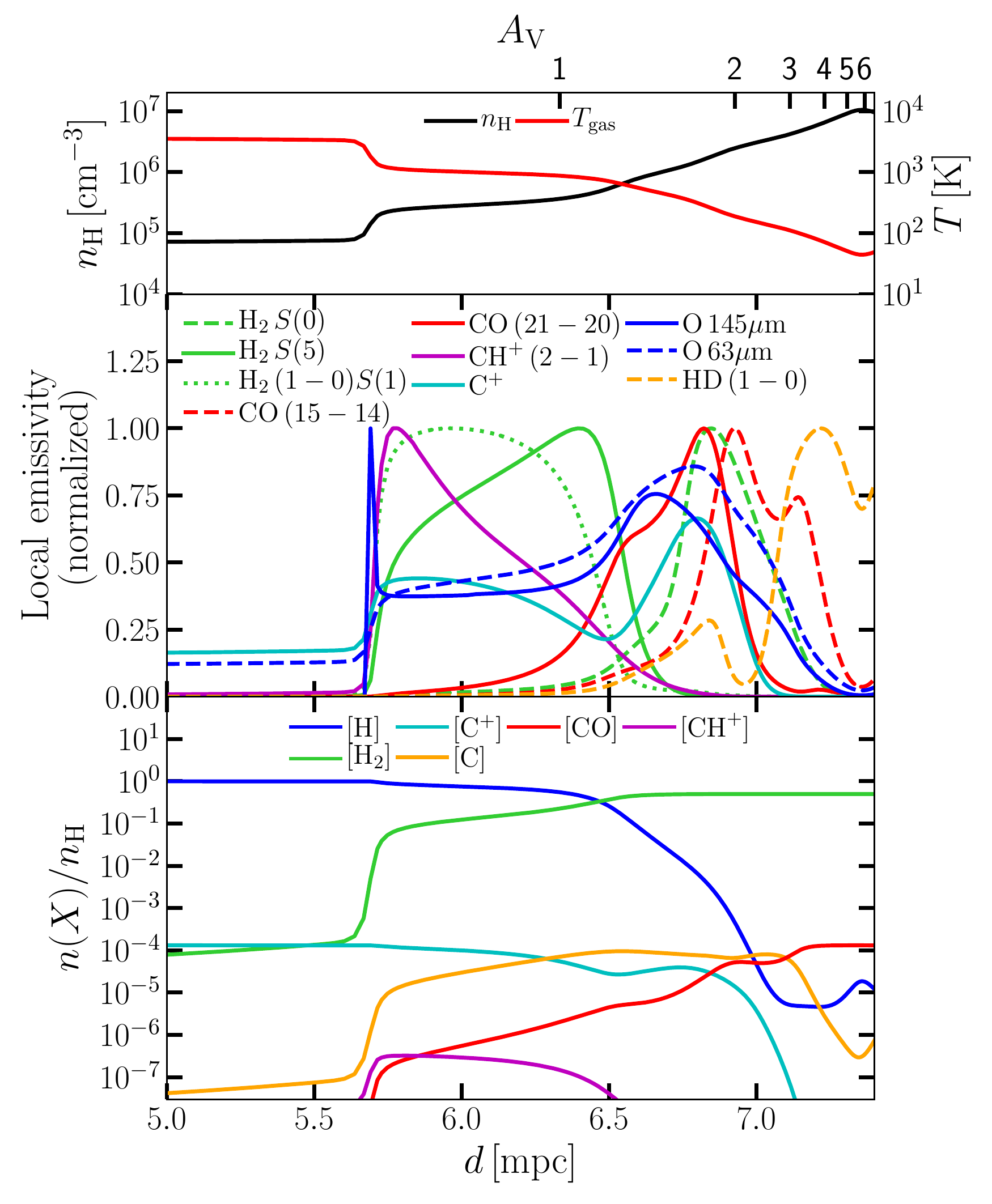}
\caption{Orion Bar PDR model ($P_{\mathrm{th}}=2.8\times10^8$ K cm$^{-3}$, global scaling factor = 1.3). Top: Evolution of the H nuclei number density and gas temperature with A$_\textrm{V}$ or distance (in 10$^{-3}$ pc). Middle: Spatial profile of the local emissivities of the main tracers. The emissivities have been scaled so that their maximum in the cloud is 1. Bottom: Spatial profiles of the abundances of the species of interest in the model. \label{Fig:OrionBar_Best_model_locem}
}
\end{figure*}


\subsubsection{NGC 7023}
\label{Sec:model_results_N7023}

The best fit was found for a model with $P_{\mathrm{th}}=10^8$ K cm$^{-3}$ and with a global scaling factor of $f=0.7$. Comparison of line intensities computed by this model to observed values is presented in Fig. \ref{Fig:N7023_Best_model_12CO}. This model shows excellent agreement with the high-J $^{12}$CO lines (above J$_{up} = 10$), with the H$_2$ pure rotational lines (except the S(3) line), and with the CH$^+$ lines (except the (1-0) line). We also note that the model is able to fit both the ortho- and para-H$_2$ lines and therefore can account for the observed non-equilibrium ortho-to-para ratio \citep[][see Sect.\,\ref{subsec:excitation} for further discussion]{Fuente99}.
The discrepancy obtained for the H$_2$ S(3) line (a factor of two brighter in the model than in the observations) likely indicates that the actual dust extinction in the line of sight
towards NGC\,7023 has a stronger silicate feature than assumed in our model (cf. Tab.\,\ref{Tab:InputParam}).

Further evidence of the adequacy of the model is given by its success in reproducing additional lines that were not used in the adjustment of the model (see also Fig. \ref{Fig:N7023_Best_model_12CO}). The $^{12}$CO lines below J$_{up} = 11$ show an excellent agreement and the H$_2$ vibrational lines (1-0) S(1) and (1-0) S(2) are reproduced within a factor $< 2$. 
For HCO$^+$, the model reproduces the J=1-0 line within a factor of 2, but underestimates the J=6-5 line by a factor $\sim 4$. This line is very sensitive to the density due to the high dipole moment of HCO$^+$. For instance for a gas kinetic temperature of T=50\,K a change of a factor of 2 in the density n(H$_2$) leads to an increase of the J=6-5 line by a factor 2 while the J=1-0 remains practically unchanged. This deviation suggests that the density provided by the isobaric model is good within a factor of a few.
The [OI] $145\,\mu\mathrm{m}$, [CI] $609\,\mu\mathrm{m}$ and HD J=1-0 lines are well reproduced. Other lines on this same figure exhibit stronger differences of at most a factor of $\sim6$: the [CII] $158\,\mu\mathrm{m}$ line is underestimated and the [OI] $63\,\mu \mathrm{m}$ line is overestimated. 
Herschel observations have shown that the [CII] emission comes not only from the H$_2$ filaments but also from the surrounding more diffuse gas strongly emitting in the PAH features \citep{Joblin10}.
The adopted beam dilution factor is in this case not justified and this correction could account for most of the discrepancy between the observed and calculated fluxes.
The [OI] 63\,$\mu$m line is overestimated in the model but this line is known to be optically thick and affected by self-absorption in most cases. Its study therefore requires a detailed analysis in velocity components. 
Finally, the model underestimates the $^{13}$CO lines by a factor of $\sim 3$.  This is consistent with the arguments raised in Sec.\,\ref{Sec:CO_rot}, in particular that the  $^{13}$CO emission arises from other regions inside the beam than the bright sharp interface.

Overall, we can conclude that the model is able to reproduce 17 observational constraints with only two free parameters, which is indicative of the adequacy of the model and of the physics it contains to describe the warm molecular edge region of the PDR. The grey lines in Fig.\,\ref{Fig:N7023_Best_model_12CO} demonstrate that the gas thermal pressure is best constrained by the high-J $^{12}$CO lines.

\subsubsection{Orion Bar}
\label{Sec:model_results_OBar}

The best fit was found for $P_{\mathrm{th}}=2.8\times 10^8$ K cm$^{-3}$, and a global scaling factor of $f=1.3$. The results of this  model compared to the observations are presented in Table \ref{Tab:Orion_obs_model} and Fig.~\ref{Fig:OrionBar_Best_model_12CO}. The model provides a satisfactory fit of the high-J $^{12}$CO lines,  CH$^+$ lines,  pure rotational lines of H$_2$ (with maximal differences by a factor of 2 except for the H$_2$ S(0) line that is underestimated by a factor of $\sim 4$). We thus obtained a reasonable agreement, within a factor of a few, for all the tracers used in the fitting procedure.

Most of the other tracers (not used in the fitting procedure) show similar differences. The low lying  $^{12}$CO lines (J$_{up} < 11$) are underestimated by up to a factor of $\sim 3$. The OH, HD, [OI] and [CI] lines are reproduced within a factor of 3-4. Some tracers show however stronger discrepancies such as the $^{13}$CO lines, which are underestimated by up to a factor of $\sim8$ except for the highest line, and the [CII] $158\,\mu$m line that is underestimated by a factor of 16. For these lines at least part of the discrepancy is related to the correction by the beam dilution factor. We have discussed previously that some of the emission from  $^{13}$CO is expected to arise from the surrounding molecular cloud  (cf. Sec.\,\ref{Sec:CO_rot}). Emission in the [CII] line is also known to be extended in this region \citep[see][]{Goicoechea15b}.
In the case of the vibrational H$_2$ lines, the (1-0) S(1) line is strongly overestimated (factor of $\sim 6-7$) while the (2-1) S(1) line is underestimated by a factor of $\sim$2 only.

To summarize, the agreement of the model with observations is not as good for Orion Bar as for NGC\,7023 and leaves some areas of concern.  On average, from the spatial structure that is reported in the next section we can derive that the emission lines coming from the most external layers are overpredicted by the model (CH$^+$, H$_2$ lines except low H$_2$ rotational lines), whereas those arising from more internal layers (low- and mid-J CO lines, low H$_2$ rotational lines) are underpredicted.
This suggests that a single pressure is not sufficient to describe the evolution of gas density and temperature across the hot/warm irradiated interface (see also the grey curves in Fig.~\ref{Fig:OrionBar_Best_model_12CO}).

\subsubsection{Spatial structure}
\label{Sec:SpatialStructure}

In Figs.\,\ref{Fig:N7023_Best_model_locem} and \ref{Fig:OrionBar_Best_model_locem}, we show the spatial structure of the calculated PDRs in terms of the gas temperature and H nuclei number density (upper pannels), of the abundances of the different species (lower panels), and of the emission regions of the different lines (middle panels). Our calculations were performed on a slab of gas of A$_\textrm{V}$=10. However, we provide here information for the region up to A$_\textrm{V}\sim$5 in which the tracers we selected emit. Although the thickness and absolute position of the transitions depend on the characteristics of each PDR, the general trends can be summarized as follows:  the atomic part has a density of a few  $10^4$\,$\mathrm{cm}^{-3}$ (close to $10^5$\,$\mathrm{cm}^{-3}$ in Orion Bar) followed by a hot/warm ($T=1000-100\,\mathrm{K}$) and relatively dense (few $10^5$ to few $10^6\,\mathrm{cm}^{-3}$) molecular part. In the Orion Bar model, the atomic region is found to be significantly more extended ($\sim 6\times10^{-3}$\,pc) than in NGC~7023 ($\sim 0.5\times10^{-3}$\,pc). It has been truncated on Fig.\,\ref{Fig:OrionBar_Best_model_locem} for better readability.
In both PDRs, the hot/warm molecular region is found to start significantly before the H/H$_2$ transition. Indeed a sharp increase of the H$_2$ abundance is seen at $0.6\times10^{-3}$ pc whereas the H/H$_2$ transition occurs around $1.5\times 10^{-3}$ pc, in NGC\,7023. For the Orion Bar, these values are $5.7\times10^{-3}$ pc and $6.5\times10^{-3}$ pc, respectively. Note that this increase in H$_2$ abundance impacts the penetration of UV photons, which leads to a significant decrease of the heating rate and therefore of the temperature in the PDR. 
The size of the hot/warm molecular region is $\sim 1.5\times10^{-3}$ pc in Orion Bar, which is a factor of 2 lower than in NGC~7023. Taking into account the distances to the studied objects this would correspond to a thickness of $\sim 2"$  for NGC\,7023 and $\sim 0.8"$  for the Orion Bar, which agrees well with the observations of the filaments in NGC\,7023 and the dilution procedure we adopted to correct the observed intensities.

Emission in the excited CO lines peaks at the back end of the warm molecular region, close to the C$^+$/C/CO transition, with higher-J lines peaking closer to the surface. The emission profiles of the highest lines (e. g. $J=$19-18 or 21-20) exhibit a wide left tail extending towards the H/H$_2$ transition and accounting for roughly half of the total emission. This fraction of the emission comes from a low fraction (fractional abundance $\sim 10^{-6}$) of CO existing before the C$^+$/C/CO transition. In this region, we find CO formation to be initiated by CH$^+$ formation as explained in Sec.\,\ref{subsec:excitation}. Emission in the H$_2$ rotational lines spans the whole warm molecular region, with S(5) peaking close to the H/H$_2$ transition and S(0) peaking closer to the high-J CO emission peaks. Vibrational H$_2$ emission (e.g (1-0) S(1) line) peaks at the very edge of the hot/warm molecular region, where the H$_2$ abundance starts to increase steeply due to self-shielding processes. CH$^{+}$ emission coincides with the higher H$_2$ rotational lines (such as S(5)) and the vibrational line (1-0) S(1).
Finally, the fine structure lines [CII] (158 $\mu$m) and [OI] (63 and 145\,$\mu$m) arise in the whole hot/warm molecular region. The [CII] 158\,$\mu$m and [OI] 63\,$\mu$m lines also show significant emission in the atomic region in our model.

\section{Discussion}
\label{Sec:Discussion}

\begin{figure*} [ht]
\begin{center}
\includegraphics[width=1\textwidth]{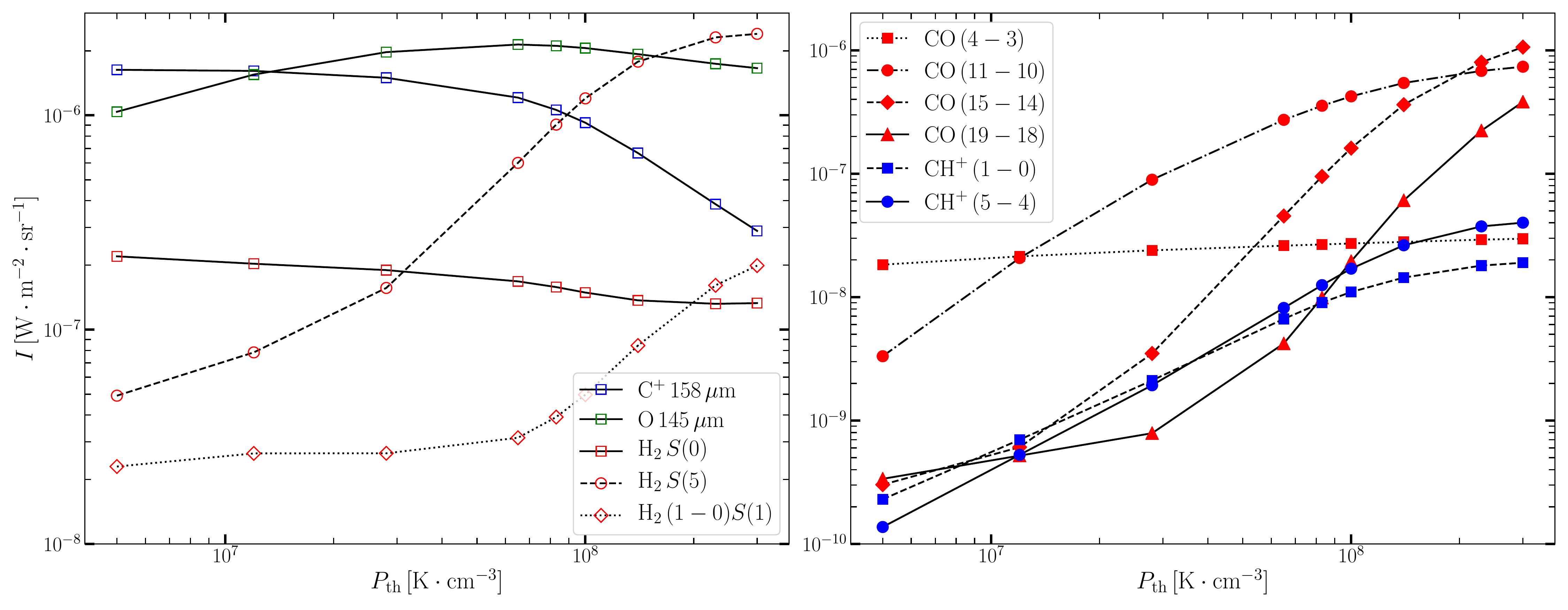}
\end{center}
\caption{Variation of the intensity of different lines of interest with the gas thermal pressure for the illumination conditions of NGC~7023~NW, as computed by the PDR model.
\label{Fig:Pressure_effect}}
\end{figure*}

\subsection{Formation and excitation of CO at the PDR bright edges}
\label{subsec:excitation}
The main objective of this paper is to analyse the emission in high-J CO lines in PDRs. This emission, as observed in the two template PDRs, NGC\,7023 and Orion Bar, points either to an enhanced formation of CO in the PDR warm external layers or to an increased temperature in the layers containing CO. As will be discussed in more details below, this question is intimately related to another important question, which is the formation rate of H$_2$ in PDRs. Several authors have earlier studied H$_2$ rotational emission in PDRs and found that a significant fraction of H$_2$ has an excitation temperature in the range 400-700 K, much higher than expected from models \citep{Parmar91, Draine99, Allers05, Habart11}. The authors suggested that this is due either to an increase of the photoelectric heating efficiency or a larger H$_2$ formation rate on grains at high temperatures or to dynamical effects such as advection. The effect of an advancing photodissociation front was also invoked in  NGC~7023\,NW to account for the non-equilibrium values of the ortho-to-para ratio \citep{Lemaire99, Fuente99,Fleming10, Le17}. Since our models of NGC\,7023 and Orion Bar are successful to account for both CO high-J and H$_2$ line intensities (as well as the ortho-to-para ratio), it is worth to investigate the underlying reason for that.

In our models, we use the prescription of \cite{LeBourlot12} that considers the formation on grains via the Langmuir-Hinshelwood (LH) and the Eley Rideal (ER) mechanisms. The resulting formation rate at the edge of the NGC\,7023 and Orion Bar models are found to be 3-4 times higher than the rate determined for the diffuse gas of $\sim 3\times10^{-17}$ cm$^3$ s$^{-1}$ \citep{Jura74, Gry02}. The ER mechanism is found to dominate at the edge of the PDR \citep{Rollig13} because the grains are warm. We note that \cite{Bron14} have shown that the LH mechanism can also be efficient because of the temperature fluctuations of small grains due to photon absorption, although this mechanism has not been taken into account in our models.  Thanks to the high H$_2$ formation rate, the atomic/molecular transition is shifted towards the edge of the cloud, i.e. in a hotter gas where H$_2$ can be more excited. In our models, we also used a fully efficient conversion between ortho- and para-H$_2$ on dust grains, i.e. all H$_2$ molecules adsorbing on grains are converted before desorbing, which happens to be a good prescription to account for the observed relative H$_2$ line intensities and therefore the observed ortho-to-para ratio. This efficient conversion on grains is justified by the temperature fluctuations of the grains as shown in \cite{Bron16}.

The presence of H$_2$ in the hot/warm gas impacts the chemistry including the formation of CO since several routes initiated by the formation of CH$^+$ become efficient. The C$^+$ + H$_2$ reaction has an activation threshold of $\sim 4500$\,K that can be overcome first, by the kinetic energy of the reactants in hot gas, and second, by the internal energy of FUV-pumped H$_2$.
In our models, we compute this reaction rate with the prescription by \cite{Agundez10} that takes into account H$_2$ excitation. H$_2$ ro-vibrational level populations of the ground electronic state are computed explicitly taking into account collisional excitation and de-excitation, radiative emission and pumping, UV pumping in electronically excited states followed by radiative de-excitation in the ground electronic state, excitation at formation following \cite{Sizun10} and chemical destruction. This detailed treatment of H$_2$ excitation was already found to provide a good agreement between predicted CH$^+$ and SH$^+$ line intensities and observations in the Orion Bar \citep{Agundez10, Nagy13,Goicoechea17}. Once CH$^+$ is present in the gas, efficient ion-neutral reactions take place to produce species such as the CH$^+$, CH$_2^+$, CH$_3^+$ chain. Then CH$_3^+$ reacts with O to give HCO$^+$ followed by electronic recombination that gives CO. So the formation of CO is found to occur via this hot chemistry at the edge of PDRs as soon as H$_2$ starts to self-shield \citep[see also][]{Goicoechea16}.

As can be seen in Figs.\,\ref{Fig:N7023_Best_model_locem} and \ref{Fig:OrionBar_Best_model_locem}, CO emits in high rotational transitions as soon as H$_2$ is formed efficiently in the PDR. In the frame of our stationary PDR models, $\sim$~50\% of the J=19-18 line intensity originates behind the H/H$_2$ transition and before the C$^+$/C/CO one. The remaining 50\% are produced at the C$^+$/C/CO transition. In our models, the gas temperature between the two transition layers (H/H$_2$ and C$^+$/C/CO) is a few hundreds Kelvin, enough to excite collisionally CO in high rotational states. The main heating mechanism of the gas is the photo-electric effect on grains. \cite{Allers05} modified both the UV dust extinction and the photoelectric efficiency in order to increase the gas temperature and therefore account for the H$_2$ emission in the Orion Bar. This is in line with the fact that the abundance of polycyclic aromatic hydrocarbons (PAHs) is found to increase at the border of PDRs \citep[see][and references therein]{Pilleri12}. Since PAHs are major contributors to the photoelectric heating, we implemented in the code several ad-hoc profiles for the photoelectric heating that describe at best the observed abundance variation. However we could not conclude that these modified models were necessary to account for the observations since the new implementation of H$_2$ formation was already leading to a sufficient warm molecular layer. In fact the analysis of mid-IR observations shows that the increased abundance of PAHs is found at A$_\textrm{V}$$\leq$1 whereas our model predicts that excited CO emission is found at a different depth of A$_\textrm{V}$$\sim$2--3 (Figs.\,\ref{Fig:N7023_Best_model_locem} and \ref{Fig:OrionBar_Best_model_locem}). We therefore did not include a profile in the photo-electric heating and used our standard implementation based on \cite{Bakes94} in the final analysis presented here.

Finally, approximations in the computation of the self and mutual shielding of pre-dissociating UV lines can affect the calculated CO line intensities. In our models, we use the Federman et al. (1979) prescription to estimate the self-shielding. This formalism considers only the self-shielding of lines by themselves. UV absorption lines of CO and its isotopologues are only shifted by hundredths of Angstrom. Shielding of CO lines by H$_2$ lines, and mutual line shielding of CO and its isotopologues can reduce the photo-destruction rates of CO and its isotopologues. The Meudon PDR code allows to compute exactly such shieldings but at the price of a very significant computing time  \citep{Goicoechea07}. We performed some test calculations since a complete grid exploration was out of reach. We found that in the NGC~7023 and Orion Bar parameter range, this effect contributes to a maximum of a factor 3 on high-J  $^{13}$CO lines, such as the 15-14 transition, and is not sufficient to explain the mismatch between the modelled and observed $^{13}$CO line intensities. This supports that most of this mismatch arises from the mixing of components in the beam, as discussed in Sec.\,\ref{Sec:CO_rot}.

\subsection{Structure of the brightest PDR interfaces}\label{sec:structure_discussion}

As presented in Sec.\,\ref{Sec:model_results} our models manage to account for most of the observed lines that trace the bright PDR interface within a factor of a few (even better in the case of NGC\,7023).

We can then conclude that, to the first order, it is possible to explain the line intensities emitted at the edge of bright PDRs (G$_0$ $\sim 10^3 - 10^4$) with a single interface modelled by a stationary isobaric PDR model at high thermal pressure ($\sim$ 10$^8$ K cm$^{-3}$). A PDR model at P$_{th}$=10$^8$ K cm$^{-3}$ has been used for the Orion Bar in studies related to the one presented here but only for a limited data set and without exploring the role of the parameters \citep{Nagy13,Cuadrado15}. This idea of using such isobaric models has been raised previously by several authors \citep{Marconi98, Lemaire99, Allers05} but, most of the times, the alternative clump/interclump scenario has been used or invoked \citep{Stutzki90, Parmar91, Meixner93, Tauber94, Hogerheijde95, VanDerWerf96, Usuda96, Andree-Labsch17}.  In this scenario, clumps with density of $\sim10^6-10^7\,\mathrm{cm}^{-3}$ that include about 10\% of the gas, are embedded in a more diffuse gas $\sim 10^4-10^5\,\mathrm{cm}^{-3}$, i.e. the difference of density between the two components is a factor $\sim$ 100.
In our isobaric model at high pressure ($\sim$ few 10$^8$ K cm$^{-3}$) a comparable gradient in density naturally arises from the hypothesis of constant pressure. It is clear that the assumption of constant pressure equilibrium for our two PDRs is an approximation but the strength of our isobaric model is, however, that it can describe the structure by a single parameter while the clumpy models have to select or fit a very large and partially arbitrary parameter space. Some lines are found to be more responsive to the pressure than others as shown in Fig.\,\ref{Fig:Pressure_effect}. This is the case of the  0-0 S(5) H$_2$ line and the high-J CO lines. On the opposite, the [CII] line is rather constant at 5\,10$^6$ < P$_{th}$ < 10$^8$ K cm$^{-3}$ and its intensity then decreases significantly at higher pressures.

Still, a stationary model is  not expected to capture all the complexity of a strongly irradiated PDR, such as the Orion Bar, where out-of-equilibrium effects take place \citep{Bertoldi96}. However there is no point in refining our model considering the poor spatial resolution in most tracers. The pressure we derived at the Orion Bar edge, is likely an effective pressure providing the best compromise between fitting the emission of surface tracers (excited H$_2$ lines, CH$^+$ lines) and that of CO (mid-/high-J lines mainly emitted at A$_\textrm{V}$$\sim$2--4).
The computed value of the H$_2$~1-0\,S(1)/2-1\,S(1) ratio, which is significantly too high compared to the observations (a factor $\sim$10) suggests that the gas density is too high at the surface of the PDR, considering that the 1-0\,S(1) line intensity increases with collisional relaxation of higher UV-pumped vibrational levels. The existence of a pressure gradient is in line with the work of 
\cite{Goicoechea16} who showed that the gas kinematics derived from the ALMA maps suggests gas flowing from the high-pressure molecular layers (P$_{\mathrm{th}}\sim 2\times10^{8}\,\mathrm{K~cm}^{-3}$) to the atomic layers (P$_{\mathrm{th}}\sim 5\times10^{7}\,\mathrm{K~cm}^{-3}$). The advection of molecular gas through the PDR edge would impact both the chemical and thermal structures and therefore the calculated line intensities \citep{Bertoldi96}. In this case one can expect a lower value for the H$_2$~1-0S(1)/2-1S(1) ratio since more emission will arise from UV-pumped levels.
\citet{Stoerzer98} calculated that on average the H$_2$ lines are affected by a factor of 3.

\subsection{A $P_{th}$ - $G_0$ relation in PDRs}
\label{subsec:P-G0}

Our work shows that the emission of the warm molecular gas at the PDR edge can be attributed to a slab at a high thermal pressure of $10^8$ and $3\times10^8\mathrm{K~cm}^{-3}$ in NGC7023~NW and Orion Bar, respectively. There is a trend that this pressure increases with the $G_0$ value. To further explore this relation, we compiled data from the literature. While doing so, one needs to be sure that the collected values are consistent with our study, which means that this pressure was derived from relevant tracers. In particular we avoided studies that were only based on the analysis of the rotational lines of H$_2$.

\begin{figure}
\begin{center}
\includegraphics[width=0.5\textwidth]{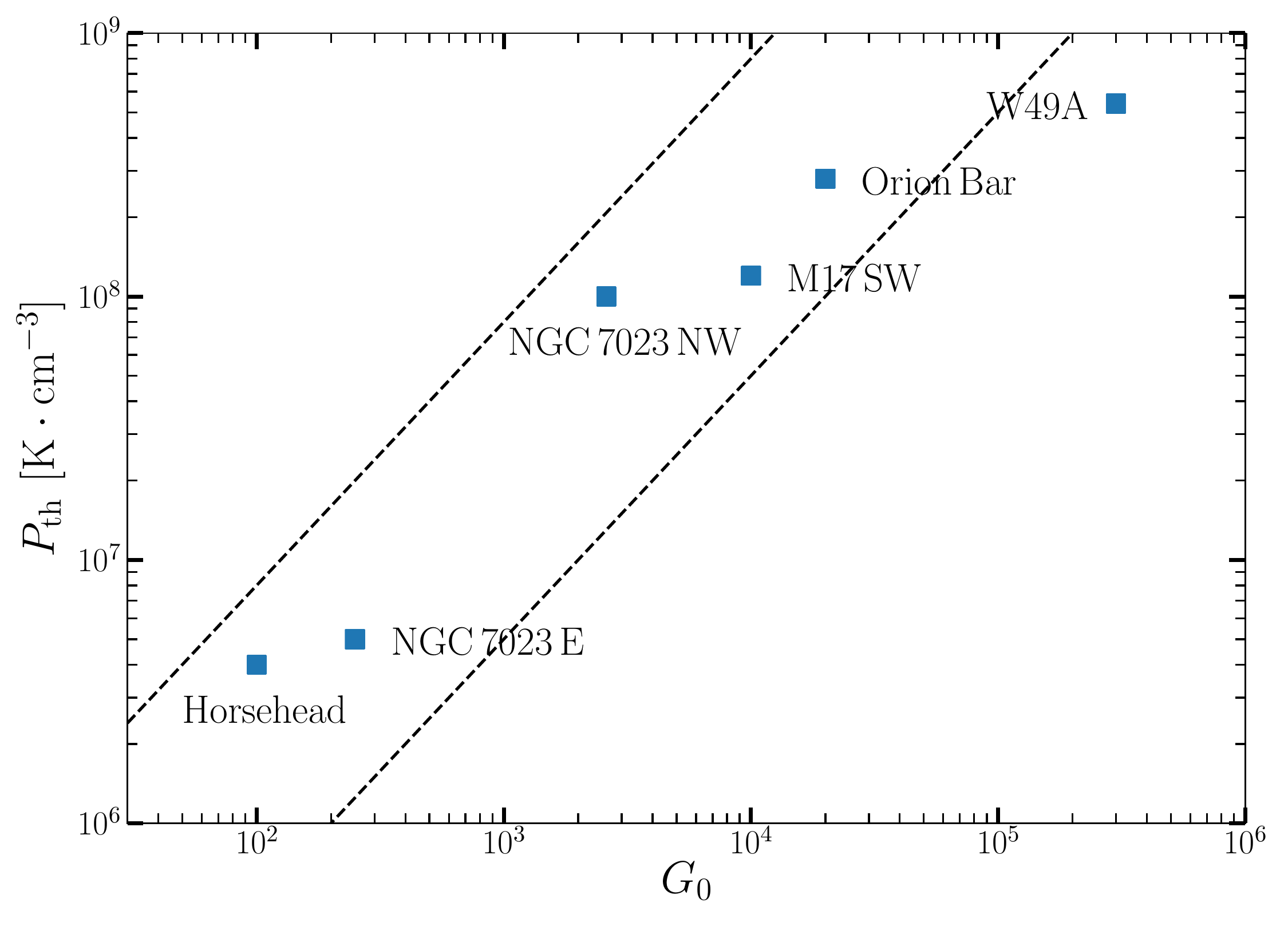}
\end{center}
\caption{Relation between the thermal pressure in the dense structures of PDRs and the UV intensity $G_0$. See the text for references. The dashed lines show the range of values obtained by \cite{Bron18} in their photoevaporating PDR models.
\label{Fig:Pressure_G0}}
\end{figure}

The data presented in Fig.\ref{Fig:Pressure_G0} is coming from the following studies. \cite{Habart05} have studied in details the Horsehead PDR and have shown that there is a density gradient at the interface with the \hii region, which can be modelled by a thermal pressure equilibrium at P$_{\mathrm{th}}\sim4\times10^{6}\,\mathrm{K~cm}^{-3}$. The value of P$_{\mathrm{th}}=5\times10^{6}\,\mathrm{K~cm}^{-3}$ for NGC~7023~E corresponds to the maximum value derived by \cite{Koehler14} from their combined analysis of dust and excited CO lines. \cite{Perez10} included some mid-J CO lines in their analysis of the warm gas in M17~SW and concluded that the high-density gas (n$_\textrm{H}$ = $5\times10^5\mathrm{cm}^{-3}$) has a maximum temperature of 230~K. We therefore used these values to derive a pressure of P$_{\mathrm{th}}=1.2\times10^{8}\,\mathrm{K~cm}^{-3}$. Finally, we included the massive star forming region W49A in which \cite{Nagy12} studied the warm molecular gas and derived a kinetic temperature map using H$_2$, CO excited lines and a volume density map using HCN excited lines. The maximum thermal pressure can be derived at the center of the map with a value of P$_{\mathrm{th}}=5.4\times10^{8}\,\mathrm{K~cm}^{-3}$.

Figure \ref{Fig:Pressure_G0} shows that $P_{\mathrm{th}}$ indeed increases with $G_0$. Considering the very uncertain error bars, we do not provide here a fitting of the reported points that would provide a more quantitative scaling of $P_{\mathrm{th}}$ with $G_0$. Nevertheless, a visual inspection of Fig.\, \ref{Fig:Pressure_G0} leads to $P_{\mathrm{th}}$/$G_0$=1-4 10$^4 \,\mathrm{K~cm}^{-3}$ except for W49A that falls below this range.
It is interesting to note that this graph can help rationalizing the results presented in \citet{Stock15}. In the two PDRs studied, S~106 and IRAS~23133+6050, the CO SLEDs are close to those measured in Orion Bar, which could be explained by a UV radiation field, G$_0$, of a few 10$^4$. We can also comment on the results obtained by 
\cite{Indriolo17} on the CO SLEDs in prototypical massive star-forming regions in which both a high-UV field and shocks are thought to excite the gas. Our study suggests that excitation by UV photons also plays a major role in the case of Orion\,S and W49N. Indeed both objects have similar CO SLEDs and W49N/A is found to follow to some extent the P$_{th}$-G$_0$ trend shown in Fig\,\ref{Fig:Pressure_G0}. In these objects, shocks are however also involved; they are likely the major driver for emission in CO lines with J$_{up}$>25 and are revealed when line profiles can be resolved. For instance, \cite{Tahani16} showed that for the CO J=16-15 line towards Orion S, there is a narrow (4\,$\mathrm{km~s}^{-1}$) component associated with the PDR and a broad (15\,$\mathrm{km~s}^{-1}$) component associated with shock excitation, both having similar integrated intensities. More detailed modelling would be necessary to disentangle the contribution of both excitations on the CO SLEDs.

The obtained  P$_{th}$-G$_0$ relation can also give us further insights into the (unclear) origin of the density structures that are found at the edge of \hii regions (case of Orion Bar) or of atomic regions (case of NGC 7023 in which no  \hii region is present).
It suggests that the UV radiation field plays a major role in the compression of the PDR. As the pressures found in the PDRs are significantly higher than the pressures found in the H\textsc{ii} regions (e.g. $P_{\mathrm{th}}=6\times 10^7$ K cm$^{-3}$ in the Orion Bar, \citealt{Goicoechea16}), pressurization by the thermal pressure of the H\textsc{ii} region (e.g. in an expanding H\textsc{ii} region) is not sufficient to explain the trend. Photoevaporation, in which photoheated gas at the ionization and dissociation fronts expands into the central cavity and exerts by reaction a force on the neutral and/or molecular part of the cloud \citep{Bertoldi89,Bertoldi96}, could induce  compression of the molecular part of the PDR and explain the pressure difference with the central ionized/atomic cavity. In addition, the tight correlation with $G_0$ independently of the presence of an ionization front (case of NGC\,7023) close to the PDR seems to indicate that non-ionizing (FUV) photons can be at least as efficient as ionizing photons for this photoevaporation process.
These considerations have found theoretical support in a recent study by \cite{Bron18}. The authors found that photoevaporation of the illuminated edge of the molecular cloud can indeed lead to high pressures and account for the P$_{th}$-G$_0$ trend. To illustrate this result, we show in Fig.\,\ref{Fig:Pressure_G0} the range of values (dashed lines representing $P_{\mathrm{th}}$/$G_0$= $5\times 10^3$ and $8\times 10^4$\,K cm$^{-3}$, respectively) obtained by the authors using their time-dependent hydrodynamical PDR code. The agreement with the observations is striking and opens new perspectives to study dynamical evolution of the strongly illuminated edges of molecular clouds in massive star-forming regions.

\section{Conclusion}
\label{Sec:Conclusion}
Thanks to {\it Herschel}, we have measured the CO SLEDs in two prototypical PDRs: NGC 7023 NW  (observed $^{12}$CO lines from J$_{up}$=4 to 19) and the Orion Bar (observed $^{12}$CO lines from J$_{up}$=4 to 23). The excitation temperature deduced for J$_{up}\geq$ 15 from the rotational diagrams are 112 and 147\,K, respectively, showing the presence of warm CO gas at the irradiated PDR edge. We have used the Meudon PDR code and more specifically stationary isobaric PDR models to account for high-J $^{12}$CO lines as well as for H$_2$ and CH$^+$ lines. Only the thermal pressure value and a global scaling factor were used as free parameters. The best models were obtained for a gas thermal pressure of $P_{\mathrm{th}}\sim$ \,10$^8$\,K~cm$^{-3}$ and provide a good agreement with the observed values. The prediction made by these models for other lines from HCO$^+$, O, C, C$^+$, HD and OH has also been found to be compatible with the observed values.

Compared to previous works, we found that simulating in detail the H$_2$ formation process on grains, its level excitation and considering state-to-state chemistry for key reactions, it is possible to explain line emission of molecules at the edge of PDRs with a stationary model without the introduction of ad-hoc hypothesis as clumps or shocks. One of the key mechanisms to account for warm CO in PDRs is the high efficiency of H$_2$ formation at the PDR edge, which brings the H/H$_2$ transition closer to the interface. 
The gas temperature ranges from 1000 K where H$_2$ starts to self-shield to 100 K at the C$^+$/C/CO transition. 
CO starts to form as soon as H$_2$ appears in the PDR. This is due to the fact that the warm temperature, high density and presence of FUV-pumped H$_2$ allow the formation of CH$^+$ via the H$_2$ + C$^+$ reaction, opening a hot chemistry channel that leads to the formation of CO. As a consequence, in the frame of these stationary models, a significant fraction ($\sim$50 \% in the models of the two PDRs presented here) of high-J CO emission is produced before the C$^+$/C/CO transition. In any case, the separation between the  H/H$_2$ and C$^+$/C/CO transition layers is predicted to be very small (less than one arsec. at the distance of Orion) and this is in line with ALMA observations \citep{Goicoechea16}.

Our results impact our view of the irradiated edge of star-forming regions and the feedback of star formation on its parental cloud. 
In the two prototypical PDRs, NGC 7023 NW and Orion Bar, we found that the FUV photons from nearby massive stars have enough energy to explain CO excitation in mid- and high-J levels. No additional energy source as mechanical heating is required. A comparison of NGC 7023 NW and Orion Bar with other typical PDRs shows a correlation between the thermal pressure at the edge of PDRs and the intensity of the UV radiation field.  A similar correlation was recently reported by \cite{Wu18} in their spatial study of the Carina nebula.
This seems to indicate that the UV radiation field is the main driver of the high pressure at the edge of PDRs.
This high-pressure edge is very sharp of the order of a few $10^{-3}$\,pc (one arcsec at the distance of Orion) and is expected to evolve in the surrounding gas at lower pressure. The presence of pressure gradients and advection processes is invoked but could not be further exploited due to the limited spatial resolution of the observations. 

Implication for extragalactic studies has been discussed by \cite{Indriolo17} and is supported by our study. The high-pressure edges being thin structures, their observation suffers from beam dilution effects.
The availability of high-spatial resolution observations is therefore crucial to refine PDR models and is only feasible while studying Galactic PDRs.
Perspectives in this topic include the coming James Webb Space Telescope that will soon give us access to subarcsecond-resolution observations for a large number of H$_2$ rotational and ro-vibrational lines. Observations of high-J CO lines with high spatial resolution are however currently out of reach and this could be the case for quite some time. The use of alternative species, which could be observed by ALMA, such as HCO$^+$ or HCN, has to be explored. From the modelling side, further improvement should include a better description of the grain populations and their properties since the penetration of the UV field plays a central role in the energetics and chemistry of PDRs. An additional big step would consist in coupling the chemistry and energetics with dynamical processes that can describe the evolution of these high-pressure edges \citep{Gorti02}.

\begin{acknowledgements}
This work was in part supported by the Programme National ''Physique et Chimie du Milieu Interstellaire'' (PCMI) of CNRS/INSU with INC/INP, and co-funded by CEA and CNES. It was supported by the German \emph{Deut\-sche For\-schungs\-ge\-mein\-schaft, DFG\/} project number SFB 956, C1.
J.R.G and J.C. thank Spanish MINECO for funding supports under grants AYA2012-32032 and AYA2017-85111-P.
E. B. thanks the ERC grant ERC-2013-Syg-610256-NANOCOSMOS for support. 
\end{acknowledgements}

\bibliography{PDR7023Orion}

\begin{thebibliography}{144}
\expandafter\ifx\csname natexlab\endcsname\relax\def\natexlab#1{#1}\fi

\bibitem[{{Ag{\'u}ndez} {et~al.}(2010){Ag{\'u}ndez}, {Goicoechea},
  {Cernicharo}, {Faure}, \& {Roueff}}]{Agundez10}
{Ag{\'u}ndez}, M., {Goicoechea}, J.~R., {Cernicharo}, J., {Faure}, A., \&
  {Roueff}, E. 2010, \apj, 713, 662

\bibitem[{{Alecian} {et~al.}(2008){Alecian}, {Catala}, {Wade}, {Donati},
  {Petit}, {Landstreet}, {B{\"o}hm}, {Bouret}, {Bagnulo}, {Folsom}, {Grunhut},
  \& {Silvester}}]{Alecian08}
{Alecian}, E., {Catala}, C., {Wade}, G.~A., {et~al.} 2008, \mnras, 385, 391

\bibitem[{{Allers} {et~al.}(2005){Allers}, {Jaffe}, {Lacy}, {Draine}, \&
  {Richter}}]{Allers05}
{Allers}, K.~N., {Jaffe}, D.~T., {Lacy}, J.~H., {Draine}, B.~T., \& {Richter},
  M.~J. 2005, \apj, 630, 368

\bibitem[{{An} \& {Sellgren}(2003)}]{An03}
{An}, J.~H. \& {Sellgren}, K. 2003, \apj, 599, 312

\bibitem[{{Andree-Labsch} {et~al.}(2017){Andree-Labsch}, {Ossenkopf-Okada}, \&
  {R{\"o}llig}}]{Andree-Labsch17}
{Andree-Labsch}, S., {Ossenkopf-Okada}, V., \& {R{\"o}llig}, M. 2017, \aap,
  598, A2

\bibitem[{{Bakes} \& {Tielens}(1994)}]{Bakes94}
{Bakes}, E.~L.~O. \& {Tielens}, A.~G.~G.~M. 1994, \apj, 427, 822

\bibitem[{{Balakrishnan} {et~al.}(2002){Balakrishnan}, {Yan}, \&
  {Dalgarno}}]{Balakrishnan02}
{Balakrishnan}, N., {Yan}, M., \& {Dalgarno}, A. 2002, \apj, 568, 443

\bibitem[{{Benisty} {et~al.}(2013){Benisty}, {Perraut}, {Mourard}, {Stee},
  {Lima}, {Le Bouquin}, {Borges Fernandes}, {Chesneau}, {Nardetto},
  {Tallon-Bosc}, {McAlister}, {Ten Brummelaar}, {Ridgway}, {Sturmann},
  {Sturmann}, {Turner}, {Farrington}, \& {Goldfinger}}]{Benisty13}
{Benisty}, M., {Perraut}, K., {Mourard}, D., {et~al.} 2013, \aap, 555, A113

\bibitem[{{Bergin} {et~al.}(2010){Bergin}, {Phillips}, {Comito}, {Crockett},
  {Lis}, {Schilke}, {Wang}, {Bell}, {Blake}, {Bumble}, {Caux}, {Cabrit},
  {Ceccarelli}, {Cernicharo}, {Daniel}, {de Graauw}, {Dubernet},
  {Emprechtinger}, {Encrenaz}, {Falgarone}, {Gerin}, {Giesen}, {Goicoechea},
  {Goldsmith}, {Gupta}, {Hartogh}, {Helmich}, {Herbst}, {Joblin}, {Johnstone},
  {Kawamura}, {Langer}, {Latter}, {Lord}, {Maret}, {Martin}, {Melnick},
  {Menten}, {Morris}, {M{\"u}ller}, {Murphy}, {Neufeld}, {Ossenkopf}, {Pagani},
  {Pearson}, {P{\'e}rault}, {Plume}, {Roelfsema}, {Qin}, {Salez}, {Schlemmer},
  {Stutzki}, {Tielens}, {Trappe}, {van der Tak}, {Vastel}, {Yorke}, {Yu}, \&
  {Zmuidzinas}}]{Bergin10}
{Bergin}, E.~A., {Phillips}, T.~G., {Comito}, C., {et~al.} 2010, \aap, 521, L20

\bibitem[{{Bernard-Salas} {et~al.}(2012){Bernard-Salas}, {Habart}, {Arab},
  {Abergel}, {Dartois}, {Martin}, {Bontemps}, {Joblin}, {White}, {Bernard}, \&
  {Naylor}}]{Bernard-Salas12}
{Bernard-Salas}, J., {Habart}, E., {Arab}, H., {et~al.} 2012, \aap, 538, A37

\bibitem[{{Bernard-Salas} {et~al.}(2015){Bernard-Salas}, {Habart},
  {K{\"o}hler}, {Abergel}, {Arab}, {Lebouteiller}, {Pinto}, {van der Wiel},
  {White}, \& {Hoffmann}}]{Bernard-Salas15}
{Bernard-Salas}, J., {Habart}, E., {K{\"o}hler}, M., {et~al.} 2015, \aap, 574,
  A97

\bibitem[{{Bern{\'e}} {et~al.}(2007){Bern{\'e}}, {Joblin}, {Deville}, {Smith},
  {Rapacioli}, {Bernard}, {Thomas}, {Reach}, \& {Abergel}}]{Berne07}
{Bern{\'e}}, O., {Joblin}, C., {Deville}, Y., {et~al.} 2007, \aap, 469, 575

\bibitem[{{Bertoldi}(1989)}]{Bertoldi89}
{Bertoldi}, F. 1989, \apj, 346, 735

\bibitem[{{Bertoldi} \& {Draine}(1996)}]{Bertoldi96}
{Bertoldi}, F. \& {Draine}, B.~T. 1996, \apj, 458, 222

\bibitem[{{Bohlin} {et~al.}(1978){Bohlin}, {Savage}, \& {Drake}}]{Bohlin78}
{Bohlin}, R.~C., {Savage}, B.~D., \& {Drake}, J.~F. 1978, \apj, 224, 132

\bibitem[{{Bron} {et~al.}(2018){Bron}, {Ag{\'u}ndez}, {Goicoechea}, \&
  {Cernicharo}}]{Bron18}
{Bron}, E., {Ag{\'u}ndez}, M., {Goicoechea}, J.~R., \& {Cernicharo}, J. 2018,
  ArXiv e-prints

\bibitem[{{Bron} {et~al.}(2014){Bron}, {Le Bourlot}, \& {Le Petit}}]{Bron14}
{Bron}, E., {Le Bourlot}, J., \& {Le Petit}, F. 2014, \aap, 569, A100

\bibitem[{{Bron} {et~al.}(2016){Bron}, {Le Petit}, \& {Le Bourlot}}]{Bron16}
{Bron}, E., {Le Petit}, F., \& {Le Bourlot}, J. 2016, \aap, 588, A27

\bibitem[{{Burton} {et~al.}(1990){Burton}, {Hollenbach}, \&
  {Tielens}}]{Burton90}
{Burton}, M.~G., {Hollenbach}, D.~J., \& {Tielens}, A.~G.~G.~M. 1990, \apj,
  365, 620

\bibitem[{{Cardelli} {et~al.}(1989){Cardelli}, {Clayton}, \&
  {Mathis}}]{Cardelli89}
{Cardelli}, J.~A., {Clayton}, G.~C., \& {Mathis}, J.~S. 1989, \apj, 345, 245

\bibitem[{{Chokshi} {et~al.}(1988){Chokshi}, {Tielens}, {Werner}, \&
  {Castelaz}}]{Chokshi88}
{Chokshi}, A., {Tielens}, A.~G.~G.~M., {Werner}, M.~W., \& {Castelaz}, M.~W.
  1988, \apj, 334, 803

\bibitem[{{Cuadrado} {et~al.}(2015){Cuadrado}, {Goicoechea}, {Pilleri},
  {Cernicharo}, {Fuente}, \& {Joblin}}]{Cuadrado15}
{Cuadrado}, S., {Goicoechea}, J.~R., {Pilleri}, P., {et~al.} 2015, \aap, 575,
  A82

\bibitem[{{de Graauw} {et~al.}(1996){de Graauw}, {Haser}, {Beintema},
  {Roelfsema}, {van Agthoven}, {Barl}, {Bauer}, {Bekenkamp}, {Boonstra},
  {Boxhoorn}, {Cote}, {de Groene}, {van Dijkhuizen}, {Drapatz}, {Evers},
  {Feuchtgruber}, {Frericks}, {Genzel}, {Haerendel}, {Heras}, {van der Hucht},
  {van der Hulst}, {Huygen}, {Jacobs}, {Jakob}, {Kamperman}, {Katterloher},
  {Kester}, {Kunze}, {Kussendrager}, {Lahuis}, {Lamers}, {Leech}, {van der
  Lei}, {van der Linden}, {Luinge}, {Lutz}, {Melzner}, {Morris}, {van Nguyen},
  {Ploeger}, {Price}, {Salama}, {Schaeidt}, {Sijm}, {Smoorenburg}, {Spakman},
  {Spoon}, {Steinmayer}, {Stoecker}, {Valentijn}, {Vandenbussche}, {Visser},
  {Waelkens}, {Waters}, {Wensink}, {Wesselius}, {Wiezorrek}, {Wieprecht},
  {Wijnbergen}, {Wildeman}, \& {Young}}]{Degraauw96}
{de Graauw}, T., {Haser}, L.~N., {Beintema}, D.~A., {et~al.} 1996, \aap, 315,
  L49

\bibitem[{{Demyk} {et~al.}(2007){Demyk}, {M{\"a}der}, {Tercero}, {Cernicharo},
  {Demaison}, {Margul{\`e}s}, {Wegner}, {Keipert}, \& {Sheng}}]{Demyk07}
{Demyk}, K., {M{\"a}der}, H., {Tercero}, B., {et~al.} 2007, \aap, 466, 255

\bibitem[{{Draine} \& {Bertoldi}(1999)}]{Draine99}
{Draine}, B.~T. \& {Bertoldi}, F. 1999, in ESA Special Publication, Vol. 427,
  The Universe as Seen by ISO, ed. P.~{Cox} \& M.~{Kessler}, 553

\bibitem[{{Etxaluze} {et~al.}(2013){Etxaluze}, {Goicoechea}, {Cernicharo},
  {Polehampton}, {Noriega-Crespo}, {Molinari}, {Swinyard}, {Wu}, \&
  {Bally}}]{Etxaluze13}
{Etxaluze}, M., {Goicoechea}, J.~R., {Cernicharo}, J., {et~al.} 2013, \aap,
  556, A137

\bibitem[{{Federman} {et~al.}(1979){Federman}, {Glassgold}, \&
  {Kwan}}]{Federman79}
{Federman}, S.~R., {Glassgold}, A.~E., \& {Kwan}, J. 1979, \apj, 227, 466

\bibitem[{{Fitzpatrick} \& {Massa}(1986)}]{Fitzpatrick86}
{Fitzpatrick}, E.~L. \& {Massa}, D. 1986, \apj, 307, 286

\bibitem[{{Fitzpatrick} \& {Massa}(1990)}]{Fitzpatrick90}
{Fitzpatrick}, E.~L. \& {Massa}, D. 1990, \apjs, 72, 163

\bibitem[{{Fleming} {et~al.}(2010){Fleming}, {France}, {Lupu}, \&
  {McCandliss}}]{Fleming10}
{Fleming}, B., {France}, K., {Lupu}, R.~E., \& {McCandliss}, S.~R. 2010, \apj,
  725, 159

\bibitem[{{Fuente} {et~al.}(1992){Fuente}, {Martin-Pintado}, {Cernicharo},
  {Brouillet}, \& {Duvert}}]{Fuente92}
{Fuente}, A., {Martin-Pintado}, J., {Cernicharo}, J., {Brouillet}, N., \&
  {Duvert}, G. 1992, \aap, 260, 341

\bibitem[{{Fuente} {et~al.}(1996{\natexlab{a}}){Fuente}, {Martin-Pintado},
  {Neri}, {Rogers}, \& {Moriarty-Schieven}}]{Fuente96}
{Fuente}, A., {Martin-Pintado}, J., {Neri}, R., {Rogers}, C., \&
  {Moriarty-Schieven}, G. 1996{\natexlab{a}}, \aap, 310, 286

\bibitem[{{Fuente} {et~al.}(1999){Fuente}, {Mart{\'{\i}}n-Pintado},
  {Rodr{\'{\i}}guez-Fern{\'a}ndez}, {Rodr{\'{\i}}guez-Franco}, {de Vicente}, \&
  {Kunze}}]{Fuente99}
{Fuente}, A., {Mart{\'{\i}}n-Pintado}, J., {Rodr{\'{\i}}guez-Fern{\'a}ndez},
  N.~J., {et~al.} 1999, \apjl, 518, L45

\bibitem[{{Fuente} {et~al.}(1998){Fuente}, {Martin-Pintado},
  {Rodriguez-Franco}, \& {Moriarty-Schieven}}]{Fuente98}
{Fuente}, A., {Martin-Pintado}, J., {Rodriguez-Franco}, A., \&
  {Moriarty-Schieven}, G.~D. 1998, \aap, 339, 575

\bibitem[{{Fuente} {et~al.}(1996{\natexlab{b}}){Fuente}, {Rodriguez-Franco}, \&
  {Martin-Pintado}}]{Fuente96_Orion}
{Fuente}, A., {Rodriguez-Franco}, A., \& {Martin-Pintado}, J.
  1996{\natexlab{b}}, \aap, 312, 599

\bibitem[{{Gerin} {et~al.}(1998){Gerin}, {Phillips}, {Keene}, {Betz}, \&
  {Boreiko}}]{Gerin98}
{Gerin}, M., {Phillips}, T.~G., {Keene}, J., {Betz}, A.~L., \& {Boreiko}, R.~T.
  1998, \apj, 500, 329

\bibitem[{{Giard} {et~al.}(1994){Giard}, {Bernard}, {Lacombe}, {Normand}, \&
  {Rouan}}]{Giard94}
{Giard}, M., {Bernard}, J.~P., {Lacombe}, F., {Normand}, P., \& {Rouan}, D.
  1994, \aap, 291, 239

\bibitem[{{Goicoechea} {et~al.}(2015{\natexlab{a}}){Goicoechea},
  {Chavarr{\'{\i}}a}, {Cernicharo}, {Neufeld}, {Vavrek}, {Bergin}, {Cuadrado},
  {Encrenaz}, {Etxaluze}, {Melnick}, \& {Polehampton}}]{Goicoechea15}
{Goicoechea}, J.~R., {Chavarr{\'{\i}}a}, L., {Cernicharo}, J., {et~al.}
  2015{\natexlab{a}}, \apj, 799, 102

\bibitem[{{Goicoechea} {et~al.}(2017){Goicoechea}, {Cuadrado}, {Pety}, {Bron},
  {Black}, {Cernicharo}, {Chapillon}, {Fuente}, \& {Gerin}}]{Goicoechea17}
{Goicoechea}, J.~R., {Cuadrado}, S., {Pety}, J., {et~al.} 2017, \aap, 601, L9

\bibitem[{{Goicoechea} {et~al.}(2013){Goicoechea}, {Etxaluze}, {Cernicharo},
  {Gerin}, {Neufeld}, {Contursi}, {Bell}, {De Luca}, {Encrenaz}, {Indriolo},
  {Lis}, {Polehampton}, \& {Sonnentrucker}}]{Goicoechea13}
{Goicoechea}, J.~R., {Etxaluze}, M., {Cernicharo}, J., {et~al.} 2013, \apjl,
  769, L13

\bibitem[{{Goicoechea} {et~al.}(2011){Goicoechea}, {Joblin}, {Contursi},
  {Bern{\'e}}, {Cernicharo}, {Gerin}, {Le Bourlot}, {Bergin}, {Bell}, \&
  {R{\"o}llig}}]{Goicoechea11}
{Goicoechea}, J.~R., {Joblin}, C., {Contursi}, A., {et~al.} 2011, \aap, 530,
  L16

\bibitem[{{Goicoechea} \& {Le Bourlot}(2007)}]{Goicoechea07}
{Goicoechea}, J.~R. \& {Le Bourlot}, J. 2007, \aap, 467, 1

\bibitem[{{Goicoechea} {et~al.}(2016){Goicoechea}, {Pety}, {Cuadrado},
  {Cernicharo}, {Chapillon}, {Fuente}, {Gerin}, {Joblin}, {Marcelino}, \&
  {Pilleri}}]{Goicoechea16}
{Goicoechea}, J.~R., {Pety}, J., {Cuadrado}, S., {et~al.} 2016, \nat, 537, 207

\bibitem[{{Goicoechea} {et~al.}(2015{\natexlab{b}}){Goicoechea}, {Teyssier},
  {Etxaluze}, {Goldsmith}, {Ossenkopf}, {Gerin}, {Bergin}, {Black},
  {Cernicharo}, {Cuadrado}, {Encrenaz}, {Falgarone}, {Fuente}, {Hacar}, {Lis},
  {Marcelino}, {Melnick}, {M{\"u}ller}, {Persson}, {Pety}, {R{\"o}llig},
  {Schilke}, {Simon}, {Snell}, \& {Stutzki}}]{Goicoechea15b}
{Goicoechea}, J.~R., {Teyssier}, D., {Etxaluze}, M., {et~al.}
  2015{\natexlab{b}}, \apj, 812, 75

\bibitem[{{Gonzalez Garcia} {et~al.}(2008){Gonzalez Garcia}, {Le Bourlot}, {Le
  Petit}, \& {Roueff}}]{Gonzalez08}
{Gonzalez Garcia}, M., {Le Bourlot}, J., {Le Petit}, F., \& {Roueff}, E. 2008,
  \aap, 485, 127

\bibitem[{{Gorti} \& {Hollenbach}(2002)}]{Gorti02}
{Gorti}, U. \& {Hollenbach}, D. 2002, \apj, 573, 215

\bibitem[{{Greve} {et~al.}(2014){Greve}, {Leonidaki}, {Xilouris}, {Wei{\ss}},
  {Zhang}, {van der Werf}, {Aalto}, {Armus}, {D{\'{\i}}az-Santos}, {Evans},
  {Fischer}, {Gao}, {Gonz{\'a}lez-Alfonso}, {Harris}, {Henkel}, {Meijerink},
  {Naylor}, {Smith}, {Spaans}, {Stacey}, {Veilleux}, \& {Walter}}]{Greve14}
{Greve}, T.~R., {Leonidaki}, I., {Xilouris}, E.~M., {et~al.} 2014, \apj, 794,
  142

\bibitem[{{Gry} {et~al.}(2002){Gry}, {Boulanger}, {Nehm{\'e}}, {Pineau des
  For{\^e}ts}, {Habart}, \& {Falgarone}}]{Gry02}
{Gry}, C., {Boulanger}, F., {Nehm{\'e}}, C., {et~al.} 2002, \aap, 391, 675

\bibitem[{{Habart} {et~al.}(2011){Habart}, {Abergel}, {Boulanger}, {Joblin},
  {Verstraete}, {Compi{\`e}gne}, {Pineau Des For{\^e}ts}, \& {Le
  Bourlot}}]{Habart11}
{Habart}, E., {Abergel}, A., {Boulanger}, F., {et~al.} 2011, \aap, 527, A122

\bibitem[{{Habart} {et~al.}(2005){Habart}, {Abergel}, {Walmsley}, {Teyssier},
  \& {Pety}}]{Habart05}
{Habart}, E., {Abergel}, A., {Walmsley}, C.~M., {Teyssier}, D., \& {Pety}, J.
  2005, \aap, 437, 177

\bibitem[{{Habart} {et~al.}(2004){Habart}, {Boulanger}, {Verstraete},
  {Walmsley}, \& {Pineau des For{\^e}ts}}]{Habart04}
{Habart}, E., {Boulanger}, F., {Verstraete}, L., {Walmsley}, C.~M., \& {Pineau
  des For{\^e}ts}, G. 2004, \aap, 414, 531

\bibitem[{{Habart} {et~al.}(2010){Habart}, {Dartois}, {Abergel}, {Baluteau},
  {Naylor}, {Polehampton}, {Joblin}, {Ade}, {Anderson}, {Andr{\'e}}, {Arab},
  {Bernard}, {Blagrave}, {Bontemps}, {Boulanger}, {Cohen}, {Compiegne}, {Cox},
  {Davis}, {Emery}, {Fulton}, {Gry}, {Huang}, {Jones}, {Kirk}, {Lagache},
  {Lim}, {Madden}, {Makiwa}, {Martin}, {Miville-Desch{\^e}nes}, {Molinari},
  {Moseley}, {Motte}, {Okumura}, {Pinheiro Gon{\c c}alves}, {Rodon}, {Russeil},
  {Saraceno}, {Sidher}, {Spencer}, {Swinyard}, {Ward-Thompson}, {White}, \&
  {Zavagno}}]{Habart10}
{Habart}, E., {Dartois}, E., {Abergel}, A., {et~al.} 2010, \aap, 518, L116

\bibitem[{{Habing}(1968)}]{Habing68}
{Habing}, H.~J. 1968, \bain, 19, 421

\bibitem[{{Hailey-Dunsheath} {et~al.}(2012){Hailey-Dunsheath}, {Sturm},
  {Fischer}, {Sternberg}, {Graci{\'a}-Carpio}, {Davies},
  {Gonz{\'a}lez-Alfonso}, {Mark}, {Poglitsch}, {Contursi}, {Genzel}, {Lutz},
  {Tacconi}, {Veilleux}, {Verma}, \& {de Jong}}]{Hailey12}
{Hailey-Dunsheath}, S., {Sturm}, E., {Fischer}, J., {et~al.} 2012, \apj, 755,
  57

\bibitem[{{Haykal} {et~al.}(2014){Haykal}, {Carvajal}, {Tercero}, {Kleiner},
  {L{\'o}pez}, {Cernicharo}, {Motiyenko}, {Huet}, {Guillemin}, \&
  {Margul{\`e}s}}]{Haykal14}
{Haykal}, I., {Carvajal}, M., {Tercero}, B., {et~al.} 2014, \aap, 568, A58

\bibitem[{{Hogerheijde} {et~al.}(1995){Hogerheijde}, {Jansen}, \& {van
  Dishoeck}}]{Hogerheijde95}
{Hogerheijde}, M.~R., {Jansen}, D.~J., \& {van Dishoeck}, E.~F. 1995, \aap,
  294, 792

\bibitem[{{Hollenbach} \& {Tielens}(1999)}]{Hollenbach99}
{Hollenbach}, D.~J. \& {Tielens}, A.~G.~G.~M. 1999, Reviews of Modern Physics,
  71, 173

\bibitem[{{Houck} {et~al.}(2004){Houck}, {Roellig}, {van Cleve}, {Forrest},
  {Herter}, {Lawrence}, {Matthews}, {Reitsema}, {Soifer}, {Watson}, {Weedman},
  {Huisjen}, {Troeltzsch}, {Barry}, {Bernard-Salas}, {Blacken}, {Brandl},
  {Charmandaris}, {Devost}, {Gull}, {Hall}, {Henderson}, {Higdon}, {Pirger},
  {Schoenwald}, {Sloan}, {Uchida}, {Appleton}, {Armus}, {Burgdorf},
  {Fajardo-Acosta}, {Grillmair}, {Ingalls}, {Morris}, \& {Teplitz}}]{Houck04}
{Houck}, J.~R., {Roellig}, T.~L., {van Cleve}, J., {et~al.} 2004, \apjs, 154,
  18

\bibitem[{{Indriolo} {et~al.}(2017){Indriolo}, {Bergin}, {Goicoechea},
  {Cernicharo}, {Gerin}, {Gusdorf}, {Lis}, \& {Schilke}}]{Indriolo17}
{Indriolo}, N., {Bergin}, E.~A., {Goicoechea}, J.~R., {et~al.} 2017, \apj, 836,
  117

\bibitem[{{Indriolo} {et~al.}(2015){Indriolo}, {Neufeld}, {Gerin}, {Schilke},
  {Benz}, {Winkel}, {Menten}, {Chambers}, {Black}, {Bruderer}, {Falgarone},
  {Godard}, {Goicoechea}, {Gupta}, {Lis}, {Ossenkopf}, {Persson},
  {Sonnentrucker}, {van der Tak}, {van Dishoeck}, {Wolfire}, \&
  {Wyrowski}}]{Indriolo15}
{Indriolo}, N., {Neufeld}, D.~A., {Gerin}, M., {et~al.} 2015, \apj, 800, 40

\bibitem[{{Joblin} {et~al.}(2010){Joblin}, {Pilleri}, {Montillaud}, {Fuente},
  {Gerin}, {Bern{\'e}}, {Ossenkopf}, {Le Bourlot}, {Teyssier}, {Goicoechea},
  {Le Petit}, {R{\"o}llig}, {Akyilmaz}, {Benz}, {Boulanger}, {Bruderer},
  {Dedes}, {France}, {G{\"u}sten}, {Harris}, {Klein}, {Kramer}, {Lord},
  {Martin}, {Martin-Pintado}, {Mookerjea}, {Okada}, {Phillips}, {Rizzo},
  {Simon}, {Stutzki}, {van der Tak}, {Yorke}, {Steinmetz}, {Jarchow},
  {Hartogh}, {Honingh}, {Siebertz}, {Caux}, \& {Colin}}]{Joblin10}
{Joblin}, C., {Pilleri}, P., {Montillaud}, J., {et~al.} 2010, \aap, 521, L25

\bibitem[{{Jura}(1974)}]{Jura74}
{Jura}, M. 1974, \apj, 191, 375

\bibitem[{{Kamenetzky} {et~al.}(2014){Kamenetzky}, {Rangwala}, {Glenn},
  {Maloney}, \& {Conley}}]{Kamenetzky14}
{Kamenetzky}, J., {Rangwala}, N., {Glenn}, J., {Maloney}, P.~R., \& {Conley},
  A. 2014, \apj, 795, 174

\bibitem[{{Karska} {et~al.}(2013){Karska}, {Herczeg}, {van Dishoeck},
  {Wampfler}, {Kristensen}, {Goicoechea}, {Visser}, {Nisini}, {San
  Jos{\'e}-Garc{\'{\i}}a}, {Bruderer}, {{\'S}niady}, {Doty}, {Fedele},
  {Y{\i}ld{\i}z}, {Benz}, {Bergin}, {Caselli}, {Herpin}, {Hogerheijde},
  {Johnstone}, {J{\o}rgensen}, {Liseau}, {Tafalla}, {van der Tak}, \&
  {Wyrowski}}]{Karska13}
{Karska}, A., {Herczeg}, G.~J., {van Dishoeck}, E.~F., {et~al.} 2013, \aap,
  552, A141

\bibitem[{{Kaufman} {et~al.}(1999){Kaufman}, {Wolfire}, {Hollenbach}, \&
  {Luhman}}]{Kaufman99}
{Kaufman}, M.~J., {Wolfire}, M.~G., {Hollenbach}, D.~J., \& {Luhman}, M.~L.
  1999, \apj, 527, 795

\bibitem[{{Kazandjian} {et~al.}(2012){Kazandjian}, {Meijerink}, {Pelupessy},
  {Israel}, \& {Spaans}}]{Kazandjian12}
{Kazandjian}, M.~V., {Meijerink}, R., {Pelupessy}, I., {Israel}, F.~P., \&
  {Spaans}, M. 2012, \aap, 542, A65

\bibitem[{{Kazandjian} {et~al.}(2015){Kazandjian}, {Meijerink}, {Pelupessy},
  {Israel}, \& {Spaans}}]{Kazandjian15}
{Kazandjian}, M.~V., {Meijerink}, R., {Pelupessy}, I., {Israel}, F.~P., \&
  {Spaans}, M. 2015, \aap, 574, A127

\bibitem[{{K{\"o}hler} {et~al.}(2014){K{\"o}hler}, {Habart}, {Arab},
  {Bernard-Salas}, {Ayasso}, {Abergel}, {Zavagno}, {Polehampton}, {van der
  Wiel}, {Naylor}, {Makiwa}, {Dassas}, {Joblin}, {Pilleri}, {Bern{\'e}},
  {Fuente}, {Gerin}, {Goicoechea}, \& {Teyssier}}]{Koehler14}
{K{\"o}hler}, M., {Habart}, E., {Arab}, H., {et~al.} 2014, \aap, 569, A109

\bibitem[{{Kramer} {et~al.}(2008){Kramer}, {Cubick}, {R{\"o}llig}, {Sun},
  {Yonekura}, {Aravena}, {Bensch}, {Bertoldi}, {Bronfman}, {Fujishita},
  {Fukui}, {Graf}, {Hitschfeld}, {Honingh}, {Ito}, {Jakob}, {Jacobs}, {Klein},
  {Koo}, {May}, {Miller}, {Miyamoto}, {Mizuno}, {Onishi}, {Park}, {Pineda},
  {Rabanus}, {Sasago}, {Schieder}, {Simon}, {Stutzki}, {Volgenau}, \&
  {Yamamoto}}]{Kramer08}
{Kramer}, C., {Cubick}, M., {R{\"o}llig}, M., {et~al.} 2008, \aap, 477, 547

\bibitem[{{Kurucz}(1993)}]{Kurucz93}
{Kurucz}, R.~L. 1993, VizieR Online Data Catalog, 6039

\bibitem[{{Laor} \& {Draine}(1993)}]{Laor93}
{Laor}, A. \& {Draine}, B.~T. 1993, \apj, 402, 441

\bibitem[{{Le} {et~al.}(2017){Le}, {Pak}, {Kaplan}, {Mace}, {Lee}, {Pavel},
  {Jeong}, {Oh}, {Lee}, {Chun}, {Yuk}, {Pyo}, {Hwang}, {Kim}, {Park}, {Sok Oh},
  {Yu}, {Park}, {Minh}, \& {Jaffe}}]{Le17}
{Le}, H.~A.~N., {Pak}, S., {Kaplan}, K., {et~al.} 2017, \apj, 841, 13

\bibitem[{{Le Bourlot} {et~al.}(2012){Le Bourlot}, {Le Petit}, {Pinto},
  {Roueff}, \& {Roy}}]{LeBourlot12}
{Le Bourlot}, J., {Le Petit}, F., {Pinto}, C., {Roueff}, E., \& {Roy}, F. 2012,
  \aap, 541, A76

\bibitem[{{Le Petit} {et~al.}(2006){Le Petit}, {Nehm{\'e}}, {Le Bourlot}, \&
  {Roueff}}]{LePetit06}
{Le Petit}, F., {Nehm{\'e}}, C., {Le Bourlot}, J., \& {Roueff}, E. 2006, \apjs,
  164, 506

\bibitem[{{Le Petit} {et~al.}(2004){Le Petit}, {Roueff}, \&
  {Herbst}}]{LePetit04}
{Le Petit}, F., {Roueff}, E., \& {Herbst}, E. 2004, \aap, 417, 993

\bibitem[{{Lemaire} {et~al.}(1996){Lemaire}, {Field}, {Gerin}, {Leach}, {Pineau
  des Forets}, {Rostas}, \& {Rouan}}]{Lemaire96}
{Lemaire}, J.~L., {Field}, D., {Gerin}, M., {et~al.} 1996, \aap, 308, 895

\bibitem[{{Lemaire} {et~al.}(1999){Lemaire}, {Field}, {Maillard}, {Pineau des
  For{\^e}ts}, {Falgarone}, {Pijpers}, {Gerin}, \& {Rostas}}]{Lemaire99}
{Lemaire}, J.~L., {Field}, D., {Maillard}, J.~P., {et~al.} 1999, \aap, 349, 253

\bibitem[{{Lis} \& {Schilke}(2003)}]{Lis03}
{Lis}, D.~C. \& {Schilke}, P. 2003, \apjl, 597, L145

\bibitem[{{Manoj} {et~al.}(2013){Manoj}, {Watson}, {Neufeld}, {Megeath},
  {Vavrek}, {Yu}, {Visser}, {Bergin}, {Fischer}, {Tobin}, {Stutz}, {Ali},
  {Wilson}, {Di Francesco}, {Osorio}, {Maret}, \& {Poteet}}]{Manoj13}
{Manoj}, P., {Watson}, D.~M., {Neufeld}, D.~A., {et~al.} 2013, \apj, 763, 83

\bibitem[{{Marconi} {et~al.}(1998){Marconi}, {Testi}, {Natta}, \&
  {Walmsley}}]{Marconi98}
{Marconi}, A., {Testi}, L., {Natta}, A., \& {Walmsley}, C.~M. 1998, \aap, 330,
  696

\bibitem[{{Martini} {et~al.}(1999){Martini}, {Sellgren}, \&
  {DePoy}}]{Martini99}
{Martini}, P., {Sellgren}, K., \& {DePoy}, D.~L. 1999, \apj, 526, 772

\bibitem[{{Martini} {et~al.}(1997){Martini}, {Sellgren}, \& {Hora}}]{Martini97}
{Martini}, P., {Sellgren}, K., \& {Hora}, J.~L. 1997, \apj, 484, 296

\bibitem[{{Mashian} {et~al.}(2015){Mashian}, {Sturm}, {Sternberg}, {Janssen},
  {Hailey-Dunsheath}, {Fischer}, {Contursi}, {Gonz{\'a}lez-Alfonso},
  {Graci{\'a}-Carpio}, {Poglitsch}, {Veilleux}, {Davies}, {Genzel}, {Lutz},
  {Tacconi}, {Verma}, {Wei{\ss}}, {Polisensky}, \& {Nikola}}]{Mashian15}
{Mashian}, N., {Sturm}, E., {Sternberg}, A., {et~al.} 2015, \apj, 802, 81

\bibitem[{{Mathis} {et~al.}(1983){Mathis}, {Mezger}, \& {Panagia}}]{Mathis83}
{Mathis}, J.~S., {Mezger}, P.~G., \& {Panagia}, N. 1983, \aap, 128, 212

\bibitem[{{Mathis} {et~al.}(1977){Mathis}, {Rumpl}, \& {Nordsieck}}]{Mathis77}
{Mathis}, J.~S., {Rumpl}, W., \& {Nordsieck}, K.~H. 1977, \apj, 217, 425

\bibitem[{{McCall} {et~al.}(1999){McCall}, {Geballe}, {Hinkle}, \&
  {Oka}}]{McCall99}
{McCall}, B.~J., {Geballe}, T.~R., {Hinkle}, K.~H., \& {Oka}, T. 1999, \apj,
  522, 338

\bibitem[{{Meixner} {et~al.}(1992){Meixner}, {Haas}, {Tielens}, {Erickson}, \&
  {Werner}}]{Meixner92}
{Meixner}, M., {Haas}, M.~R., {Tielens}, A.~G.~G.~M., {Erickson}, E.~F., \&
  {Werner}, M. 1992, \apj, 390, 499

\bibitem[{{Meixner} \& {Tielens}(1993)}]{Meixner93}
{Meixner}, M. \& {Tielens}, A.~G.~G.~M. 1993, \apj, 405, 216

\bibitem[{{Menten} {et~al.}(2007){Menten}, {Reid}, {Forbrich}, \&
  {Brunthaler}}]{Menten07}
{Menten}, K.~M., {Reid}, M.~J., {Forbrich}, J., \& {Brunthaler}, A. 2007, \aap,
  474, 515

\bibitem[{{Meyer} {et~al.}(1997){Meyer}, {Cardelli}, \& {Sofia}}]{Meyer97}
{Meyer}, D.~M., {Cardelli}, J.~A., \& {Sofia}, U.~J. 1997, \apjl, 490, L103

\bibitem[{{Meyer} {et~al.}(1998){Meyer}, {Jura}, \& {Cardelli}}]{Meyer98}
{Meyer}, D.~M., {Jura}, M., \& {Cardelli}, J.~A. 1998, \apj, 493, 222

\bibitem[{{Nagy} {et~al.}(2017){Nagy}, {Choi}, {Ossenkopf-Okada}, {van der
  Tak}, {Bergin}, {Gerin}, {Joblin}, {R{\"o}llig}, {Simon}, \&
  {Stutzki}}]{Nagy17}
{Nagy}, Z., {Choi}, Y., {Ossenkopf-Okada}, V., {et~al.} 2017, \aap, 599, A22

\bibitem[{{Nagy} {et~al.}(2012){Nagy}, {van der Tak}, {Fuller}, {Spaans}, \&
  {Plume}}]{Nagy12}
{Nagy}, Z., {van der Tak}, F.~F.~S., {Fuller}, G.~A., {Spaans}, M., \& {Plume},
  R. 2012, \aap, 542, A6

\bibitem[{{Nagy} {et~al.}(2013){Nagy}, {Van der Tak}, {Ossenkopf}, {Gerin}, {Le
  Petit}, {Le Bourlot}, {Black}, {Goicoechea}, {Joblin}, {R{\"o}llig}, \&
  {Bergin}}]{Nagy13}
{Nagy}, Z., {Van der Tak}, F.~F.~S., {Ossenkopf}, V., {et~al.} 2013, \aap, 550,
  A96

\bibitem[{{Okada} {et~al.}(2013){Okada}, {Pilleri}, {Bern{\'e}}, {Ossenkopf},
  {Fuente}, {Goicoechea}, {Joblin}, {Kramer}, {R{\"o}llig}, {Teyssier}, \& {van
  der Tak}}]{Okada13}
{Okada}, Y., {Pilleri}, P., {Bern{\'e}}, O., {et~al.} 2013, \aap, 553, A2

\bibitem[{{Oliveira} \& {H{\'e}brard}(2006)}]{Oliveira06}
{Oliveira}, C.~M. \& {H{\'e}brard}, G. 2006, \apj, 653, 345

\bibitem[{{Omodaka} {et~al.}(1994){Omodaka}, {Hayashi}, {Hasegawa}, \&
  {Hayashi}}]{Omodaka94}
{Omodaka}, T., {Hayashi}, M., {Hasegawa}, T., \& {Hayashi}, S.~S. 1994, \apj,
  430, 256

\bibitem[{{Ossenkopf} {et~al.}(2011){Ossenkopf}, {R{\"o}llig}, {Kramer},
  {Okada}, {Fuente}, {Akyilmaz Yabaci}, {Benz}, {Bern{\'e}}, {Boulanger},
  {Bruderer}, {Dedes}, {France}, {Gerin}, {Goicoechea}, {Gusdorf},
  {G{\"u}sten}, {Harris}, {Joblin}, {Klein}, {Latter}, {Le Petit}, {Lord},
  {Martin}, {Pilleri}, {Martin-Pintado}, {Mookerjea}, {Neufeld}, {Phillips},
  {Rizzo}, {Simon}, {Stutzki}, {van der Tak}, {Teyssier}, \&
  {Yorke}}]{Ossenkopf11}
{Ossenkopf}, V., {R{\"o}llig}, M., {Kramer}, C., {et~al.} 2011, in EAS
  Publications Series, Vol.~52, EAS Publications Series, ed. M.~{R{\"o}llig},
  R.~{Simon}, V.~{Ossenkopf}, \& J.~{Stutzki}, 181--186

\bibitem[{{Ossenkopf} {et~al.}(2013){Ossenkopf}, {R{\"o}llig}, {Neufeld},
  {Pilleri}, {Lis}, {Fuente}, {van der Tak}, \& {Bergin}}]{Ossenkopf13}
{Ossenkopf}, V., {R{\"o}llig}, M., {Neufeld}, D.~A., {et~al.} 2013, \aap, 550,
  A57

\bibitem[{{Padovani} {et~al.}(2013){Padovani}, {Hennebelle}, \&
  {Galli}}]{Padovani13}
{Padovani}, M., {Hennebelle}, P., \& {Galli}, D. 2013, \aap, 560, A114

\bibitem[{{Parikka} {et~al.}(2017){Parikka}, {Habart}, {Bernard-Salas},
  {Goicoechea}, {Abergel}, {Pilleri}, {Dartois}, {Joblin}, {Gerin}, \&
  {Godard}}]{Parikka17}
{Parikka}, A., {Habart}, E., {Bernard-Salas}, J., {et~al.} 2017, \aap, 599, A20

\bibitem[{{Parmar} {et~al.}(1991){Parmar}, {Lacy}, \& {Achtermann}}]{Parmar91}
{Parmar}, P.~S., {Lacy}, J.~H., \& {Achtermann}, J.~M. 1991, \apjl, 372, L25

\bibitem[{{P{\'e}rez-Beaupuits} {et~al.}(2010){P{\'e}rez-Beaupuits}, {Spaans},
  {Hogerheijde}, {G{\"u}sten}, {Baryshev}, \& {Boland}}]{Perez10}
{P{\'e}rez-Beaupuits}, J.~P., {Spaans}, M., {Hogerheijde}, M.~R., {et~al.}
  2010, \aap, 510, A87

\bibitem[{{Pilbratt} {et~al.}(2010){Pilbratt}, {Riedinger}, {Passvogel},
  {Crone}, {Doyle}, {Gageur}, {Heras}, {Jewell}, {Metcalfe}, {Ott}, \&
  {Schmidt}}]{Pilbratt10}
{Pilbratt}, G.~L., {Riedinger}, J.~R., {Passvogel}, T., {et~al.} 2010, \aap,
  518, L1

\bibitem[{{Pilleri} {et~al.}(2012){Pilleri}, {Montillaud}, {Bern{\'e}}, \&
  {Joblin}}]{Pilleri12}
{Pilleri}, P., {Montillaud}, J., {Bern{\'e}}, O., \& {Joblin}, C. 2012, \aap,
  542, A69

\bibitem[{{Rangwala} {et~al.}(2011){Rangwala}, {Maloney}, {Glenn}, {Wilson},
  {Rykala}, {Isaak}, {Baes}, {Bendo}, {Boselli}, {Bradford}, {Clements},
  {Cooray}, {Fulton}, {Imhof}, {Kamenetzky}, {Madden}, {Mentuch}, {Sacchi},
  {Sauvage}, {Schirm}, {Smith}, {Spinoglio}, \& {Wolfire}}]{Rangwala11}
{Rangwala}, N., {Maloney}, P.~R., {Glenn}, J., {et~al.} 2011, \apj, 743, 94

\bibitem[{{Roelfsema} {et~al.}(2012){Roelfsema}, {Helmich}, {Teyssier},
  {Ossenkopf}, {Morris}, {Olberg}, {Shipman}, {Risacher}, {Akyilmaz},
  {Assendorp}, {Avruch}, {Beintema}, {Biver}, {Boogert}, {Borys}, {Braine},
  {Caris}, {Caux}, {Cernicharo}, {Coeur-Joly}, {Comito}, {de Lange},
  {Delforge}, {Dieleman}, {Dubbeldam}, {de Graauw}, {Edwards}, {Fich},
  {Flederus}, {Gal}, {di Giorgio}, {Herpin}, {Higgins}, {Hoac}, {Huisman},
  {Jarchow}, {Jellema}, {de Jonge}, {Kester}, {Klein}, {Kooi}, {Kramer},
  {Laauwen}, {Larsson}, {Leinz}, {Lord}, {Lorenzani}, {Luinge}, {Marston},
  {Mart{\'{\i}}n-Pintado}, {McCoey}, {Melchior}, {Michalska}, {Moreno},
  {M{\"u}ller}, {Nowosielski}, {Okada}, {Orlea{\'n}ski}, {Phillips}, {Pearson},
  {Rabois}, {Ravera}, {Rector}, {Rengel}, {Sagawa}, {Salomons},
  {S{\'a}nchez-Su{\'a}rez}, {Schieder}, {Schl{\"o}der}, {Schm{\"u}lling},
  {Soldati}, {Stutzki}, {Thomas}, {Tielens}, {Vastel}, {Wildeman}, {Xie},
  {Xilouris}, {Wafelbakker}, {Whyborn}, {Zaal}, {Bell}, {Bjerkeli}, {De Beck},
  {Cavali{\'e}}, {Crockett}, {Hily-Blant}, {Kama}, {Kaminski}, {Lefl{\'o}ch},
  {Lombaert}, {de Luca}, {Makai}, {Marseille}, {Nagy}, {Pacheco}, {van der
  Wiel}, {Wang}, \& {Y{\i}ld{\i}z}}]{Roelfsema12}
{Roelfsema}, P.~R., {Helmich}, F.~P., {Teyssier}, D., {et~al.} 2012, \aap, 537,
  A17

\bibitem[{{Rogers} {et~al.}(1995){Rogers}, {Heyer}, \& {Dewdney}}]{Rogers95}
{Rogers}, C., {Heyer}, M.~H., \& {Dewdney}, P.~E. 1995, \apj, 442, 694

\bibitem[{{R{\"o}llig} {et~al.}(2007){R{\"o}llig}, {Abel}, {Bell}, {Bensch},
  {Black}, {Ferland}, {Jonkheid}, {Kamp}, {Kaufman}, {Le Bourlot}, {Le Petit},
  {Meijerink}, {Morata}, {Ossenkopf}, {Roueff}, {Shaw}, {Spaans}, {Sternberg},
  {Stutzki}, {Thi}, {van Dishoeck}, {van Hoof}, {Viti}, \&
  {Wolfire}}]{Rollig07}
{R{\"o}llig}, M., {Abel}, N.~P., {Bell}, T., {et~al.} 2007, \aap, 467, 187

\bibitem[{{R{\"o}llig} {et~al.}(2006){R{\"o}llig}, {Ossenkopf}, {Jeyakumar},
  {Stutzki}, \& {Sternberg}}]{Rollig06}
{R{\"o}llig}, M., {Ossenkopf}, V., {Jeyakumar}, S., {Stutzki}, J., \&
  {Sternberg}, A. 2006, \aap, 451, 917

\bibitem[{{R{\"o}llig} {et~al.}(2013){R{\"o}llig}, {Szczerba}, {Ossenkopf}, \&
  {Gl{\"u}ck}}]{Rollig13}
{R{\"o}llig}, M., {Szczerba}, R., {Ossenkopf}, V., \& {Gl{\"u}ck}, C. 2013,
  \aap, 549, A85

\bibitem[{{Rollins} \& {Rawlings}(2012)}]{Rollins12}
{Rollins}, R.~P. \& {Rawlings}, J.~M.~C. 2012, \mnras, 427, 2328

\bibitem[{{Rosenberg} {et~al.}(2015){Rosenberg}, {van der Werf}, {Aalto},
  {Armus}, {Charmandaris}, {D{\'{\i}}az-Santos}, {Evans}, {Fischer}, {Gao},
  {Gonz{\'a}lez-Alfonso}, {Greve}, {Harris}, {Henkel}, {Israel}, {Isaak},
  {Kramer}, {Meijerink}, {Naylor}, {Sanders}, {Smith}, {Spaans}, {Spinoglio},
  {Stacey}, {Veenendaal}, {Veilleux}, {Walter}, {Wei{\ss}}, {Wiedner}, {van der
  Wiel}, \& {Xilouris}}]{Rosenberg15}
{Rosenberg}, M.~J.~F., {van der Werf}, P.~P., {Aalto}, S., {et~al.} 2015, \apj,
  801, 72

\bibitem[{{Savage} \& {Sembach}(1996)}]{Savage96}
{Savage}, B.~D. \& {Sembach}, K.~R. 1996, \araa, 34, 279

\bibitem[{{Sellgren} {et~al.}(1992){Sellgren}, {Werner}, \&
  {Dinerstein}}]{Sellgren92}
{Sellgren}, K., {Werner}, M.~W., \& {Dinerstein}, H.~L. 1992, \apj, 400, 238

\bibitem[{{Sim{\'o}n-D{\'{\i}}az} {et~al.}(2006){Sim{\'o}n-D{\'{\i}}az},
  {Herrero}, {Esteban}, \& {Najarro}}]{SimonDiaz06}
{Sim{\'o}n-D{\'{\i}}az}, S., {Herrero}, A., {Esteban}, C., \& {Najarro}, F.
  2006, \aap, 448, 351

\bibitem[{{Sizun} {et~al.}(2010){Sizun}, {Bachellerie}, {Aguillon}, \&
  {Sidis}}]{Sizun10}
{Sizun}, M., {Bachellerie}, D., {Aguillon}, F., \& {Sidis}, V. 2010, Chemical
  Physics Letters, 498, 32

\bibitem[{{Smith} {et~al.}(2007){Smith}, {Armus}, {Dale}, {Roussel}, {Sheth},
  {Buckalew}, {Jarrett}, {Helou}, \& {Kennicutt}}]{Smith07}
{Smith}, J.~D.~T., {Armus}, L., {Dale}, D.~A., {et~al.} 2007, \pasp, 119, 1133

\bibitem[{{Sternberg} \& {Dalgarno}(1989)}]{Sternberg89}
{Sternberg}, A. \& {Dalgarno}, A. 1989, \apj, 338, 197

\bibitem[{{Stock} {et~al.}(2015){Stock}, {Wolfire}, {Peeters}, {Tielens},
  {Vandenbussche}, {Boersma}, \& {Cami}}]{Stock15}
{Stock}, D.~J., {Wolfire}, M.~G., {Peeters}, E., {et~al.} 2015, \aap, 579, A67

\bibitem[{{Stoerzer} {et~al.}(1995){Stoerzer}, {Stutzki}, \&
  {Sternberg}}]{Stoerzer95}
{Stoerzer}, H., {Stutzki}, J., \& {Sternberg}, A. 1995, \aap, 296, L9

\bibitem[{{St{\"o}rzer} \& {Hollenbach}(1998)}]{Stoerzer98}
{St{\"o}rzer}, H. \& {Hollenbach}, D. 1998, \apj, 495, 853

\bibitem[{{Stutzki} \& {Guesten}(1990)}]{Stutzki90}
{Stutzki}, J. \& {Guesten}, R. 1990, \apj, 356, 513

\bibitem[{{Stutzki} {et~al.}(1988){Stutzki}, {Stacey}, {Genzel}, {Harris},
  {Jaffe}, \& {Lugten}}]{Stutzki88}
{Stutzki}, J., {Stacey}, G.~J., {Genzel}, R., {et~al.} 1988, \apj, 332, 379

\bibitem[{{Tahani} {et~al.}(2016){Tahani}, {Plume}, {Bergin}, {Tolls},
  {Phillips}, {Caux}, {Cabrit}, {Goicoechea}, {Goldsmith}, {Johnstone}, {Lis},
  {Pagani}, {Menten}, {M{\"u}ller}, {Ossenkopf-Okada}, {Pearson}, \& {van der
  Tak}}]{Tahani16}
{Tahani}, K., {Plume}, R., {Bergin}, E.~A., {et~al.} 2016, \apj, 832, 12

\bibitem[{{Tauber} {et~al.}(1994){Tauber}, {Tielens}, {Meixner}, \&
  {Goldsmith}}]{Tauber94}
{Tauber}, J.~A., {Tielens}, A.~G.~G.~M., {Meixner}, M., \& {Goldsmith}, P.~F.
  1994, \apj, 422, 136

\bibitem[{{Tielens} \& {Hollenbach}(1985)}]{Tielens85}
{Tielens}, A.~G.~G.~M. \& {Hollenbach}, D. 1985, \apj, 291, 722

\bibitem[{{Tielens} {et~al.}(1993){Tielens}, {Meixner}, {van der Werf},
  {Bregman}, {Tauber}, {Stutzki}, \& {Rank}}]{Tielens93}
{Tielens}, A.~G.~G.~M., {Meixner}, M.~M., {van der Werf}, P.~P., {et~al.} 1993,
  Science, 262, 86

\bibitem[{{Usuda} {et~al.}(1996){Usuda}, {Sugai}, {Kawabata}, {Inoue},
  {Kataza}, \& {Tanaka}}]{Usuda96}
{Usuda}, T., {Sugai}, H., {Kawabata}, H., {et~al.} 1996, \apj, 464, 818

\bibitem[{{van den Ancker} {et~al.}(1997){van den Ancker}, {The}, {Tjin A
  Djie}, {Catala}, {de Winter}, {Blondel}, \& {Waters}}]{VanDenAncker97}
{van den Ancker}, M.~E., {The}, P.~S., {Tjin A Djie}, H.~R.~E., {et~al.} 1997,
  \aap, 324, L33

\bibitem[{{van der Tak} {et~al.}(2007){van der Tak}, {Black}, {Sch{\"o}ier},
  {Jansen}, \& {van Dishoeck}}]{VanDerTak07}
{van der Tak}, F.~F.~S., {Black}, J.~H., {Sch{\"o}ier}, F.~L., {Jansen}, D.~J.,
  \& {van Dishoeck}, E.~F. 2007, \aap, 468, 627

\bibitem[{{van der Werf} {et~al.}(1996){van der Werf}, {Stutzki}, {Sternberg},
  \& {Krabbe}}]{VanDerWerf96}
{van der Werf}, P.~P., {Stutzki}, J., {Sternberg}, A., \& {Krabbe}, A. 1996,
  \aap, 313, 633

\bibitem[{{van Leeuwen}(2007)}]{VanLeeuwen07}
{van Leeuwen}, F. 2007, \aap, 474, 653

\bibitem[{{Visser} {et~al.}(2012){Visser}, {Kristensen}, {Bruderer}, {van
  Dishoeck}, {Herczeg}, {Brinch}, {Doty}, {Harsono}, \& {Wolfire}}]{Visser12}
{Visser}, R., {Kristensen}, L.~E., {Bruderer}, S., {et~al.} 2012, \aap, 537,
  A55

\bibitem[{{Walmsley} {et~al.}(2000){Walmsley}, {Natta}, {Oliva}, \&
  {Testi}}]{Walmsley00}
{Walmsley}, C.~M., {Natta}, A., {Oliva}, E., \& {Testi}, L. 2000, \aap, 364,
  301

\bibitem[{{Weingartner} \& {Draine}(2001)}]{Weingartner01}
{Weingartner}, J.~C. \& {Draine}, B.~T. 2001, \apjs, 134, 263

\bibitem[{{Werner} {et~al.}(2004){Werner}, {Roellig}, {Low}, {Rieke}, {Rieke},
  {Hoffmann}, {Young}, {Houck}, {Brandl}, {Fazio}, {Hora}, {Gehrz}, {Helou},
  {Soifer}, {Stauffer}, {Keene}, {Eisenhardt}, {Gallagher}, {Gautier}, {Irace},
  {Lawrence}, {Simmons}, {Van Cleve}, {Jura}, {Wright}, \&
  {Cruikshank}}]{Werner04a}
{Werner}, M.~W., {Roellig}, T.~L., {Low}, F.~J., {et~al.} 2004, \apjs, 154, 1

\bibitem[{{Witt} {et~al.}(2006){Witt}, {Gordon}, {Vijh}, {Sell}, {Smith}, \&
  {Xie}}]{Witt06}
{Witt}, A.~N., {Gordon}, K.~D., {Vijh}, U.~P., {et~al.} 2006, \apj, 636, 303

\bibitem[{{Wu} {et~al.}(2018){Wu}, {Bron}, {Onaka}, {Le Petit}, {Galliano},
  {Languignon}, {Nakamura}, \& {Okada}}]{Wu18}
{Wu}, R., {Bron}, E., {Onaka}, T., {et~al.} 2018, \aap

\bibitem[{{Wu} {et~al.}(2015){Wu}, {Madden}, {Galliano}, {Wilson},
  {Kamenetzky}, {Lee}, {Schirm}, {Hony}, {Lebouteiller}, {Spinoglio},
  {Cormier}, {Glenn}, {Maloney}, {Pereira-Santaella}, {R{\'e}my-Ruyer}, {Baes},
  {Boselli}, {Bournaud}, {De Looze}, {Hughes}, {Panuzzo}, \& {Rangwala}}]{Wu15}
{Wu}, R., {Madden}, S.~C., {Galliano}, F., {et~al.} 2015, \aap, 575, A88

\bibitem[{{Wyrowski} {et~al.}(1997){Wyrowski}, {Schilke}, {Hofner}, \&
  {Walmsley}}]{Wyrowski97}
{Wyrowski}, F., {Schilke}, P., {Hofner}, P., \& {Walmsley}, C.~M. 1997, \apjl,
  487, L171

\bibitem[{{Yang} {et~al.}(2010){Yang}, {Stancil}, {Balakrishnan}, \&
  {Forrey}}]{Yang10}
{Yang}, B., {Stancil}, P.~C., {Balakrishnan}, N., \& {Forrey}, R.~C. 2010,
  \apj, 718, 1062

\bibitem[{{Y{\i}ld{\i}z} {et~al.}(2010){Y{\i}ld{\i}z}, {van Dishoeck},
  {Kristensen}, {Visser}, {J{\o}rgensen}, {Herczeg}, {van Kempen},
  {Hogerheijde}, {Doty}, {Benz}, {Bruderer}, {Wampfler}, {Deul}, {Bachiller},
  {Baudry}, {Benedettini}, {Bergin}, {Bjerkeli}, {Blake}, {Bontemps}, {Braine},
  {Caselli}, {Cernicharo}, {Codella}, {Daniel}, {di Giorgio}, {Dominik},
  {Encrenaz}, {Fich}, {Fuente}, {Giannini}, {Goicoechea}, {de Graauw},
  {Helmich}, {Herpin}, {Jacq}, {Johnstone}, {Larsson}, {Lis}, {Liseau}, {Liu},
  {Marseille}, {McCoey}, {Melnick}, {Neufeld}, {Nisini}, {Olberg}, {Parise},
  {Pearson}, {Plume}, {Risacher}, {Santiago-Garc{\'{\i}}a}, {Saraceno},
  {Shipman}, {Tafalla}, {Tielens}, {van der Tak}, {Wyrowski}, {Dieleman},
  {Jellema}, {Ossenkopf}, {Schieder}, \& {Stutzki}}]{Yildiz10}
{Y{\i}ld{\i}z}, U.~A., {van Dishoeck}, E.~F., {Kristensen}, L.~E., {et~al.}
  2010, \aap, 521, L40

\bibitem[{{Young Owl} {et~al.}(2000){Young Owl}, {Meixner}, {Wolfire},
  {Tielens}, \& {Tauber}}]{YoungOwl00}
{Young Owl}, R.~C., {Meixner}, M.~M., {Wolfire}, M., {Tielens}, A.~G.~G.~M., \&
  {Tauber}, J. 2000, \apj, 540, 886

\end{thebibliography}
\bibliographystyle{aa}

\end{document}